\documentclass[a4paper,11pt]{article}
\pdfoutput=1 

\usepackage{jheppub} 

\usepackage[T1]{fontenc} 

\usepackage{axodraw4j}
\usepackage{color}
\usepackage{latexsym}
\usepackage{mathrsfs}
\usepackage{amssymb}
\usepackage{amsmath}
\usepackage{amsfonts, epsf,epsfig,color}
\usepackage{graphicx}
\usepackage{tensor}
\usepackage{manfnt}
\usepackage[all]{xy}

\newcommand{\R}{\mathbb{R}}

\newcommand{\p}{\partial}

\newcommand{\rd}{\, \mathrm{d}}

\newcommand{\be}{\begin{equation}\label}
\newcommand{\ee}{\end{equation}}
\newcommand{\bea}{\begin{eqnarray}\label}
\newcommand{\eea}{\end{eqnarray}}

\title{\boldmath Lattice Gerbe Theory}

\author[a]{Arthur E. Lipstein}
\author[b]{and Ronald A. Reid-Edwards}

\affiliation[a]{Mathematical Institute, Andrew Wiles Building,\\Woodstock Road, Oxford, OX2 6CG, UK}
\affiliation[b]{Department of Physics and Mathematics, University of Hull,\\ Cottingham Road, Hull, HU6 7RX, UK}

\abstract{We formulate the theory of a 2-form gauge field on a Euclidean spacetime lattice. In this approach, the fundamental degrees of freedom live on the faces of the lattice, and the action can be constructed from the sum over Wilson surfaces associated with each fundamental cube of the lattice. If we take the gauge group to be $U(1)$, the theory reduces to the well-known abelian gerbe theory in the continuum limit. We also explore a very simple and natural non-abelian generalization with gauge group $U(N) \times U(N)$. In the classical continuum limit, it reduces to a free theory, but at non-zero lattice spacing it is an interacting theory which gives rise to $U(N)$ Yang-Mills theory upon dimensional reduction. Formulating the theory on a lattice has several other advantages. In particular, it is possible to compute many observables, such as the expectation value of Wilson surfaces, analytically at strong coupling and numerically for any value of the coupling.}

\dedicated{For Keir.}

\begin{document} 
\maketitle
\flushbottom

\section{Introduction}

Understanding whether or not it is possible to formulate an interacting theory of gerbes is an important question in theoretical physics.  In the context of string theory, such a theory is needed to describe objects known as M5-branes. In particular, the low-energy theory of a stack of coincident M5-branes should be a six-dimensional superconformal theory containing a self-dual gerbe, five scalars, and eight fermions, which can be encoded in a $(2,0)$ tensor multiplet. For the case of a single M5-brane, the theory is abelian and was constructed in \cite{Perry:1996mk,Pasti:1997gx}. For two or more coincident M5-branes, the conjectured theory is non-abelian and is very challenging to formulate. One major obstacle is that it is unclear how to define a nonabelian 2-form gauge field. Another challenge is that since the gerbe is self-dual, the theory cannot be effectively probed using perturbative methods. Furthermore, it is not possible to construct a conventional interacting Lagrangian which respects (2,0) superconformal symmetry using only $(2,0)$ tensor multiplets, as one can easily verify using dimensional analysis. On the other hand, it is conceivable that one can construct a Lagrangian which does not exhibit all of the expected symmetries classically but gives rise to the correct theory quantum mechanically. Although quantum corrections usually break classical symmetries via anomalies, there are examples where symmetries actually become enhanced in the quantum theory. Indeed, this is the case for the M2-brane theory, whose supersymmetry becomes enhanced at strong coupling \cite{Aharony:2008ug}.

If there is any hope of writing down a classical Lagrangian which describes the $(2,0)$ theory quantum mechanically, it appears that we must relax some assumption about the properties that this classical theory is expected to have. There have been many attempts to make progress in understanding this theory and we shall not atempt to provide a comprehensive survey of the many approaches currently under investgation. A somewhat selective example is \cite{ArkaniHamed:2001ie,Lambert:2010wm,Douglas:2010iu,Lambert:2010iw,Ho:2011ni,Chu:2012um,Samtleben:2012mi,Saemann:2012uq,Bonetti:2012st,Kim:2012tr,Samtleben:2012fb,Saemann:2013pca}. One approach is to dimensionally reduce the theory to five dimensions. In particular, it has been conjectured that the $(2,0)$ theory compactified on a circle is equivalent to five-dimensional maximal super-Yang-Mills theory once non-perturbative effects are taken into account \cite{Douglas:2010iu,Lambert:2010iw}. Although this conjecture has passed some tests \cite{Tachikawa:2011ch,Kim:2011mv,Young:2011aa,Kim:2012ava,Kallen:2012zn,Kim:2012qf}, it is unclear whether five-dimensional super Yang-Mills is UV finite \cite{Bern:2012di,Papageorgakis:2014dma}, which is an implication of the conjecture. Regardless of the ultimate fate of this line of inquiry, considering dimensional reductions of the $(2,0)$ theory to five or fewer dimensions has provided a great deal of insight, particularly into electric-magnetic duality of four-dimensional supersymmetric theories \cite{Gaiotto:2009we}. One major motivation for studying electric-magnetic duality of such theories is that  it may ultimately provide insight into the dynamics of non-supersymmetric QCD. Indeed, Mandelstam \cite{Mandelstam:1974pi}, Polyakov \cite{Polyakov:1975rs,Polyakov:1976fu}, and 't Hooft \cite{'tHooft:1977hy,'tHooft:1981ht} suggested long ago that the phenomenon of confinement in QCD is equivalent to monopole condensation, and this mechanism was later demonstrated by Seiberg and Witten for non-abelian Yang-Mills theories with $\mathcal{N}=2$ supersymmetry \cite{Seiberg:1994rs}.

Our approach is not to attempt to construct a theory of gerbes on a conventional space-time, but instead to attempt to construct such a theory on a lattice. The justification for this approach is ultimately that one may make concrete progress, although an ulterior motivation is the idea that the correct model of space-time required to formulate a theory of interacting gerbes may not be a smooth manifold but an, as yet unidentified, structure that reduces to a conventional manifold in some limit. In the absence of a concrete proposal for such a structure, we speculate that it is possible that a lattice approximation retains key features that the continuum approximation lacks that are crucial to the successful construction of certain theories. We are not suggesting that the space-time of M-theory is a lattice; however, we do feel that some of the ideas of lattice gauge theory may provide novel inspiration for tackling problems in string theory that have thus far remained impervious to our best efforts.

In this paper, we will construct a theory of gerbes on a Euclidean hypercubic lattice in $D \geq 3$ dimensions. For simplicity, we will not consider self-duality or supersymmetry, although we beleive it may be possible to incorporate these properties into our construction. The analogous formulation for Yang-Mills theory was pioneered by Kenneth Wilson \cite{Wilson:1974sk}. In Wilson's formulation, the basic degrees of freedom can be taken to be unitary matrices which live on the links of the lattice and the action is obtained by summing over the Wilson loop associated with each fundamental square of the lattice (known as a plaquette). As the lattice spacing goes to zero, this action reduces to the continuum action of Yang-Mills theory. On the other hand, at non-zero lattice spacing, the quantum theory is completely well-defined and it is possible to compute many physical observables at strong coupling both analytically and numerically. There are many good references on lattice Yang-Mills, for example \cite{Creutz, Rothe}. 

Wilson's formulation of non-abelian Maxwell theory can be naturally extended to gerbes. Since the gerbe correponds to a 2-form gauge field, the basic degrees of freedom are associated with the faces of the lattice and the gauge transformations are associated with the links\footnote{The gauge transformations themselves are defined up to gauge transformations which are associated with the lattice points.}. Furthermore, the natural gauge-invariant observables are closed Wilson surfaces constructed from these faces, and the action is simply given by summing over the Wilson surface associated with each fundamental cube, which we call a \emph{cubet}. 

If the face variables are elements of $U(1)$, then the theory reduces to the well-known abelian gerbe theory in the continuum limit  \footnote{Abelian p-forms on the lattice were previously considered in \cite{Frohlich:1982gf,Omero:1982hp}.}. Furthermore, this constuction has a very natural non-abelian generalization. Recall that in lattice Yang-Mills, the link variables are unitary matrices whose indices are associated with the endpoints of the links. This suggests that in non-abelian lattice gerbe theory, the indices of the face variables should be associated with the \emph{edges} of the faces. Hence, it is natural to take the face variables to be four-index objects which transform in some representation of $U(N)\times U(N)$, although more general gauge groups may also be possible \footnote{Non-abelian face variables with four color indices were first proposed in \cite{Nepomechie:1982rb}. Subsequently, \cite{Orland:1982fv,Orland:1984pt,Orland:1984bi} considered nonabelian face variables with eight color indices coupled to four-dimensional lattice QCD.  Dimensional reduction and strong coupling of gebres on a lattice have also been considered independently in \cite{Rey:2010uz}. We learnt of those references after this paper was completed and we are grateful to the authors of those papers for bringing them to our attention.}. Remarkably, this theory gives rise to $U(N)$ Yang-Mills theory after dimensional reduction. On the other hand, if we take the continuum limit of the classical theory before dimensional reduction, we obtain a non-interacting theory. Hence, we find that dimensional reduction does not commute with taking the naive continuum limit of the classical theory.

Although the classical non-abelian lattice gerbe theory becomes non-interacting in the continuum limit, the situation may change in the quantum theory and we will explore this possibility elsewhere. At non-zero lattice spacing, the lattice gerbe theory is interacting and it is possible to compute surface operators both analytically and numerically by adapting techniques from lattice Yang-Mills. Using these techniques, we demonstrate that Wilson surfaces in three-dimensional lattice gerbe theory obey a volume law for all values of the coupling. By contrast, Wilson surfaces in the six-dimensional lattice gerbe obey an \emph{area} law at weak coupling and a \emph{volume} law at strong coupling. This suggests the possibility of a phase transition in the six-dimensional theory and goes to the heart of understanding the continuum limit of the quantum theory. Gerbes naturally couple to strings and so, if it is possible to incorporate dynamical strings into the lattice gerbe theory, these results suggest that they will become confined at strong coupling. The phase structure of abelian 2-form gauge fields and string confinement were previously studied in \cite{Frohlich:1982gf,Orland:1981ku,Orland:1994qt,Quevedo:1996uu,Polyakov:1996nc,Diamantini:1996vf}.

This paper is organized as follows. In section \ref{review}, we review some basics of lattice Yang-Mills theory. In section \ref{abelian}, we describe the abelian lattice gerbe theory. In section \ref{nonabelian}, we describe a non-abelian generalization of lattice gerbe theory, which appears to be very simple and natural. In section \ref{continuum}, we expand the non-abelian theory to cubic order in the fields and describe the classical continuum limit. In section \ref{reduction}, we describe dimensional reduction and demonstrate that it gives rise to $U(N)$ Yang-Mills theory. In section \ref{analytics}, we present various analytical results for Wilson surfaces in both the abelian and non-abelian gerbe theories.  In section \ref{numerics}, we demonstrate Monte Carlo techniques for numerically computing observables in the abelian lattice gerbe theory. Finally, in section \ref{conclusion}, we describe conclusions and future directions.        

\section{Review of Lattice Yang-Mills Theory} \label{review}

\begin{figure}[h] 
       \centering
       \includegraphics[width=3.6in]{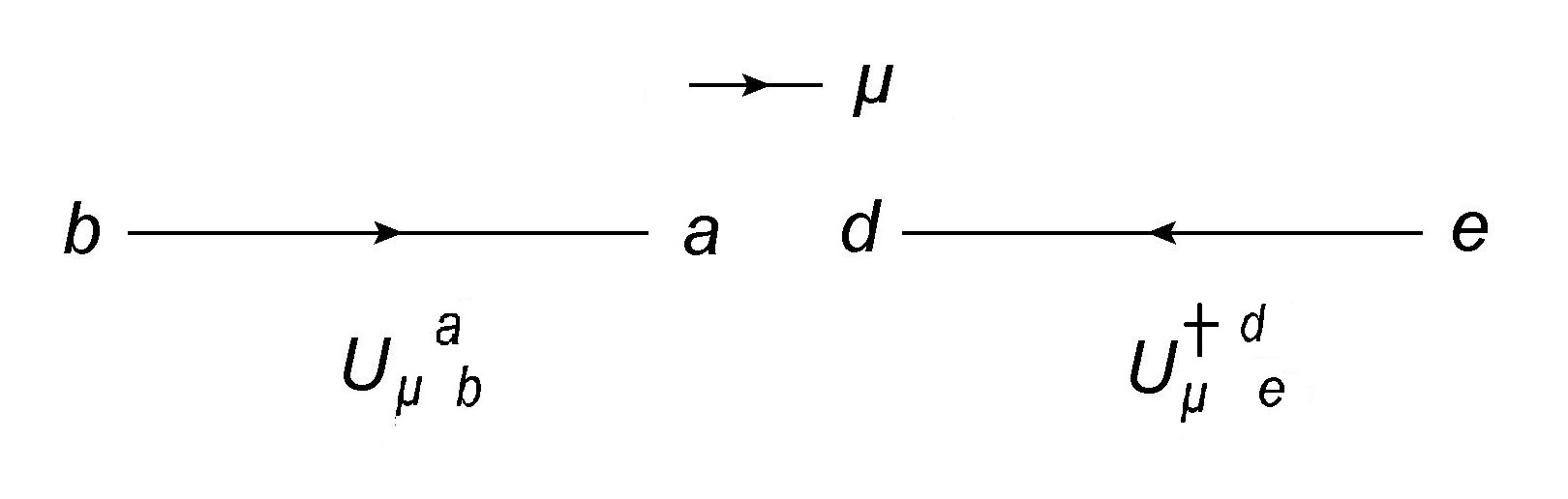}  
    \caption{A link variable and its conjugate.}
    \label{fig:link1}
    \end{figure}

\begin{figure}[htbp] 
       \centering
       \includegraphics[width=2.4in]{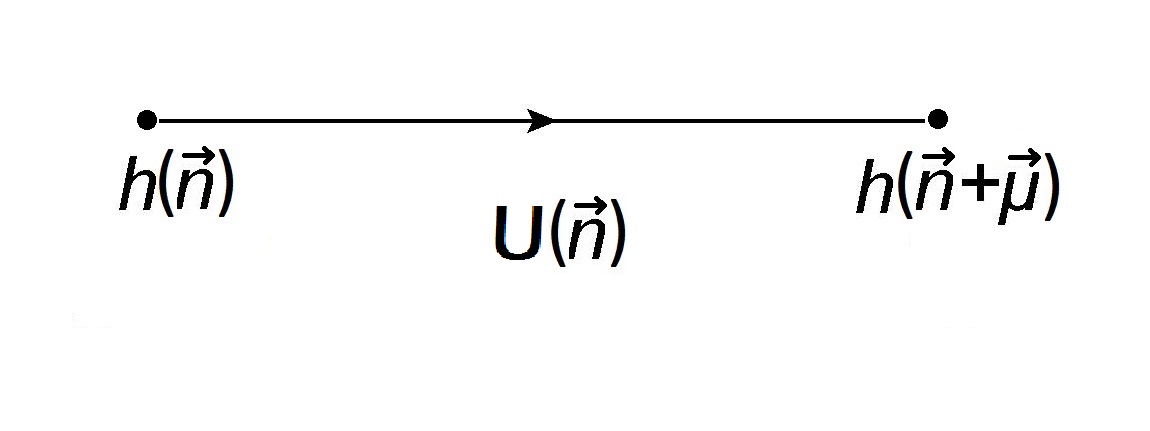}  
    \caption{Gauge transformation of Wilson line residing on a link.}
    \label{edgeg}
    \end{figure}

In this section, we will review some basic ideas of lattice Yang-Mills theory, which we will make use of in this paper. We cannot do the subject justice here, but there are many good textbooks to which the interested reader may turn, see for example \cite{Creutz,Rothe}. The theory lives on a Euclidean hypercubic lattice. The basic degrees
of freedom are matrices which live on the links of the lattice. We
will take these matrices to be in the group $U(N)$. Physically, these
link variables correspond to Wilson lines along the links and therefore
have an orientation. It is therefore natural to assign arrows to the
link variables, as depicted in Figure \ref{fig:link1}.
In particular, if we change the orientation of an arrow, relative to the coordinate axis, the variable $U$ is replaced by its adjoint $U^{\dagger}$. Under a $U(N)$ gauge transformation,
the link variables transform as follows: 
$$
U^{a}{}_{b}(\vec{n})\rightarrow h^{\dagger a}{}_{a'}(\vec{n}+\vec{\mu})\;U^{a'}{}_{b'}(\vec{n})\;h^{b'}{}_{b}(\vec{n})
$$
where\footnote{Where $\vec{\mu}$ is a lattice vector of length $\alpha$.}
$$
h^{\dagger a}{}_{a'}({\vec {n}}+\vec{\mu})=\exp\left(-i\alpha\varphi^{a}{}_{a'}({\vec {n}}+\vec{\mu})\right),	\qquad	h^{b'}{}_{b}({\vec {n}})=\exp\left(i\alpha\varphi^{b'}{}_{b}({\vec{n}})\right),
$$
and $\alpha$ is the lattice spacing and $\varphi^a{}_{a'}(\vec{n})$ is a scalar field located at the site $\vec{n}$. The raised and lowered indices are in the $R$ and $\bar{R}$ representation of $U(N)$ respectively. In the absence of coupling to matter, all non-singlet representations are equivalent to the adjoint representation in the continuum limit. We shall therefore  take $R$ to correspond to the fundamental representation from now on. Similar considerations will apply to lattice gerbe theory. Note that the gauge transformations
are associated with the vertices of the link, as depicted in Figure \ref{edgeg}.

\begin{figure}[htbp] 
       \centering
       \includegraphics[width=3.5in]{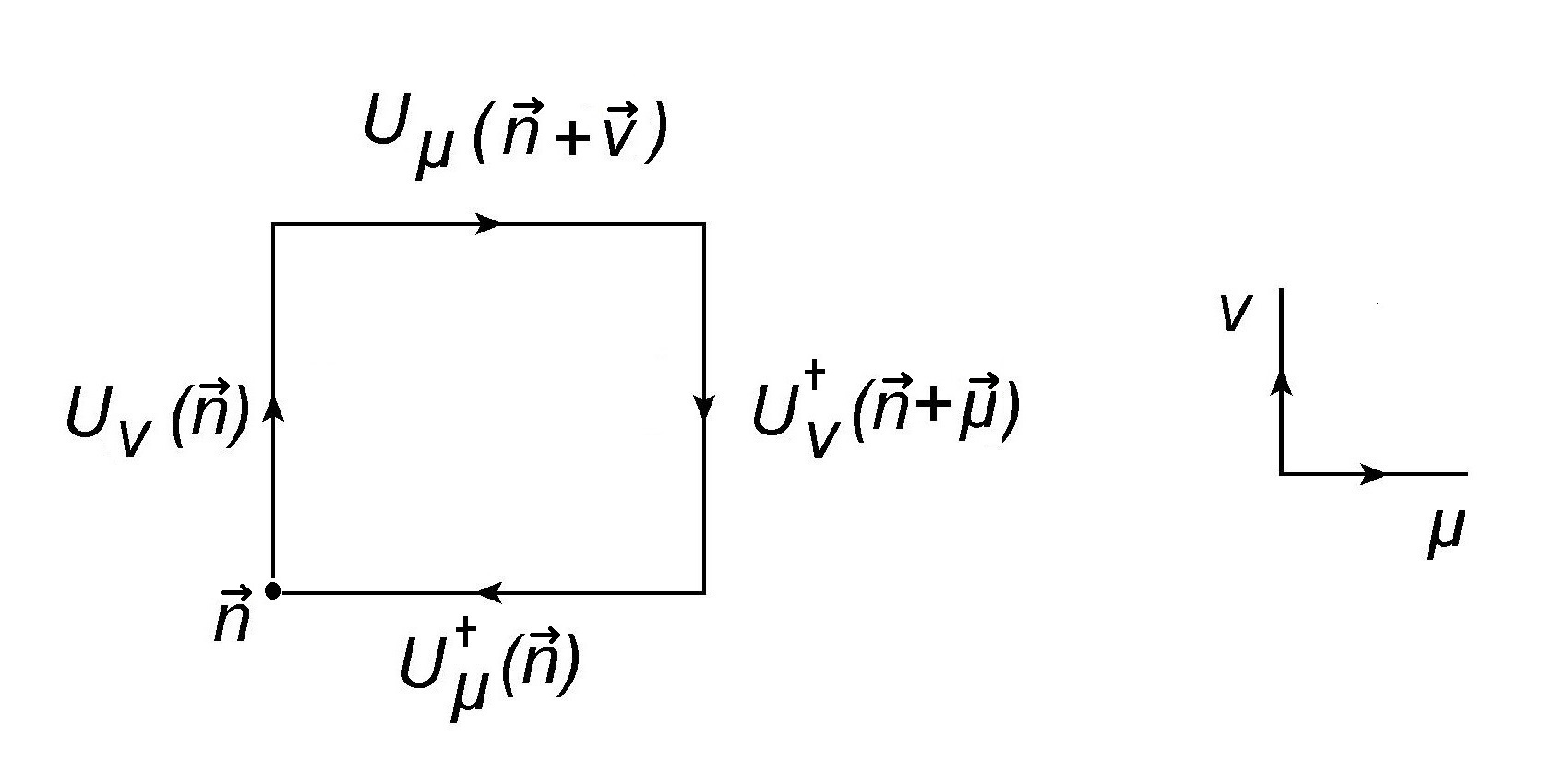}  
    \caption{Wilson loop associated with a plaquette.}
    \label{plaquettewl}
    \end{figure}

Now consider a set of link variables which form a closed loop. If
we take the product of the link variables and sum over repeated indices,
we obtain a gauge invariant observable known as a Wilson loop. Wilson
loops are the basic observables of lattice Yang-Mills theory. In particular,
the action for lattice Yang-Mills comes from adding up the Wilson
surfaces associated with each fundamental square, or plaquette. To
see this, let us compute the Wilson loop associated with the plaquette
in Figure  \ref{plaquettewl}. The link variables can be parameterized as follows: 

\[
U_{\nu}\left(\vec{n}\right)=e^{iA_{\nu}\left(\vec{n}\right)\alpha},\,\,\, U_{\mu}\left(\vec{n}+\vec{\nu}\right)=e^{iA_{\mu}\left(\vec{n}+\vec{\nu}\right)\alpha},\,\,\, U_{\nu}^{\dagger}\left(\vec{n}+\vec{\mu}\right)=e^{-iA_{\nu}\left(\vec{n}+\vec{\mu}\right)\alpha},\,\,\, U_{\mu}^{\dagger}\left(\vec{n}\right)=e^{-iA_{\mu}\left(\vec{n}\right)\alpha},
\]
where $\alpha$ is the lattice spacing, $A_{\mu}$ is a Lie-algebra
valued 1-form, and $\vec{\mu}$ and $\vec{\nu}$ are lattice vectors of length $\alpha$. The Wilson loop is then given by

\[
W_{\mu\nu}(\vec{n})=Tr\left[U_{\nu}\left(\vec{n}\right)U_{\mu}\left(\vec{n}+\vec{\nu}\right)U_{\nu}^{\dagger}\left(\vec{n}+\vec{\mu}\right)U_{\mu}^{\dagger}\left(\vec{n}\right)\right].
\]
After a bit of algebra, one finds that 
\begin{equation}
W_{\mu\nu}\left(\vec{n}\right)=Tr\left[\exp\left(-i\alpha^{2}F_{\mu\nu}(\vec{n})+\mathcal{O}\left(\alpha^{4}\right)\right)\right]
\label{wf}
\end{equation}
where
\[
F_{\mu\nu}=\Delta_{\mu}A_{\nu}-\Delta_{\nu}A_{\mu}-i\left[A_{\mu},A_{\nu}\right]
\]
and we have introduced the lattice derivative
\begin{equation}
\Delta_{\mu}f(\vec{n})=\frac{f\left(\vec{n}+\vec{\mu}\right)-f(\vec{n})}{\alpha}.
\label{derivative}
\end{equation}

The action of the lattice gauge theory is given by
\begin{equation}
S=\beta\sum_{{\bf n}}\sum_{\mu<\nu}\left(1-\frac{1}{N}\Re W_{\mu\nu}(\vec{n})\right)\label{eq:lgaction}
\end{equation}
where $\beta$ is an arbitrary real parameter and $\Re W_{\mu\nu}$ denotes the real part of $W_{\mu\nu}$. From the point of view
of statistical mechanics, $\beta$ can be thought of as the inverse
temperature. We will see shortly that it can also be thought of as
the inverse Yang-Mills coupling. In particular, using (\ref{wf}), we see that in the classical continuum
limit, the action reduces to that of $U(N)$ Yang-Mills theory in
four dimensions:

\[
\lim_{\alpha\rightarrow0}S=\frac{1}{4g^{2}}\int \rd^{4}x\;Tr\left(F_{\mu\nu}F^{\mu\nu}\right)
\]
where we identified the $\beta=N/2g^{2}$, neglected terms of
$\mathcal{O}\left(\alpha^{6}\right)$, and taken 
\[
\sum_{\vec{n}}\alpha^{4}\rightarrow\int \rd^{4}x.
\] 
By classical continuum limit,
we mean that we neglect renormalization due to quantum effects as
we take the lattice spacing to zero. Although lattice Yang-Mills
naturally lives in four dimensions, it can be defined in $D\geq 2$ by absorbing
powers of the lattice spacing into the definition of $\beta$.

The expectation value of an observable $\mathcal{O}$ which is a function of
link variables $U$ is defined by

\begin{equation}
\left\langle \mathcal{O}[U]\right\rangle =\frac{\int\mathcal{D}U\;\mathcal{O}[U]\;e^{-S[U]}}{\int\mathcal{D}U\;e^{-S[U]}}\label{eq:expectationvalue},
\end{equation}
where the integral is over all link variables, and $S[U]$ is the action
(\ref{eq:lgaction}). Let $\mathcal{O}[U]$ be an arbitrary Wilson
loop. At nonzero lattice spacing and strong coupling, it is not difficult to show that the expectation value of the Wilson loop obeys
an area law. We will describe this schematically. For a more detailed
explanation, see for example chapter 10 of \cite{Creutz}.
In particular, when $\beta\ll1$, we can expand the exponentials in (\ref{eq:expectationvalue}) and integrate them against the Wilson loop
using standard results from group theory. In particular, for a given
link variable, the relavant integrals are

\[
\int \rd U\;U_{b}^{a}=0,\,\,\,\int \rd U\;U_{b}^{a}U^{\dagger}{}_{d}^{c}=\frac{1}{N}\delta_{d}^{a}\delta_{b}^{c}.
\]
Using these integrals, one finds that the leading contribution
comes from tiling the interior of the Wilson loop with plaquettes and integrating
each of these plaquettes against a plaquette with the opposite orientation coming from the action. Since each plaquette in
the action is accompanied by a factor of $\beta$, we see that to
leading order, $\left\langle \mathcal{O}\right\rangle \propto\beta^{A}$, where $A$ is the number of plaquettes which tile the interior of the Wilson loop. Hence, $\ln \left\langle \mathcal{O}\right\rangle \sim A\ln\beta$,
which implies an area law at strong coupling.

\section{Abelian Lattice Gerbe Theory} \label{abelian}

Gerbes play a ubiquitous role in string theory. We briefly review the salient features from a differential-geometric perspective. Readable introductions for a physics audience include \cite{Hitchin:1999fh}, whose approach we follow here. To appreciate the continuum definition of a gerbe, it is helpful to recall the definition of a connection on a line bundle over a manifold $M$. We assume that we have an atlas of patches $\{U_{\alpha}\}$ for $M$. The curvature $F$ of the line bundle is a globally defined, closed, two-form so that $F|_{U_{\alpha}}=F|_{U_{\beta}}$. Upon restriction to a patch $U_{\alpha}$, the Poincar\'e lemma tells us that the $F$ is exact and we can write $F|_{U_{\alpha}}=\rd A_{\alpha}$. On the overlap of two patches $U_{\alpha}\cap U_{\beta}$, we have
$$
A_{\alpha}-A_{\beta}=\rd \phi_{\alpha\beta},
$$
where $f_{\alpha\beta}=\exp(i\phi_{\alpha\beta})$ is transition function defining the line bundle $f_{\alpha\beta}:U_{\alpha}\cap U_{\beta}\rightarrow S^1$.

The appropriate differential object to describe a gerbe is a globally defined, closed, three-form $H$. Restricting to a single patch $U_{\alpha}$, the gerbe is exact and we can write $H|_{U_{\alpha}}=\rd b_{\alpha}$. On the overlap of two patches $U_{\alpha}\cap U_{\beta}$, the $b$-field on each patch is related by a one-form $\lambda$
$$
b_{\alpha}-b_{\beta}=\rd \lambda_{\alpha\beta}.
$$
On the overlap of three patches
$$
\lambda_{\alpha\beta}+\lambda_{\beta\gamma}+\lambda_{\gamma\alpha}=\rd h_{\alpha\beta\gamma}.
$$
Thus, on the overlap of three patches $h:U_{\alpha}\cap U_{\beta}\cap U_{\gamma}\rightarrow \R$. If $[H]/2\pi\in H^3(M;\R)$, we can then relate this to the topological definition of a gerbe, in terms of a cocycle $g_{\alpha\beta\gamma}:U_{\alpha}\cap U_{\beta}\cap U_{\gamma}\rightarrow S^1$, by $g_{\alpha\beta\gamma}=\exp(ih_{\alpha\beta\gamma})$.

A crucial point is that, unlike a fibre bundle, a gerbe is \emph{not} a manifold and one of the main contentions of this paper is that, in order to generalise the above construction to non-abelian groups, it may be necessary to consider a more general base than a conventional manifold. Simply introducing exotic differential objects on a conventional manifold may not be sufficient. For much of this paper the lattice theories we construct will give topologically trivial continuum limits, we shall however see a glimpse of the structure outlined above in the gauge transformations of the theory.

\subsection{Wilson Surfaces}

\begin{figure}[here] 
       \centering
       \includegraphics[width=4in]{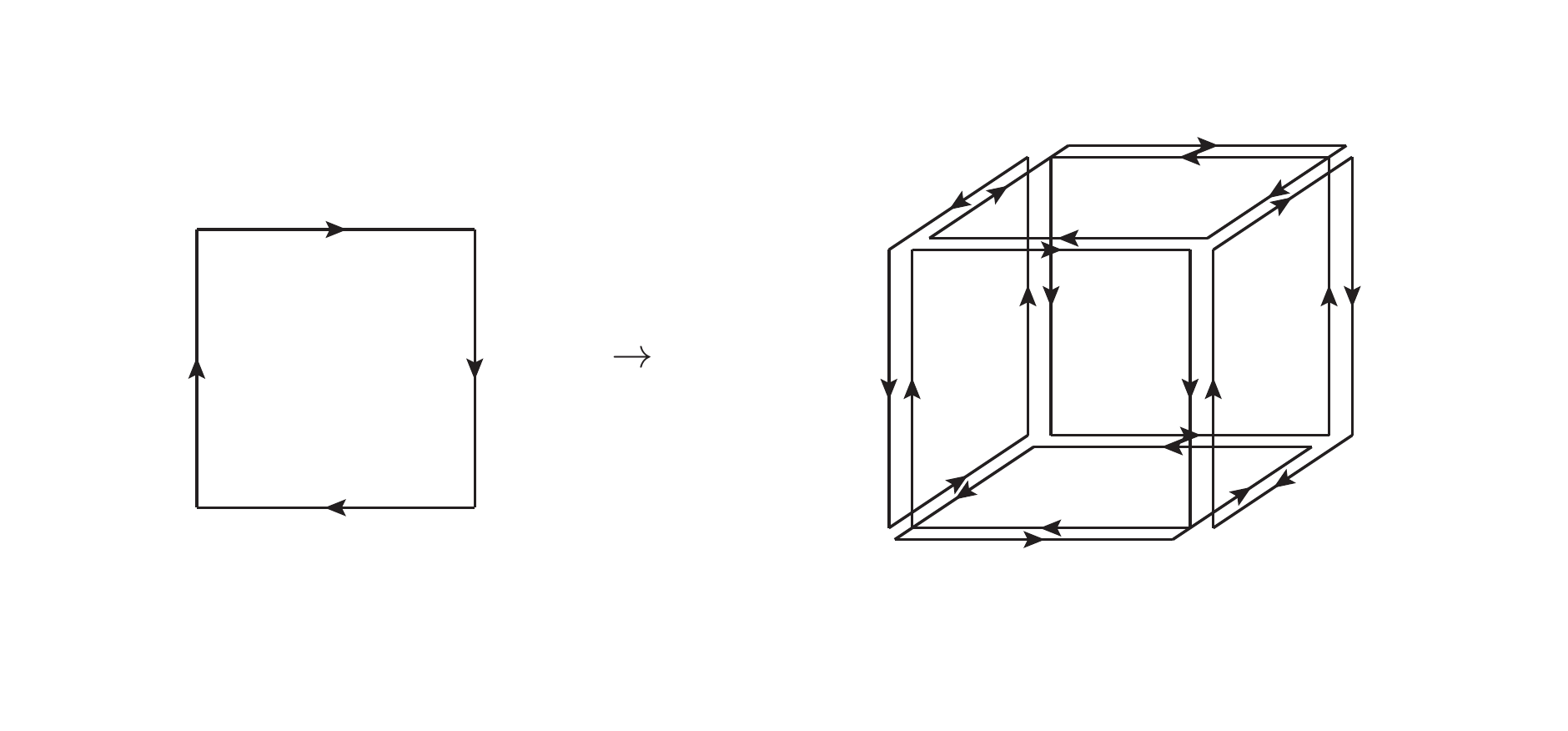}  
\caption{Fundamental cube constructed from face variables.}
\label{fig:Fig1}
    \end{figure}

In lattice Yang-Mills, the fundamental object is the Wilson loop, and the action for the theory can be defined in terms of the Wilson loops associated with each fundamental square of the lattice, which are referred to as plaquettes. In the lattice gerbe theory, the fundamental object is a Wilson surface, and the action for the theory can be defined in terms of the \emph{Wilson surface} associated with each fundamental \emph{cube} of the lattice, which we refer to as cubets (see Fig \ref{fig:Fig1}).
In lattice Yang-Mills, one associates a unitary matrix to each link of the lattice. This matrix corresponds to the Wilson line along that link. Physically, it corresponds to the phase that a quark acquires as it moves along the link. The unitary matrices which live on the links of the lattice are the basic variables of lattice Yang-Mills theory. In lattice gerbe theory, the basic variables live on the \emph{faces} (or plaquettes) of the lattice. In the abelian lattice gerbe theory, these face variables are just $U(1)$ phases. We will describe a non-abelian generalization in section \ref{nonabelian}. 

Note that each face variable can be labeled by a pair of spatial indices corresponding to the plane in which the face lies. In three dimensions, for example, there are three types of face variables, which are labeled $xy,yz,zx$. In $D$ dimensions, there are $D(D-1)/2$ different planes in which a face variable can lie. We shall denote a plaquette by\footnote{The context should clearly distinguish a plaquette from the D'Alembertian $\Box=\partial_{\mu}\partial^{\mu}$.} $\Box$. It is straightforward to systematically identify a particular plaquette by the plane it lies in and a (chosen) corner at the lattice site $\vec{n}$, as in Figure \ref{plaquettewl}. We denote the face variable associated to the plaquette with a corner at the location $\vec{n}$ and lying in the $(\mu\nu)$-plane by ${\cal W}_{\mu_1\mu_2}(\vec{n})$:
$$
{\cal W}_{\mu_1\mu_2}(\vec{n})=e^{i\alpha^2 \;b_{\mu_1\mu_2}(\vec{n})},	\qquad	{\cal W}^{\dagger}_{\mu_1\mu_2}(\vec{n})=e^{-i\alpha^2\; b_{\mu_1\mu_2}(\vec{n})}
$$
where $b_{\mu_1\mu_2}$ is the gerbe connection in this plane and $\alpha$ is the lattice spacing so that $\alpha^2$ is the area of the plaquette. This parametrisation ensures that the condition ${\cal W}_{\mu_1\mu_2}(\vec{n}){\cal W}^{\dagger}_{\mu_1\mu_2}(\vec{n})=1$ is satisfed. An important point is that $b_{\mu\nu}$ is defined over the entire plaquette $\Box_{\vec{n}}$ and not just at the point $\vec{n}$. With this understood, no confusion should arise by our denoting $b_{\mu\nu}(\Box_{\vec{n}})$ by $b_{\mu\nu}(\vec{n})$. We shall adopt this convention throughout the rest of the paper.

We take the lattice to be isotropic but the construction generalises straightforwardly. In the abelian lattice gerbe theory, the cubets can be labeled by a triplet of Lorentz indices. For example, a cubet labeled by $(\mu,\nu,\lambda)$ has faces which lie in the $(\mu,\nu)$-, $(\nu,\lambda)$-, and $(\lambda, \mu)$-planes and the Wilson surface associated to the cubet $\mbox{\mancube}$ is simply given by:   
\begin{equation}\label{Gabelian}
\Gamma(\mbox{\mancube})={\cal W}^{\dagger}_{\nu\lambda}(\vec{n})\,{\cal W}^{\dagger}_{\lambda\mu}(\vec{n})\,{\cal W}^{\dagger}_{\mu\nu}(\vec{n})\,{\cal W}_{\nu\lambda}(\vec{n}+\vec{\mu})\,{\cal W}_{\lambda\mu}(\vec{n}+\vec{\nu})\,{\cal W}_{\mu\nu}(\vec{n}+\vec{\lambda}).
\end{equation}
We shall use the term cubet to refer to the fundamental cube $\mbox{\mancube}$ and its associated Wilson surface $\Gamma(\mbox{\mancube})$. Which we are referring to should be clear from the context. Note that a particular cubet $\mbox{\mancube}_{\vec{n}}$ locally defines a three-dimensional sublattice with basis vectors along the sides of the cubet and origin at $\vec{n}$. This allows us to define a cross-product $\times$ of two lattice vectors. The orientation of each face is determined by its normal vector. For a $(\mu,\nu)$ face, the normal vector will point in the $\vec{\mu}\times\vec{\nu}$ direction, where $\vec{\mu}$ is the unit vector pointing in the positive ${\mu}$ direction. Faces whose normal vector points away from the centre of the cube are complex conjugated. From this definition, it is not difficult to see that a more general Wilson surface in the abelian theory can be computed by taking the product of all the cubets which fill the surface. For example, the Wilson surface comprising two cubets is given by the product
$\Gamma(\mbox{\mancube}\mbox{\mancube})=\Gamma(\mbox\mancube)\Gamma(\mbox{\mancube})$.

\subsection{Action}

The action is given by taking the real part of each cubet and summing over all cubets in the lattice
$$
S[{\cal W}]=\beta\sum_{\mbox{\mancube}}\left(1-\Re\left(\Gamma(\mbox{\mancube})\right)\right)
$$
where the real part of the cubet is
$$
\Re\Gamma(\mbox{\mancube})=\frac{1}{2}\left(\Gamma(\mbox{\mancube})+\Gamma^{\dagger}(\mbox{\mancube})\right).
$$
The sum over cubets may be written as
$$
\sum_{\mbox{\mancube}}=\sum_{\vec{n}}\sum_{1\leq \mu_1<\mu_2<\mu_3\leq D}
$$
where $D$ is the dimension of the lattice. The second sum sums over the $\frac{1}{6}D(D-1)(D-2)$ cubets that intersect orthogonally at the lattice point $\vec{n}$. The first sum sums over all lattice sites $\vec{n}$. The parameter $\beta$ can be interpreted as the inverse temperature of the statistical system or the inverse coupling of the field theory in the continuum limit. Adding the complex conjugate of each fundamental cube ensures that the action is real, and  is also required by parity invariance. In particular, it is not difficult to see that under a parity transformation, the Wilson surface associated with each plaquette is complex conjugated. For a trivial gerbe configuration\footnote{We define a trivial configuration as one for which the connection $b_{\mu\nu}=0$ in the region under consideration.}, all the face variables are unity and the action vanishes.

Let us focus on a particular cubet. We take $(\vec{\mu},\vec{\nu},\vec{\lambda})$ to be a basis of lattice vectors spanning the cubet at the point $\vec{n}$ and of length $\alpha$. The cubet has vertices at the points
$$
\{\;\vec{n}\;,\;\vec{n}+\vec{\mu}\;,\;\vec{n}+\vec{\nu}\;,\;\vec{n}+\vec{\lambda}\;,\;\vec{n}+\vec{\mu}+\vec{\nu}\;,\;\vec{n}+\vec{\nu}+\vec{\lambda}\;,\;\vec{n}+\vec{\mu}+\vec{\nu}+\vec{\lambda}\;\}.
$$
Our convention for labelling the face variabes is that the plaquette in the $(\mu_1,\mu_2)$-plane with vertices at $\{\vec{n},\vec{n}+\vec{\mu}_1,\vec{n}+\vec{\mu}_2,\vec{n}+\vec{\mu}_1+\vec{\mu}_2\}$ is denoted by ${\cal W}_{\mu_1\mu_2}(\vec{n})$. The Wilson surface over this cubet may then be written as (\ref{Gabelian}). In the abelian theory, the face variables are simply commuting complex functions and so we can write
$$
\Gamma(\mbox{\mancube})=\exp\left[i\alpha^2\left(b_{\mu\nu}(\vec{n})+b_{\nu\lambda}(\vec{n})+b_{\lambda\mu}(\vec{n})-b_{\mu\nu}(\vec{n}+\vec{\lambda})-b_{\nu\lambda}(\vec{n}+\vec{\mu})-b_{\lambda\mu}(\vec{n}+\vec{\nu})\right)\right].
$$
Using the definition of the lattice derivative in (\ref{derivative}), we may write
$$
b_{\nu\lambda}(\vec{n}+\vec{\mu})=b_{\nu\lambda}(\vec{n})+\alpha \Delta_{\mu}b_{\nu\lambda}(\vec{n})
$$
and similarly for cylic permutations in $(\mu,\nu,\lambda)$. It is then not hard to show that (\ref{G}) becomes
$$
\Gamma(\mbox{\mancube})=\;e^{i\alpha^3H_{\mu\nu\lambda}(\vec{n})}
$$
where
$$
H_{\mu\nu\lambda}(\vec{n}):=\Delta_{\mu} b_{\nu\lambda}(\vec{n})+\Delta_{\nu}b_{\lambda\mu}(\vec{n})+\Delta_{\lambda}b_{\mu\nu}(\vec{n}).
$$
Putting this in the action gives
$$
S[b]=\beta\sum_{\mbox{\mancube}}\left(1-\cos(\alpha^3 H(\mbox{\mancube}))\right)
$$
where $H(\mbox{\mancube})$ is the curvature for the particular fundamental cube $\mbox{\mancube}$. Expanding in powers of the lattice spacing $\alpha$ gives
$$
S[b]=\frac{\beta\alpha^6}{2}\sum_{\vec{n}, \mu,\nu\lambda}\left(H_{\mu\nu\lambda}(\vec{n})\right)^2+...
$$
where $+...$ denotes higher powers of $\alpha$. The classical continuum limit is that for which $\alpha\rightarrow 0$, whilst the number of lattice sites tends to infinity such that we make the identification
$$
\beta\sum_{\vec{n}}\alpha^6\rightarrow \widehat{\beta}\int\rd^Dx
$$
where $\widehat{\beta}=\alpha^{6-D}\beta$ and $\beta$ is scaled such that $\widehat{\beta}$ remains fixed. An interesting case is where $D=6$ and $\widehat{\beta}=\beta$ is dimensionless. This case, in which $H$ may also be self-dual, is of particular interest  and plays an important role in M-theory. Adopting the Einstein summation convention, we recover the conventional abelian gerbe action
$$
S[b]=\frac{1}{2g^2}\int\rd^Dx\; H_{\mu\nu\lambda}H^{\mu\nu\lambda},
$$
where we have introduced the coupling constant $g^2=\widehat{\beta}^{-1}$.

\subsection{Gauge transformations}

\begin{figure}[htbp] 
       \centering
       \includegraphics[width=5in]{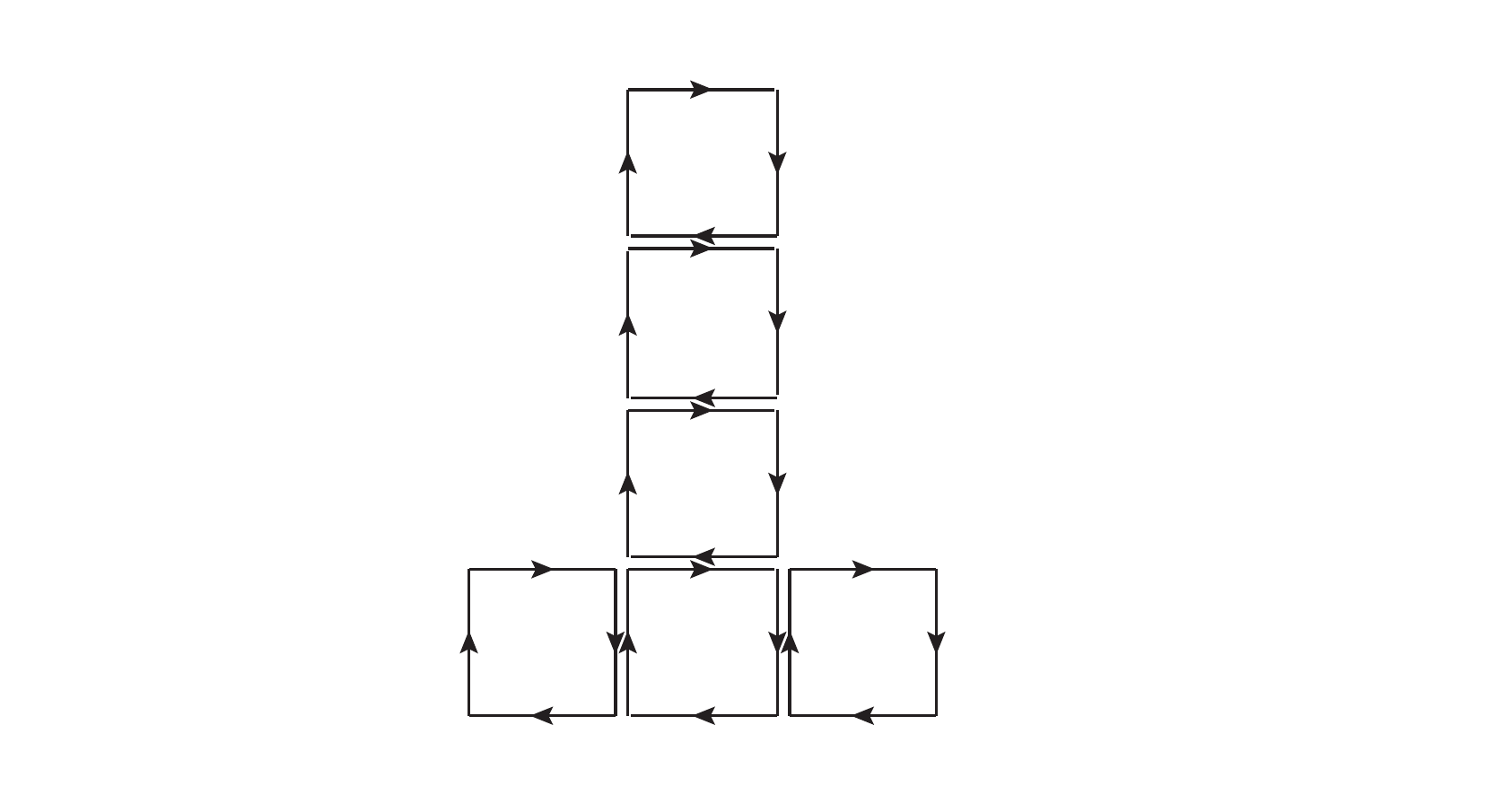}  
    \caption{The net of the fundamental cube in Fig \ref{fig:Fig1}.}
    \label{fig:Fig3}
    \end{figure}

In lattice Yang-Mills theory, the gauge transformations of a link variable are associated with the vertices of the link. By analogy, in lattice gerbe theory, the gauge transformations of a face variable ${\cal W}$ are associated with its edges. Since the edges are extended objects, this gauge transformation is necessarily \emph{non-local}. Our notion of a gauge transformations comes from the `exploded' diagram for a cubet in Fig \ref{fig:Fig3}. We have assigned an arrow to each edge. The two possible directions of an arrow along an edge are related by complex conjugation of the associated face variable. 
Note that the arrows in Figure \ref{fig:Fig3} form a closed loop around each face and if two faces share a common edge, then their arrows along that edge point in opposite directions. Once an arrow is assigned to an edge of a single face in the surface, this automatically fixes all the other arrows in the cubet. There are only two ways to assign arrows to the edges of a cubet and they are related by complex conjugation.

It is natural to define gauge transformations of the Wilson surface associated with a given plaquette as multiplication by the Wilson lines associated with the edges of that plaquette. As in lattice Yang-Mills, it is necessary to introduce an adjoint action $\dagger$ that changes the orientation of the gauge transformation along that edge. In the abelian case, $\dagger$ is simply complex conjugation. The gauge symmery is reducible and the Wilson lines associated with gauge transformations of the edges are in turn defined only up to gauge transformations associated with their end points - the corners of the fundamental cube. The fundamental cube, given by folding up the net in Fig \ref{fig:Fig3}, is manifestly gauge invariant.

The gauge transformations are given by
$$
\mathcal{W}_{\mu\nu}(\vec{n}) \rightarrow U_{\mu}(\vec{n})\;U_{\mu}^{\dagger}(\vec{n}+\vec{\nu})\;\mathcal{W}_{\mu\nu}(\vec{n})\;U_{\nu}(\vec{n}+\vec{\mu})\;U_{\nu}^{\dagger}(\vec{n})
$$
The Wilson lines which define the gauge transformation can written as 
$$
U_{\mu}(\vec{n})=e^{i\alpha \lambda_{\mu}(\vec{n})},	\quad	U_{\nu}(\vec{n}+\vec{\mu})=e^{i\alpha\lambda_{\nu}(\vec{n}+\vec{\mu})}, 	\quad
U^{\dagger}_{\mu}(\vec{n}+\vec{\nu})=e^{-i\alpha \lambda_{\nu}(\vec{n}+\vec{\nu})},	\quad	U_{\nu}^{\dagger}(\vec{n})
=e^{-i\alpha \lambda_{\nu}(\vec{n})}.
$$
Writing the shifted Wilson line potentials as
$$
\lambda_{\nu}(\vec{n}+\vec{\mu})=\lambda_{\nu}(\vec{n})+\alpha\Delta_{\mu}\lambda_{\nu}(\vec{n}), 	\qquad	\lambda_{\mu}(\vec{n}+\vec{\nu})=\lambda_{\mu}(\vec{n})+\alpha\Delta_{\nu}\lambda_{\mu}(\vec{n}),
$$
the infinitessimal gauge transformations act on the $b$-field as
$$
\delta b_{\mu\nu}(\vec{n})=\Delta_{\mu}\lambda_{\nu}(\vec{n})-\Delta_{\nu}\lambda_{\mu}(\vec{n}).
$$
Note that we have used $\vec{n}$ to denote a chosen point on a given plaquette\footnote{A more illustrative notation might be
$$
\delta b_{\mu\nu}(\Box)=\Delta_{\mu}\lambda_{\nu}(||)-\Delta_{\nu}\lambda_{\mu}(=).
$$}, but the $b$-field is defined on the entire plaquette and is not localised at the particular lattice point $\vec{n}$. In the continuum limit, this reduces to the conventional abelian gauge transformation
$$
\delta b_{\mu\nu}=\p_{\mu}\lambda_{\nu}-\p_{\nu}\lambda_{\mu},
$$
where $b$ and $\lambda$ are localised at a point in space-time. In addition to the one-form gauge transformations, there are \emph{local} scalar gauge transformations, localised on the lattice sites. These gauge transformations act on the Wilson line potentials as $\delta \lambda_{\mu}=\Delta_{\mu}\varphi$. Thus, the continuum limit of the abelian lattice gerbe theory reproduces the conventional description of a gerbe.

\section{Non-abelian Lattice Gerbe Theory} \label{nonabelian}

In this section we systematically construct a non-abelian generalisation of the lattice gerbe theory introduced in the preceding section. The fundamental variable is a generalisation of the face variables ${\cal W}(\vec{n})$ of the abelian theory to a four-index object ${\cal W}^{a\dot{a}}_{b\dot{b}}(\vec{n})$, which we will take to transform in the bi-adjiont representation of $U(N)\times U(N)$ for reasons we describe below. We describe an unambiguous way to define non-abelian Wilson surfaces, from which a gauge invariant action may be constructed. Note that questions of path ordering simply do not arise in this approach.

\subsection{Face Variables}

In lattice Yang-Mills, the basic variables are unitary matrices which are located on the links of the lattice, and the elements of each link variable can be labeled by a pair of indices which are associated with the vertices of the link, as depicted in Figure \ref{fig:link1}. 
In a lattice gerbe theory, the basic variables live on the faces of the lattice. Since each face has four edges, it is natural to assign an index to each edge, so the face variables in the non-abelian lattice gerbe theory are four-index objects, ${\cal W}^{a\dot{a}}_{b\dot{b}}$ \cite{Nepomechie:1982rb}. Throughout this paper, we shall employ the notation $(a,\dot{a},a')$ to denote colour indices associated with edges along the three orthogonal directions of the fundamental cube.

\begin{figure}[h] 
       \centering
       \includegraphics[width=5in]{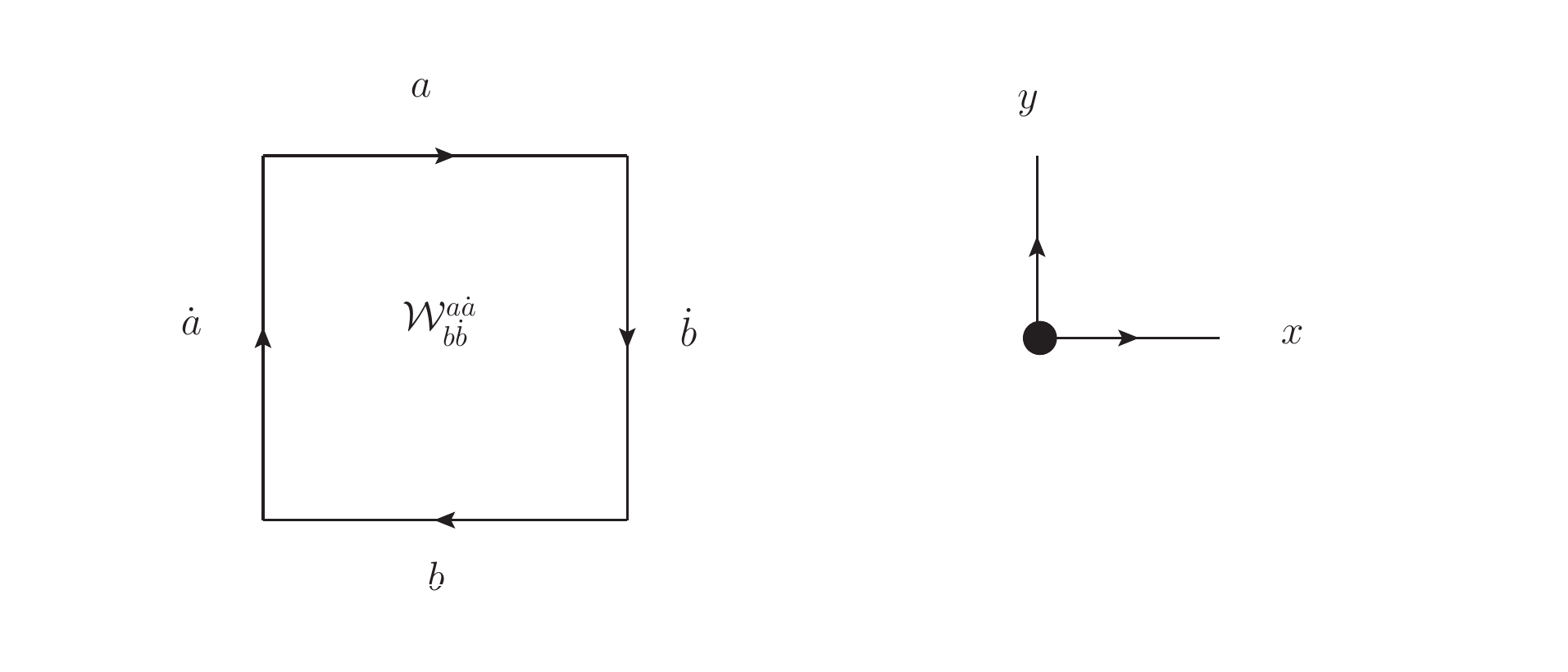}  
\caption{A non-abelain face variable.}
\label{fig:plaquette}
    \end{figure}

To understand the placement of the indices, consider the face variable in Figure \ref{fig:plaquette} which we take to lie in the $xy$-plane. Such a face variable could be one from Figure \ref{fig:Fig2}, which depicts the net of a cubet in the non-abelian gerbe theory. The Wilson surface associated with this plaquette is denoted by $\mathcal{W}^{a \dot{a}}_{b \dot{b}}$. Note that the pair of colour indices $^{a}_{b}$ are associated with links in the $x$-direction and the pair of colour indices $^{\dot{a}}_{\dot{b}}$ are associated with links in the $y$-direction. The $^{a}_{b}$  indices appear to the left of the $^{\dot{a}}_{\dot{b}}$  indices because $x$ comes before $y$ in our labelling scheme (dictated by the handedness of our chosen cordinate system). The reason for raising $a,\dot{a}$ and lowering $b,\dot{b}$ is that the arrows associated with $a$ and $\dot{a}$ point in the positive $x$ and $y$ directions, respectively, while the arrows associated with $b$ and $\dot{b}$ point in the negative $x$ and $y$ directions, respectively. We will see later that this convention dictating the placement of the colour indices greatly facilitates the construction of a gauge-invariant action. 
In lattice Yang-Mills, the constraint that each link variable is a unitary matrix corresponds to:
\begin{equation}
U_m{}^n(U^{\dagger})_n{}^p=\delta_m^p.
\label{unitarity}
\end{equation}
Physically, this unitarity constraint on the link variables ensures that if we parallel transport a particle from $m$ to $n$ and then reverse the operation, we should get back to where we started. Treating $U$ as the matrix representation of a group, the adjoint operation gives the inverse $U^{-1}=U^{\dagger}$. 

A face variable in the non-abelian gerbe theory has two types of color indices, which correspond to the two spatial directions of the face variable. In principle, we could assign a different gauge group to each spatial direction, but we will take the gauge group to be the same in all directions since this will preserve rotational symmetry (by which we mean invariance under 90 degree rotations in any pair of directions on the lattice). The analogue of (\ref{unitarity}) for the face variables is
\begin{equation}\label{constraints}
{\cal W}^{a\dot{a}}_{c\dot{c}}({\cal W}^{\dagger})^{c\dot{c}}_{b\dot{b}}=\delta^a_b\delta^{\dot{a}}_{\dot{b}},
\end{equation}
where we define a generalised adjoint operation $\dagger$ by analogy with Hermitian conjugation of square matrices
$$
(\mathcal{W}^{\dagger})_{c \dot{d}}^{a \dot{b}}=(\mathcal{W}_{a \dot{b}}^{c \dot{d}})^{*}.
$$
Note that the ordering of the pairs of indices does not change but th upper pair of indices is exchanged with the lower pair. This adjoint reverses the sense of the arrows in Fig \ref{fig:plaquette} and so is intimately related to a parity transformation. In the abelian theory ${\cal W}^{\dagger}={\cal W}^*$. 

In summary, rotational symmetry and the unitarity constraint in (\ref{constraints}) imply that the face variables transform in some representation of $U(N) \times U(N)$. In the absence of coupling to matter fields, we can choose this to be the bi-adjoint representation without loss of generality. In other words, we can choose the raised/lowered color indices of the face variables to correspond to the fundamental/anti-fundamental representation of $U(N)$. The basic construction allows for the ${\cal W}$'s to take values in a representation  $R_1\times \bar{R}_1\times R_2\times\bar{R}_2$. For the sake of concreteness and so as not to cloud the issues we wish to stress, we shall take the representations $R_i$ to be in the fundamental of $U(N)$; however, other choices of repesentation will not change the general conclusions.

Naively, ${\cal W}$ has $N^4$ complex degrees of freedom. The constraint (\ref{constraints}) gives $N^4$ real constraints, leaving $N^4$ real degrees of freedom in ${\cal W}$. A useful parameterisation of these $N^4$ real degrees of freedom is in terms of a tensor field $(B_{\mu\nu})^{ab}_{\dot{a}\dot{b}}$, related to $({\cal W}_{\mu\nu})^{ab}_{\dot{a}\dot{b}}$ by
$$
{\cal W}^{a\dot{a}}_{b\dot{b}}(\vec{n})=\exp\left(i\alpha^2B^{a\dot{a}}_{b\dot{b}}(\vec{n})\right)
$$
Since ${\cal W}^{a\dot{a}}_{b\dot{b}}$ is not a matrix, we need to be  little more careful about what we mean by this exponential of a tensor. We shall formally define
\begin{eqnarray}
{\cal W}^{ab}_{\dot{a}\dot{b}}(\vec{n})&=&\exp\left(i\alpha^2B^{ab}_{\dot{a}\dot{b}}(\vec{n})\right)\nonumber\\
&\equiv&\delta^a_b\delta^{\dot{a}}_{\dot{b}}+i\alpha^2B^{a\dot{a}}_{b\dot{b}}-\frac{\alpha^4}{2!}B^{a\dot{a}}_{c\dot{c}}B^{c\dot{c}}_{b\dot{b}}-\frac{i\alpha^6}{3!}B^{a\dot{a}}_{c\dot{c}}B^{c\dot{c}}_{d\dot{d}}B^{d\dot{d}}_{b\dot{b}}+...
\end{eqnarray}
We may write this as
$$
{\cal W}^{ab}_{\dot{a}\dot{b}}=\delta^a_b\delta^{\dot{a}}_{\dot{b}}+i\alpha^2B^{a\dot{a}}_{b\dot{b}}+\sum_{n=1}^{\infty}\frac{i^{n+1}\alpha^{2n+2}}{(n+1)!}B^{a\dot{a}}_{c_1\dot{c}_1}B^{c_1\dot{c}_1}_{c_2\dot{c}_2}...B^{c_n\dot{c}_n}_{b\dot{b}}
$$
Since the indices in these expressions contract in pairs ($a\dot{a}$ etc), we can treat ${\cal W}$ as a complex $N^2\times N^2$ matrix ${\cal W}^I{}_J$ where $I=a\dot{a}$, and we might naively think that the gauge symmetry is $U(N^2)$; however, we shall see that the action is only invariant under $U(N)\times U(N)$. The constraint (\ref{constraints}) will be satisfied if
$$
B^{a\dot{a}}_{c\dot{c}}=(B^*)_{a\dot{a}}^{c\dot{c}}
$$
were a transpose has been taken independently in the $^a_b$ and $^{\dot{a}}_{\dot{b}}$ indices, i.e. if $B$ is, what might be called, \emph{doubly Hermitian}. 

Without loss of generality, it is possible to decompose the face variables into $SU(N)$ representations as follows:
\begin{equation}\label{decomp}
(B_{\mu\nu})^{a\dot{a}}_{b\dot{b}}=b_{\mu\nu}\;\delta^a_b\,\delta^{\dot{a}}_{\dot{b}}+C_{\mu\nu}^M\,(T_M)^a_b\,\delta^{\dot{a}}_{\dot{b}}+\widetilde{C}_{\mu\nu}^{\dot{M}}\,(T_{\dot{M}})^{\dot{a}}_{\dot{b}}\,\delta^a_b+\Phi_{\mu\nu}^{M\dot{M}}\,(T_M)^a_b\,(\tilde{T}_{\dot{M}})^{\dot{a}}_{\dot{b}}
\end{equation}
where $T_M$ are generators of $SU(N)$\footnote{We thank Chris Hull for suggesting this decomposition.}. The doubly Hermitian condition $(B_{\mu\nu})^{a\dot{a}}_{b\dot{b}}=(B^*_{\mu\nu})_{a\dot{a}}^{b\dot{b}}$ means that all components are real. 
Note that pairs of colour indices are directly related to the corresponding Lorentz indices of the $B$ field, if we exchange $\mu$ and $\nu$, we must simultaneously exchange the pairs of indices $_{a}^{b}$ and $_{\dot{b}}^{\dot{a}}$. Hence, we impose the condition
\begin{equation}
(B_{\mu \nu})^{a\dot{a}}_{b\dot{b}}=-(B_{\nu \mu})^{\dot{a}a}_{\dot{b}b},
\label{antisymmetry}
\end{equation}
which reduces to the antisymmetry constraint in the abelian case. A geometric description of this condition in the continuum limit is briefly discussed in Appendix A. Putting the $SU(N)$ decomposition (\ref{decomp}) into (\ref{antisymmetry}), gives 
\begin{equation}
b_{\mu\nu}=-b_{\nu\mu},	\qquad	C_{\mu\nu}^M=-\widetilde{C}_{\nu\mu}^M,	\quad	\widetilde{C}_{\mu\nu}^{\dot{M}}=-C_{\nu\mu}^{\dot{M}},	\qquad	\Phi_{\mu\nu}^{M\dot{N}}=-\Phi_{\nu\mu}^{\dot{N}M}\;.
\label{antcomp}
\end{equation}
An important point is that $C_{\mu\nu}^M$ is not antisymmetric in the space-time indices. The singlet $b_{\mu\nu}$ is a conventional abelian gerbe. The other components of the face variable are not gerbes in the conventional sense as they are not straightforward two-forms, but have a more intricate structure in which the space-time and internal symmetries are not completely uncoupled from one another.

\subsection{Wilson Surfaces} \label{nawilson}

Recall that Wilson loops are the fundamental building blocks in Wilson's action for lattice Yang-Mills. Here we consider closed orientable Wilson surfaces on the lattice, constructed using the face variables introduced in the proceding section. The goal is to use such Wilson surfaces to define an action for lattice gerbe theory akin to that constructed by Wilson for lattice Yang-Mills theory.

The formula for a closed orientable Wilson surface in the non-abelian gerbe theory can be unambiguously constructed using the following procedure:
\begin{enumerate}
\item To each face variable on the surface, assign a colour index and an arrow to each edge such that the arrows form a closed loop. If two face variables in the surface share a common edge then their resective arrows along that edge point in opposite directions. Once an arrow is assigned to an edge of a single face variable in the surface, this automatically fixes all the other arrows in the surface. Performing a parity transformation reverses the arrows, giving the conjugate Wilson surface.
\item
To each cubet we assign a triplet of lattice basis vectors than span the three-dimensional subspace of the cubet. If the arrow associated with an edge points in the positive direction, relative to this chosen basis, then the associated colour index is raised and otherwise it is lowered. Hence, two face variables in the surface which share a common edge will share a common index, and this index will be raised in one face variable and lowered in the other. 
\item From the previous rule, a pair of opposite edges of a face variable will correspond to a pair of indices, one of which is raised and the other lowered. Assign a spatial direction to this pair of indices which corresponds to the spatial direction along which their edges lie. The order in which the two pairs of indices occurs in a face variable is then determined by the ordering of the spatial indices associated with each pair, up to cyclic permutations. For example, in three dimensions, the ordering of spatial indices is $(x,y)$, $(y,z)$, $(z,x)$.
\item Given a three-dimensional basis of the cubet, compute the normal vector to each face variable. In particular, the normal vector to a face variable which lies in the $(\mu,\nu)$-plane is given by $\vec{\mu} \times \vec{\nu}$. If the normal vector points away from the interior of the surface, then replace the face variable with its adjoint while keeping the indices in the same place.
\item
Finally, take the product of all the face variables in the surface and sum over repeated indices. For closed orientable surfaces, there are no free indices, so one obtains a gauge-invariant object.
\end{enumerate}

\begin{figure}[htbp] 
       \centering
       \includegraphics[width=2.6in]{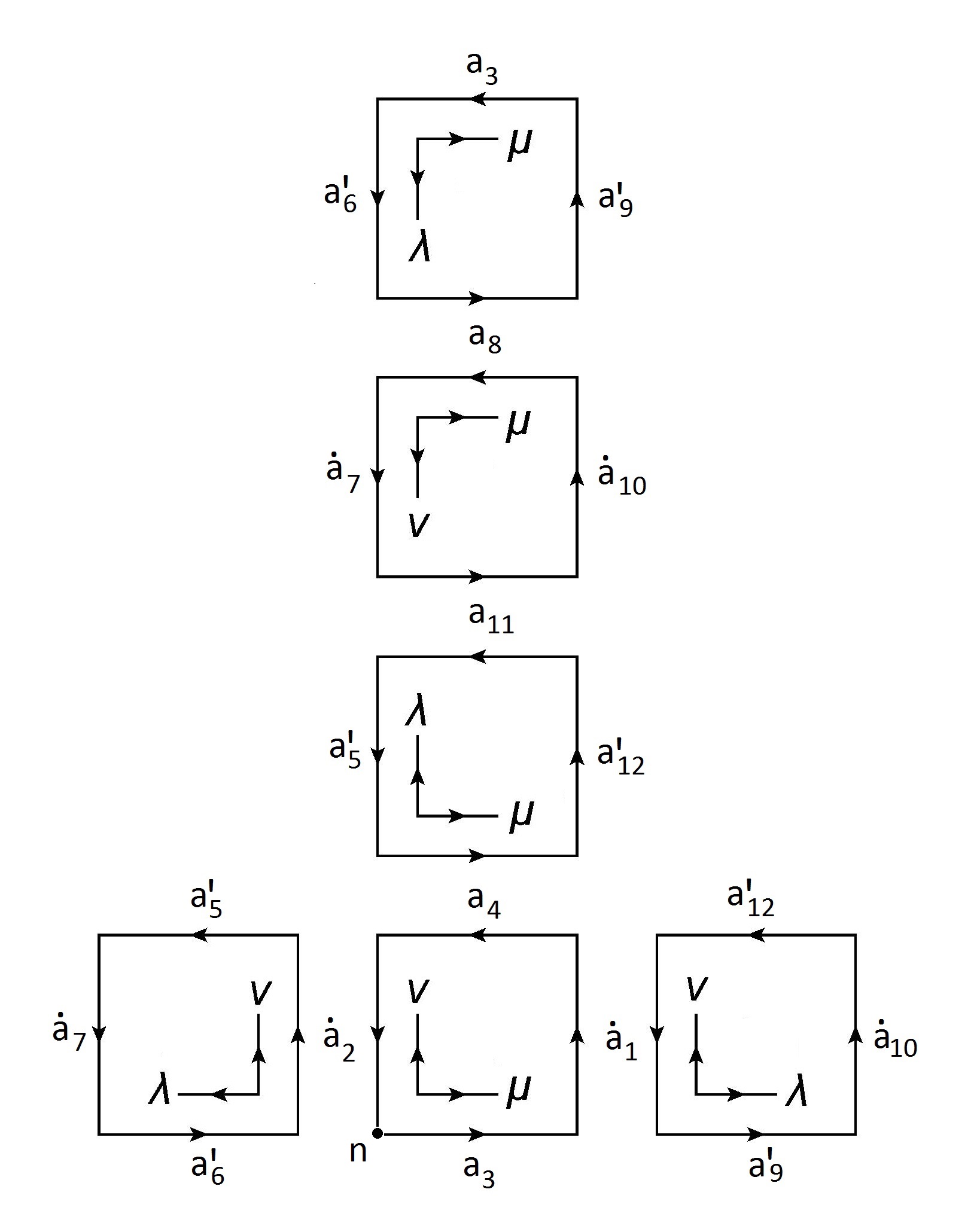}  
    \caption{Indices and arrows for a cubet.}
    \label{fig:Fig2}
    \end{figure}

\begin{figure}[htbp] 
       \centering
       \includegraphics[width=2.6in]{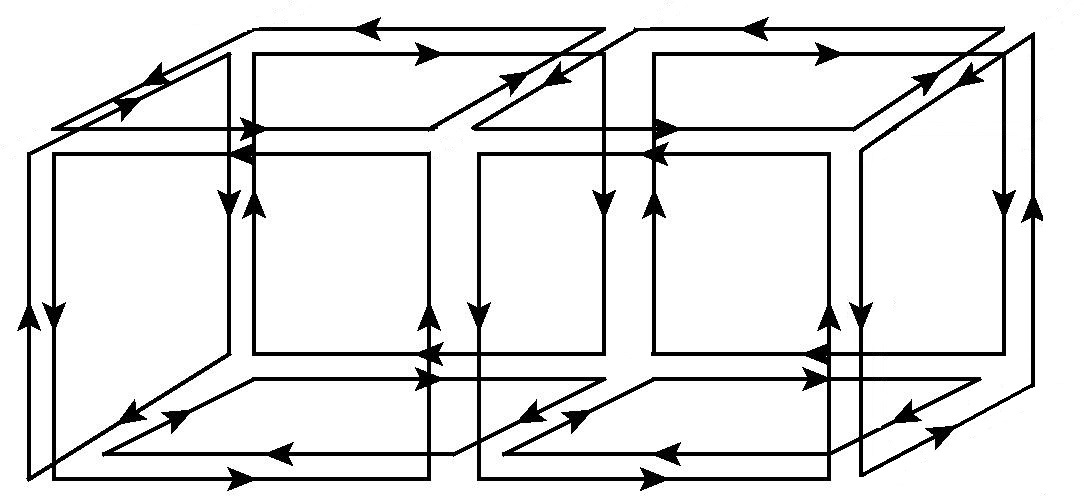}  
    \caption{$2\times1\times1$ Wilson surface.}
    \label{11}
    \end{figure}

\begin{figure}[htbp] 
       \centering
       \includegraphics[width=3.2in]{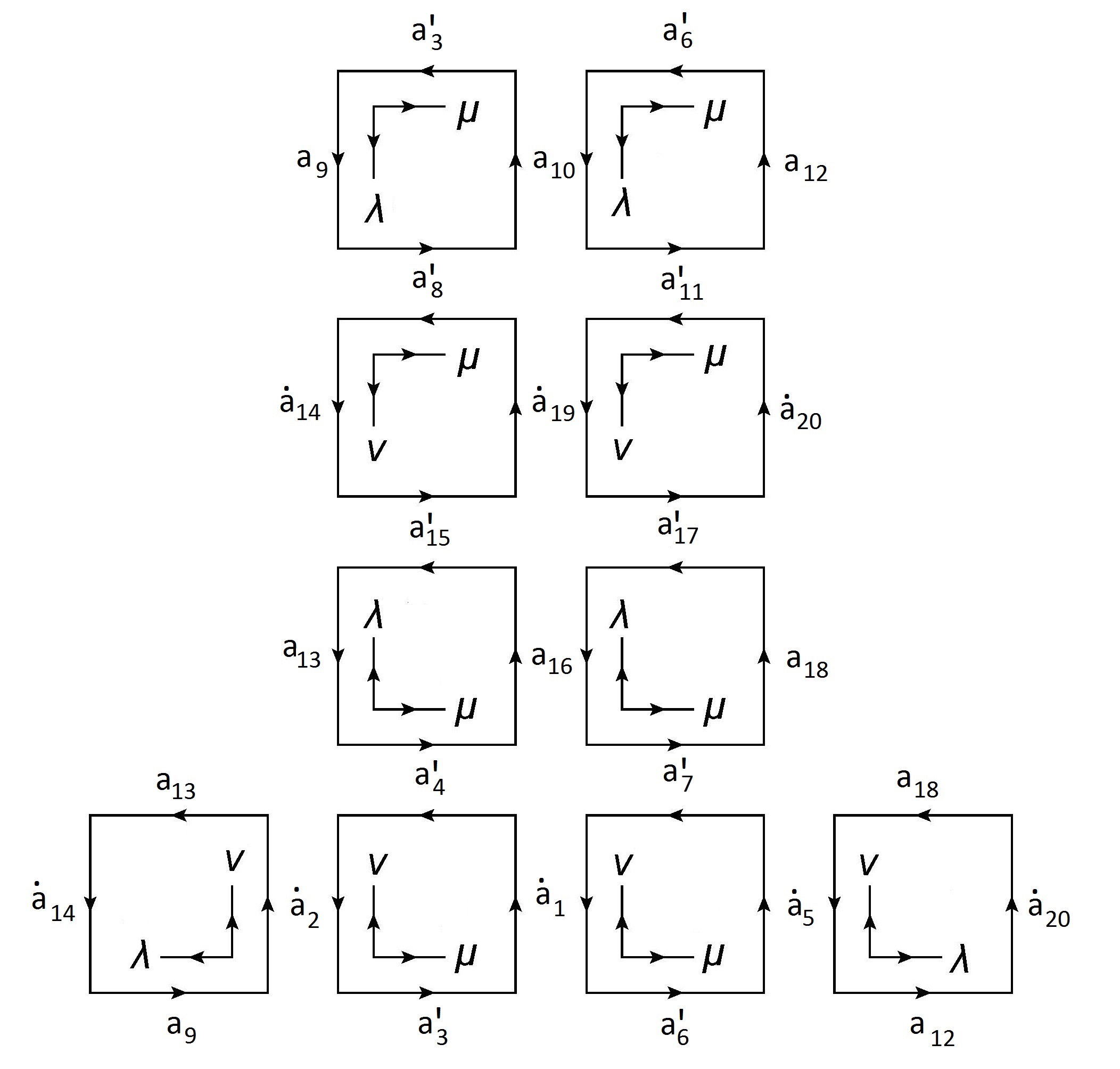}  
    \caption{Indices and arrows for a $2\times1\times1$ Wilson surface.}
    \label{211}
    \end{figure}

Let us apply the above procedure to obtain an expression for the Wilson surface correponding to a fundamental cube. We will take this cube to be spanned by the lattice vectors $(\vec{\mu},\vec{\nu},\vec{\lambda})$. For the purpose of assigning indices and arrows to the edges of the faces, it is convenient to unfold the cubet as shown in Figure \ref{fig:Fig2}, where edges with a repeated index are understood to be identified. The cubet in Figure \ref{fig:Fig2} corresponds to 
\begin{eqnarray}
\Gamma(\mbox{\mancube})=\mathcal{W}_{\mu\nu}{}_{a_{4}\dot{a}_{2}}^{a_{3}\dot{a}_{1}}(\vec{n})\mathcal{W}_{\nu\lambda}{}_{\dot{a}_{7}a_{6}'}^{\dot{a}_{2}a_{5}'}(\vec{n})\mathcal{W}_{\lambda\mu}{}_{a_{9}'a_{3}}^{a_{6}'a_{8}}(\vec{n})\mathcal{W}_{\mu\nu}^{\dagger}{}_{a_{8}\dot{a}_{10}}^{a_{11}\dot{a}_{7}}(\vec{n}+\vec{\lambda})\mathcal{W}_{\nu\lambda}^{\dagger}{}_{\dot{a}_{1}a_{12}'}^{\dot{a}_{10}a_{9}'}(\vec{n}+\vec{\mu})\mathcal{W}_{\lambda\mu}^{\dagger}{}_{a_{5}'a_{11}}^{a_{12}'a_{4}}(\vec{n}+\vec{\nu}).\nonumber\\
\label{cube}
\end{eqnarray}
In the nonabelian theory, a general closed Wilson surface is not simply given by the product of the all the cubets contained in the surface, as we found for the abelian theory. The procedure described above allows us to define closed surfaces in an unambiguous way. Path ordering issues simply do not arise in this construction. The object (\ref{cube}) is the basic building block of the theory described here. To avoid the somewhat cluttered  notation of (\ref{cube}), we shall often write (\ref{cube}) as
$$
\Gamma(\mbox{\mancube})=\mathcal{W}(\theta_1)_{a_{4}\dot{a}_{2}}^{a_{3}\dot{a}_{1}}\mathcal{W}(\theta_2)_{\dot{a}_{7}a_{6}'}^{\dot{a}_{2}a_{5}'}\mathcal{W}(\theta_3)_{a_{9}'a_{3}}^{a_{6}'a_{8}}\mathcal{W}(\theta_4)_{a_{8}\dot{a}_{10}}^{a_{11}\dot{a}_{7}}\mathcal{W}(\theta_5)_{\dot{a}_{1}a_{12}'}^{\dot{a}_{10}a_{9}'}\mathcal{W}(\theta_6)_{a_{5}'a_{11}}^{a_{12}'a_{4}},
$$
where the suppressed space-time index structure is given by
\begin{equation}
\begin{array}{ccc}
\theta_1=\alpha^2B_{\mu\nu}(\vec{n}),&\qquad\theta_2=\alpha^2B_{\nu\lambda}(\vec{n}),&\qquad \theta_3=\alpha^2B_{\lambda\mu}(\vec{n}),\\
\theta_4=-\alpha^2B_{\mu\nu}(\vec{n}+\vec{\lambda}),&\qquad \theta_5=-\alpha^2B_{\nu\lambda}(\vec{n}+\vec{\mu}),&\qquad\theta_6=-\alpha^2B_{\lambda\mu}(\vec{n}+\vec{\nu}).\nonumber
\end{array}
\end{equation}
Note that the minus signs take care of the $\dagger$ operation. As a second example, consider the Wilson surface which contains two fundamental cubes, depicted in Figure \ref{11}. For the pupose of assigning indices, it is convenient to unfold the Wilson surface as shown in Figure \ref{211}. This Wilson surface, composed of two cubets, is then given by 
\[
\Gamma(\mbox{\mancube\mancube})=\mathcal{W}{}_{a_{4}'\dot{a}_{2}}^{a_{3}'\dot{a}_{1}}\mathcal{W}{}_{a_{7}'\dot{a}_{1}}^{a_{6}'\dot{a}_{5}}\mathcal{W}{}_{a_{10}a_{3}'}^{a_{9}a_{8}'}\mathcal{W}{}_{a_{12}a_{6}'}^{a_{10}a_{11}'}\mathcal{W}_{\dot{a}_{14}a_{9}}^{\dot{a}_{2}a_{13}}\mathcal{W}^{\dagger}{}_{a_{13}a_{15}'}^{a_{16}a_{4}'}\mathcal{W}^{\dagger}{}_{a_{16}a_{17}'}^{a_{18}a_{7}'}\mathcal{W}^{\dagger}{}_{a_{8}'\dot{a}_{19}}^{a_{15}'\dot{a}_{14}}\mathcal{W}^{\dagger}{}_{a_{11}'\dot{a}_{20}}^{a_{17}'\dot{a}_{19}}\mathcal{W}^{\dagger}{}_{\dot{a}_{5}a_{18}}^{\dot{a}_{20}a_{12}}.
\]
More complicated Wilson surfaces may be constructed from simpler ones in a similar way as was seen in the abelian theory. For example\footnote{This result relies on the integration rule
$$
\int\rd{\cal W}\,({\cal W}^{\dagger})^{a\dot{a}}_{b\dot{b}}{\cal W}^{c\dot{c}}_{d\dot{d}}\sim\delta^a_d\delta^{\dot{a}}_{\dot{d}}\delta^c_b\delta^{\dot{c}}_{\dot{d}}
$$
which will be used at length in section \ref{sub:Non-abelian-theory}.}
$$
\Gamma(\mbox{\mancube}\mbox{\mancube})=\int\rd{\cal W}_c\;\Gamma(\mbox{\mancube}_L)\,\Gamma(\mbox{\mancube}_R),
$$
where the Wilson surface $\mbox{\mancube}\mbox{\mancube}$ is constructed by gluing the two cubets $\mbox{\mancube}_L$ and $\mbox{\mancube}_R$ along a common face, with face variable ${\cal W}_c$. The gluing is performed by integrating over ${\cal W}_c$. Such a construction is depicted in Figure \ref{facepairs}, where the face variable that is integrated over is shaded.

\subsection{Action} \label{naction} 

As we explained in the previous section, the face variables are four-index objects which naturally transform in some representation of $U(N) \times U(N)$, although more general gauge groups may be possible. For concreteness, we will choose the bi-adjoint representation, since the basic conclusions of the paper do not depend on the representation.  As in the abelian case, the action for the theory is essentially given by the sum of the real part of each cubet \footnote{More generally, if the raised/lowered color indices of the face variables correspond to the $R/\bar{R}$ representation of $U(N)$, one simply replaces $N$ in (\ref{action}) with the dimension of the representation $R$.}:
\begin{equation}
S[B]=\beta\sum_{\mbox{\mancube}}\left(1-\frac{1}{N^3}\Re\Gamma(\mbox{\mancube})\right)
\label{action}
\end{equation}
where $\Gamma(\mbox{\mancube})$ is given by (\ref{cube}) and $\beta$ can be interpreted as the inverse temeprature of the statistical system or the inverse coupling of the field theory in the continuum limit. The factor of $N^3$ which appears in (\ref{action}) ensures that the action vanishes for a trivial gerbe configuration, i.e. where $B=0$. Indeed, for a trivial gerbe, (\ref{cube}) reduces to $\Gamma(\mbox{\mancube})=N^{3}$.

Consider the $SU(N)$ decomposition of the $B$ field in \eqref{decomp}.
If we set the bi-adjoint field $\Phi=0$, then the face variables reduce
to a product of a $U(1)$ phase and two $SU(N)$ matrices:

\[
\mathcal{W}_{\mu\nu}{}_{a}^{b}{}_{\dot{a}}^{\dot{b}}=\mathcal{W}_{\mu\nu}\mathcal{U}_{\mu\nu}{}_{a}^{b}\mathcal{V}_{\mu\nu}{}_{\dot{a}}^{\dot{b}}
\]
where $\mathcal{W}_{\mu\nu}=\exp\left(i\alpha^{2}b_{\mu\nu}\right)$
is an abelian face variable, $\mathcal{U}_{\mu\nu}{}_{a}^{b}=\exp\left(i\alpha^{2}C_{\mu\nu}{}_{a}^{b}\right)$,
and $\mathcal{V}_{\mu\nu}{}_{\dot{a}}^{\dot{b}}=\exp\left(i\alpha^{2}\tilde{C}_{\mu\nu}{}_{\dot{a}}^{\dot{b}}\right)$.
Furthermore, the cubet in \eqref{cube} reduces to 

\begin{equation}
\Gamma_{\mu\nu\lambda}\left(\vec{n}\right)=\Gamma_{abelian}\left(\vec{n}\right)W_{\mu\nu}\left(\vec{n}\right)W_{\nu\lambda}\left(\vec{n}\right)W_{\mu\nu}\left(\vec{n}\right),
\label{phi0}
\end{equation}
where $\Gamma_{abelian}$ is the abelian cubet associated with the
field $b_{\mu\nu}$, and
\[
W_{\mu \nu}\left(\vec{n}\right)=Tr\left(\Omega_{\nu\lambda}\left(\vec{n}\right)\Omega_{\lambda\mu}\left(\vec{n}\right)\Omega_{\nu\lambda}^{\dagger}\left(\vec{n}+\vec{\mu}\right)\Omega_{\lambda\mu}^{\dagger}\left(\vec{n}+\vec{\nu}\right)\right)
\]
\[
W_{\nu \lambda}\left(\vec{n}\right)=Tr\left(\mathcal{U}_{\lambda\mu}\left(\vec{n}\right)\mathcal{U}_{\mu\nu}\left(\vec{n}\right)\mathcal{U}_{\lambda\mu}^{\dagger}\left(\vec{n}+\vec{\nu}\right)\mathcal{U}_{\mu\nu}^{\dagger}\left(\vec{n}+\vec{\lambda}\right)\right)
\]
\[
W_{\lambda\mu}\left(\vec{n}\right)=Tr\left(\mathcal{V}_{\mu\nu}\left(\vec{n}\right)\mathcal{V}_{\nu\lambda}\left(\vec{n}\right)\mathcal{V}_{\mu\nu}^{\dagger}\left(\vec{n}+\vec{\lambda}\right)\mathcal{V}_{\nu\lambda}^{\dagger}\left(\vec{n}+\vec{\mu}\right)\right).
\]
(where we denote unitary matrices with primed indices as
$\Omega$). Hence, from (\ref{phi0}) we see that if we set $\Phi=0$, the non-abelian cubet
reduces to an abelian cubet times three plaquettes, which are associated with each pair of spatial directions. From this, it is
not difficult to see that upon dimensional reduction, the non-abelian
gerbe theory will give rise to non-abelian Yang-Mills fields, as well
as other fields. We will return to dimensional reduction in section \ref{reduction}.

\subsection{Gauge Transformations} \label{gaugesection} 

\begin{figure}[htbp] 
       \centering
       \includegraphics[width=5in]{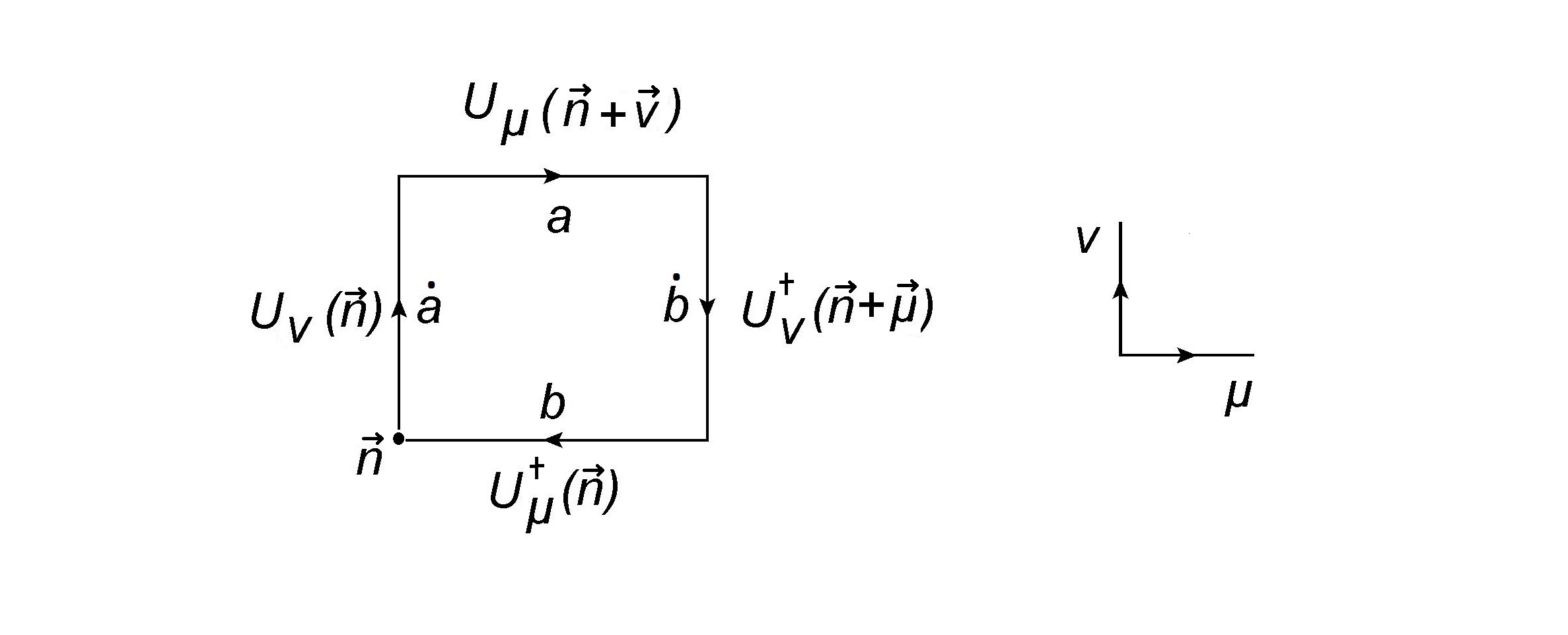}  
    \caption{Gauge transformation of nonabelian face variable.}
    \label{edgegauge}
    \end{figure}

It is natural to define gauge transformations of a face variable as multiplication by the Wilson lines along its edges. In particular, the face variable in Figure \ref{edgegauge} tansforms as follows under a gauge transformation:
\begin{equation}
({\cal W}_{\mu\nu})_{b\dot{b}}^{a\dot{a}}\rightarrow (U_{\mu})^a{}_e\;(U_{\nu})^{\dot{a}}{}_{\dot{e}}\;({\cal W}_{\mu\nu})^{e\dot{e}}_{f\dot{f}}\;(U^{\dagger}_{\mu})^f{}_b\;(U^{\dagger}_{\nu})^{\dot{f}}{}_{\dot{b}},
\label{wtrans}
\end{equation}
where
\begin{equation}
({\cal W}_{\mu\nu})^{a\dot{a}}_{b\dot{b}}(\vec{n})=\exp\Big(i\alpha^2(B_{\mu\nu})^{a\dot{a}}_{b\dot{b}}(\vec{n})\Big).
\label{wab}
\end{equation}
Here $\vec{n}$ is a lattice point at a corner of the plaquette. The gauge transformation involves Wilson lines along the $\vec{x}^{\mu}$ and $\vec{x}^{\nu}$ directions. The Wilson lines in the $\mu$-direction are
\begin{equation}
U_{\mu\,\,\, e}^{\,\,\,\, a}\left(\vec{n}+\hat{\nu}\right)=\exp\left(i\alpha\Lambda_{\mu}{}^{a}{}_{e}\left(\vec{n}+\vec{\nu}\right)\right),	\qquad	U_{\mu\,\,\, b}^{\dagger f}\left(\vec{n}\right)=\exp\left(-i\alpha\Lambda_{\mu}{}^{f}{}_{b}\left(\vec{n}\right)\right),
\label{umu}
\end{equation}
and those in the $\nu$-direction are
\begin{equation}\label{unu}
U_{\nu\,\,\,\dot{e}}^{\,\,\,\,\dot{a}}\left(\vec{n}\right)=\exp\left(i\alpha \Lambda_{\nu}{}^{\dot{a}}{}_{\dot{e}}\left(\vec{n}\right)\right),	\qquad	U_{\nu\,\,\,\dot{b}}^{\dagger\dot{f}}\left(\vec{n}+\vec{\mu}\right)=\exp\left(-i\alpha \Lambda_{\nu}{}^{\dot{f}}{}_{\dot{b}}\left(\vec{n}+\vec{\mu}\right)\right)
\end{equation}
The Wilson lines are in turn only defined up to gauge transformations associated with their vertices, as depicted in Figure \ref{edgeg}:
$$
U^{a}{}_{b}(\vec{n})\rightarrow h^{\dagger a}{}_{a'}(\vec{n}+\vec{\mu})U^{a'}{}_{b'}(\vec{n})h^{b'}{}_{b}(\vec{n}),
$$
where
$$
h^{\dagger a}{}_{a'}(\vec{n}+\vec{\mu})=\exp\left(-i\alpha\varphi^{a}{}_{a'}(\vec{n}+\vec{\mu})\right),	\qquad	h^{b'}{}_{b}(\vec{n})=\exp\left(i\alpha\varphi^{b'}{}_{b}(\vec{n})\right).
$$

\section{Classical Continuum Limit}\label{continuum}

Finding the continuum limit of a quantum theory on a lattice is a subtle issue and requires an understanding of the phase structure of the theory\footnote{The continuum limit may be taken at a phase transition, where the correlation length of the system becomes much larger than lattice spacing. See, for example, \cite{Creutz,Rothe} for details.}. In this section we will consider the non-abelian gerbe theory in terms of a small $\alpha$ expansion of the classical lattice action and discuss a thorough treatment of the quantum theory elsewhere. One reason to study the small $\alpha$ expansion of the theory is that it loosely corresponds to a derivative expansion and so the leading orders describe propagators and low-point interaction vertices that we may usefully compare with more familiar field theories such as Yang-Mills. In this section, we expand the action to cubic order in the fields and describe the classical continuum limit, where the lattice spacing goes to zero. The action is
$$
S[B]=\beta\sum_{\mbox{\mancube}}\left(1-\frac{1}{N^3}\Re(\Gamma(\mbox{\mancube}))\right)
$$
where the sum is taken over all cubets,
\begin{equation}\label{G}
\Gamma={\cal W}^{c\dot{b}}_{a\dot{c}}(\theta_1){\cal W}^{d'd}_{a'c}(\theta_2){\cal W}^{\dot{c}c'}_{\dot{d}d'}(\theta_3){\cal W}^{b\dot{d}}_{d\dot{a}}(\theta_4){\cal W}^{b'a}_{c'b}(\theta_5){\cal W}^{\dot{a}a'}_{\dot{b}b'}(\theta_6)
\end{equation}
and ${\cal W}(\theta_i)=e^{i\theta_i}$, where the $\theta_i$ are 4-index objects. Since $\theta$ is not a matrix, this is not a conventional exponential function. It may be taken as short-hand for the following
\begin{equation}\label{W}
{\cal W}^{a\dot{a}}_{b\dot{b}}=\delta^a_b\delta^{\dot{a}}_{\dot{b}}+i\theta^{a\dot{a}}_{b\dot{b}}-\frac{1}{2}\theta^{a\dot{a}}_{c\dot{c}}\theta^{c\dot{c}}_{b\dot{b}}-\frac{i}{6}\theta^{a\dot{a}}_{c\dot{c}}\theta^{c\dot{c}}_{d\dot{d}}\theta^{d\dot{d}}_{b\dot{b}}+...
\end{equation}
The $\theta$'s are given by
\begin{equation}
\begin{array}{cc}
\theta_1=-\alpha^2B_{\nu\lambda},&\qquad\theta_4=\alpha^2B_{\nu\lambda}+\alpha^3\Delta_{\mu}B_{\nu\lambda},\\
\theta_2=-\alpha^2B_{\lambda\mu},&\qquad\theta_5=\alpha^2B_{\lambda\mu}+\alpha^3\Delta_{\nu}B_{\lambda\mu},\\
\theta_3=-\alpha^2B_{\mu\nu},&\qquad\theta_6=\alpha^2B_{\mu\nu}+\alpha^3\Delta_{\lambda}B_{\mu\nu}.\nonumber
\end{array}
\end{equation}
where the lattice derivative is given by (\ref{derivative}) and we have suppressed the gauge indices which are given explicitly in (\ref{G}). The action is non-polynomial and we are interested in expanding the action up to cubic order in $\theta$. We write
$$
\Gamma=\sum_{n=0}^{\infty}\Gamma_n
$$
where $\Gamma_n$ is of order $n$ in $\theta$. It will be useful to denote the real part as $\Re\Gamma_n:={\cal V}_n$ since only the real part of $\Gamma_n$ contributes to the action. ${\cal V}_n$ will give the $n$-point vertex for the interacting gerbes. Plugging (\ref{W}) into (\ref{G}) and keeping terms only of order zero in $\theta$ we find 
${\cal V}_0=\delta^a_a\delta^{\dot{a}}_{\dot{a}}\delta^{a'}_{a'}=N^3$. This gives an overall constant shift to the action and is not physically significant. Pugging (\ref{W}) into (\ref{G}) and keeping terms only of one in $\theta$ we find
$$
\Gamma_1=iN\sum_{i=1}^6\text{Tr}_2\theta_i
$$
where $\text{Tr}_2(\theta^{a\dot{a}}_{b\dot{b}}):=\theta^{a\dot{a}}_{a\dot{a}}$ is the trace over both indices. Since ${\cal V}_1=0$ there is no contribution to the action from this term and we find tht the action may be written as
$$
S[B]=\beta\sum_{\mbox{\mancube}}\left(1-\frac{1}{N^{3}}\Re\Gamma\right)=-\frac{\beta}{N^{3}}\sum_{\mbox{\mancube}}\sum_{n=2}^{\infty}{\cal V}_{n},
$$
and we can think of ${\cal V}_n$ as the $n$-point vertex in the classical theory. As with lattice theories in general, there are an infinite number of such vertices; however, upon taking the classical continuum limit, only a finite number of these vertices survive.

\subsection{Quadratic Terms}

Plugging (\ref{W}) into (\ref{G}) and keeping terms only of second order in $\theta$ gives the quadratic part of the action
\begin{equation}\label{V2}
\frac{\beta}{N^3}\sum_{\mbox{\mancube}}{\cal V}_2=-\frac{\beta\alpha^6}{6}\sum_{\vec{n}}\sum_{1\leq \mu<\nu<\lambda\leq 6}\left(\frac{1}{2}(H_{\mu\nu\lambda})^2+\frac{1}{4N}\sum_A({\cal F}^A_{\mu\nu\lambda})^2+\frac{1}{8N^2}\sum_A({\cal G}^A_{\mu\nu\lambda})^2\right)
\end{equation}
where the components of ${\cal F}^A_{\mu\nu\lambda}$ are
$$
{\cal F}_{\mu\nu\lambda}^A=\left(\begin{array}{c}
F^M_{\mu\nu:\lambda}\\
F^{\dot{M}}_{\lambda\mu:\nu}\\
F^{M'}_{\nu\lambda:\mu}
\end{array}\right),	\qquad	{\cal G}_{\mu\nu\lambda}^A=\left(\begin{array}{c}
G^{M\dot{M}}_{\mu:\nu\lambda}\\
G^{\dot{M}M'}_{\lambda:\mu\nu}\\
G^{M'M}_{\nu:\lambda\mu}
\end{array}\right)
$$
where
$$
H_{\mu\nu\lambda}:=\Delta_{\mu}b_{\nu\lambda}+\Delta_{\lambda}b_{\mu\nu}+\Delta_{\nu}b_{\lambda\mu},
$$
$$
F^M_{\mu\nu:\lambda}:=\Delta_{\mu}C_{\nu\lambda}{}^M+\Delta_{\nu}\widetilde{C}_{\lambda\mu}^M,\qquad
F^{\dot{M}}_{\lambda\mu:\nu}:=\Delta_{\mu}\widetilde{C}_{\nu\lambda}^{\dot{M}}+\Delta_{\lambda}C_{\mu\nu}^{\dot{M}},\qquad
F^{M'}_{\nu\lambda:\mu}:=\Delta_{\nu}C_{\lambda\mu}^{M'}+\Delta_{\lambda}\widetilde{C}_{\mu\nu}^{M'},
$$
$$
G^{M\dot{M}}_{\mu:\nu\lambda}:=\Delta_{\mu}\Phi_{\nu\lambda}{}^{M\dot{M}},\qquad
G^{\dot{M}M'}_{\lambda:\mu\nu}:=\Delta_{\lambda}\Phi_{\mu\nu}{}^{\dot{M}M'}
,\qquad
G^{M'M}_{\nu:\lambda\mu}:=\Delta_{\nu}\Phi_{\lambda\mu}{}^{M'M}.
$$
Using the antisymmetry $C_{\mu\lambda}^M=-\widetilde{C}_{\lambda\mu}^M$ and similarly for other terms we can write
$$
F^M_{\mu\nu:\lambda}=2\Delta_{[\mu}C^M_{\nu]\lambda},\qquad
F^{\dot{M}}_{\lambda\mu:\nu}=2\Delta_{[\lambda}C^{\dot{M}}_{\mu]\nu},\qquad
F^{M'}_{\nu\lambda:\mu}=2\Delta_{[\nu}C^{M'}_{\lambda]\mu}.
$$
The quadratic part of the action may be written in a more compact form by introducing the field strength
\begin{eqnarray}\label{H}
({\cal H}_{\mu\nu\lambda})^{a\dot{a}a'}_{b\dot{b}b'}&:=&H_{\mu\nu\lambda}\;
\delta^a_b\,\delta^{\dot{a}}_{\dot{b}}\,\delta^{a'}_{b'}\nonumber\\
&+&F^M_{\mu\nu\lambda}\;(T_M)^a_b\,\delta^{\dot{a}}_{\dot{b}}\,\delta^{a'}_{b'}+F^{\dot{N}}_{\mu\nu\lambda}\;(T_{\dot{N}})^{\dot{a}}_{\dot{b}}\,\delta^a_b\,\delta^{a'}_{b'}+F^{P'}_{\mu\nu\lambda}\;(T_{P'})^{a'}_{b'}\,\delta^a_b\,\delta^{\dot{a}}_{\dot{b}}\nonumber\\
&+&G^{M\dot{N}}_{\mu\nu\lambda}\;(T_M)^a_b\,(T_{\dot{N}})^{\dot{a}}_{\dot{b}}\,\delta^{a'}_{b'}+G^{\dot{N}P'}_{\mu\nu\lambda}\;(T_{\dot{N}})^{\dot{a}}_{\dot{b}}\,(T_{P'})^{a'}_{b'}\,\delta^a_b+G^{P'M}_{\mu\nu\lambda}\;(T_{P'})^{a'}_{b'}\,(T_M)^a_b\,\delta^{\dot{a}}_{\dot{b}}\nonumber\\
\end{eqnarray}
The action at quadratic order may then be simply written as
$$
\frac{\beta}{N^3}\sum_{\mbox{\mancube}}{\cal V}_2=-\frac{\beta\alpha^6}{12}\sum_{\vec{n}}\sum_{1\leq \mu<\nu<\lambda\leq 6}\text{Tr}_3({\cal H}_{\mu\nu\lambda}\circ{\cal H}_{\mu\nu\lambda})
$$
where Tr$_3$ is a trace over all three pairs of free indices and the product $\circ$ is defined as
$$
({\cal H}_{\mu\nu\lambda}\circ{\cal H}_{\mu\nu\lambda})^{a\dot{a}a'}_{b\dot{b}b'}:=({\cal H}_{\mu\nu\lambda})^{a\dot{a}a'}_{c\dot{c}c'}({\cal H}_{\mu\nu\lambda})^{c\dot{c}c'}_{b\dot{b}b'}.
$$

\subsection{Cubic Terms}

The first non-trivial interactions occur at cubic order. With a little work, the real part of the cubic term may be written as ${\cal V}_{3}=V_3+\widetilde{V}_3$, where
\begin{eqnarray}
V_3&=&\frac{N^2}{2}\alpha^7\,f_{MNP}\,C_{\nu\lambda}^M\,\Delta_{\mu}C^N_{\nu\lambda}\left(C^P_{\mu\lambda}+\frac{\alpha}{2}\Delta_{\nu}C^P_{\mu\lambda}\right)\nonumber\\
&&+\frac{N^2}{2}\alpha^7\,f_{MNP}\,C_{\mu\lambda}^M\,\Delta_{\nu}C^N_{\mu\lambda}\left(C^P_{\nu\lambda}+\frac{\alpha}{2}\Delta_{\mu}C^P_{\nu\lambda}\right)+...\nonumber
\end{eqnarray}
and
\begin{eqnarray}
\widetilde{V}_3&=&\frac{N}{4}\alpha^7\delta_{\dot{M}\dot{N}}f_{MNP}\Phi_{\nu\lambda}^{M\dot{M}}\Delta_{\mu}\Phi^{N\dot{N}}_{\nu\lambda}\left(C^P_{\mu\lambda}+\frac{\alpha}{2}\Delta_{\nu}C^P_{\mu\lambda}\right)\nonumber\\
&&+\frac{N}{4}\alpha^7\delta_{M'N'}f_{MNP}\Phi_{\mu\lambda}^{MM'}\Delta_{\nu}\Phi^{NN'}_{\mu\lambda}\left(C^P_{\nu\lambda}+\frac{\alpha}{2}\Delta_{\mu}C^P_{\nu\lambda}\right)+...\nonumber
\end{eqnarray}
where the ellipsis denote similar terms that involve the structure constants $f_{\dot{M}\dot{N}\dot{P}}$ and $f_{M'N'P'}$. The full cubic expression is given in Appendix B.2, where details of this calculation may be found. We shall see in section 6 that, after dimensional reduction, this term correctly reproduces the lattice Yang-Mills interaction at cubic order, and hence conventional Yang-Mills theory in the classical continuum limit.

\subsection{Gauge transformations}

In section \ref{gaugesection}, we defined the gauge transformations for non-abelian face variables. Expanding (\ref{wtrans}) and (\ref{wab}) in powers of $\alpha$, we obtain

\begin{equation}
i\alpha^{2}B_{_{\mu\nu}\,\, b\dot{b}}^{\,\,\,\,\, a\dot{a}}\rightarrow\alpha\left[\left(\left(\Delta_{\nu}U_{\mu}\right)U_{\mu}^{\dagger}\right)_{b}^{a}\delta_{\dot{b}}^{\dot{a}}-\left(\left(\Delta_{\mu}U_{\nu}\right)U_{\nu}^{\dagger}\right)_{\dot{b}}^{\dot{a}}\delta_{b}^{a}\right]+i\alpha^{2}U_{\mu\,\,\, e}^{\,\,\,\, a}U_{\mu\,\,\, b}^{\dagger f}U_{\nu\,\,\,\dot{e}}^{\,\,\,\,\dot{a}}U_{\nu\,\,\,\dot{b}}^{\dagger\dot{f}}B_{_{\mu\nu}\,\, f\dot{f}}^{\,\,\,\,\, e\dot{e}}+\mathcal{O}\left(\alpha^{4}\right)
\label{b2b}
\end{equation}
where $U_\mu$ and $U_\nu$ are defined in (\ref{umu}) and (\ref{unu}). From this equation, it is clear that gauge transformations preserve the antisymmetry constraint (\ref{antisymmetry}).

Expanding $U_\mu$ and $U_\nu$ in powers of $\alpha$ then gives
\begin{eqnarray}
\delta(B_{\mu\nu})_{b\dot{b}}^{a\dot{a}}&=&\delta_{\dot{b}}^{\dot{a}}\left(\Delta_{\nu}\left(\Lambda_{\mu}\right){}^{a}{}_{b}\right)-\delta_{b}^{a}\left(\Delta_{\mu}\left(\Lambda_{\nu}\right){}^{\dot{a}}{}_{\dot{b}}\right)\nonumber\\
&&-i\alpha(\Lambda_{\mu})^c{}_b(B_{\mu\nu})^{a\dot{a}}_{c\dot{b}}+i\alpha(\Lambda_{\mu})^a{}_c(B_{\mu\nu})^{c\dot{a}}_{b\dot{b}}-i\alpha (B_{\mu\nu})^{a\dot{a}}_{b\dot{c}}(\Lambda_{\nu})^{\dot{c}}{}_{\dot{b}}+i\alpha(B_{\mu\nu})^{a\dot{c}}_{b\dot{b}}(\Lambda_{\nu})^{\dot{a}}{}_{\dot{c}}\nonumber\\
&&+\mathcal{O}\left(\alpha^{2}\right).\nonumber
\end{eqnarray}
Decomposing $B$ as in (\ref{decomp}) and $\Lambda$ as
$$
(\Lambda_{\mu})^a_b=\lambda_{\mu}\delta^a_b+\lambda^M_{\mu}(T_M)^a_b,
$$
where $T_M$ are generators of $SU(N)$, the gauge transformations of the component fields are given, to first order in $\alpha$, by
\begin{eqnarray}\label{finitegauge}
\delta b_{\mu\nu}&=&\Delta_{\nu}\lambda_{\mu}-\Delta_{\mu}\lambda_{\nu}\nonumber\\
\delta C_{\mu\nu}^M&=&\Delta_{\nu}\lambda^M_{\mu}+\alpha f_{NP}{}^MC^N_{\mu\nu}\lambda_{\nu}^P\nonumber\\
\delta \tilde{C}_{\mu\nu}^{\dot{M}}&=&-\Delta_{\mu}\lambda^{\dot{M}}_{\nu}+\alpha f_{\dot{N}\dot{P}}{}^{\dot{N}}\tilde{C}^{\dot{N}}_{\mu\nu}\lambda_{\mu}^{\dot{P}}\nonumber\\
\delta \Phi_{\mu\nu}^{M\dot{M}}&=&\alpha f_{NP}{}^M\lambda_{\mu}^N\Phi_{\mu\nu}^{P\dot{M}}+\alpha f_{\dot{N}\dot{P}}{}^{\dot{M}}\lambda_{\nu}^{\dot{P}}\Phi_{\mu\nu}^{M\dot{N}}
\end{eqnarray}
We have suppressed terms of order $\alpha^2$ and higher. We see that the transformations of $C_{\mu\nu}^M$ and $\Phi_{\mu\nu}^{M\dot{N}}$ resemble the gauge transformations of a Yang-Mills connection and a bi-adjoint scalar field respectively, even though both have two Lorentz indices. Note that when $\alpha\rightarrow0$, the transformations become linear. This implies that in the classical continuum limit, the theory is non-interacting. We will verify this at the level of the action in the next section.

\subsection{Classical Continuum Limit}  

In this section we look at the classical continuum limit of the non-abelian lattice gerbe theory and find that it reduces to a non-interacting theory in this limit, i.e. the interactions are suppressed by powers of the lattice spacing. Taking the naive, continuum limit involves the sending $\alpha\rightarrow 0$ and the number of lattice sites $\vec{n}$ to infinity. For simplicty we shall take $D=6$. In this double limit
$$
\sum_{\vec{n}}\alpha^6\rightarrow \int \rd^6x.
$$
In this limit the lattice derivative (\ref{derivative}) reduces to the conventional partial derivative $\Delta_{\mu}\rightarrow\partial_{\mu}$. The quadratic terms become
\begin{equation}\label{cont}
\frac{\beta}{N^3}\sum_{\mbox{\mancube}}{\cal V}_2\rightarrow-\frac{\beta}{6}\int \rd^6x\sum_{1\leq \mu<\nu<\lambda\leq 6}\left(\frac{1}{2}(H_{\mu\nu\lambda})^2+\frac{1}{4N}\sum_{i=1}^3(F^i_{\mu\nu\lambda})^2+\frac{1}{8N^2}\sum_{i=1}^3(G^i_{\mu\nu\lambda})^2\right)
\end{equation}
where
$$
H_{\mu\nu\lambda}:=3\partial_{[\mu}b_{\nu\lambda]},
$$
$$
F^M_{\mu\nu\lambda}:=\partial_{[\mu}C^M_{\nu]\lambda},\qquad
F^{\dot{N}}_{\mu\nu\lambda}:=\partial_{[\lambda}C^{\dot{N}}_{\mu]\nu},\qquad
F^{P'}_{\mu\nu\lambda}:=\partial_{[\nu}C^{P'}_{\lambda]\mu},
$$
$$
G^{M\dot{N}}_{\mu\nu\lambda}:=\partial_{\mu}\Phi^{M\dot{N}}_{\nu\lambda},\qquad
G^{\dot{N}P'}_{\mu\nu\lambda}:=\partial_{\lambda}\Phi^{\dot{N}P'}_{\mu\nu}
,\qquad
G^{P'M}_{\mu\nu\lambda}:=\partial_{\nu}\Phi^{P'M}_{\lambda\mu}.
$$
The contributions from ${\cal V}_n$, with $n>2$ all vanish in the $\alpha\rightarrow 0$ limit, as may be seen explicitly in the cubic terms (\ref{K1}) and (\ref{K2}). Thus, the classical theory is free in the continuum limit. Moreover, in the absence of non-linear terms, the gauge transformations become abelian
$$
\delta b_{\mu\nu}=\partial_{\mu}\lambda_{\nu}-\partial_{\nu}\lambda_{\mu},	\qquad	\delta C^M_{\mu\nu}=\partial_{\mu}\lambda^M_{\nu},	\qquad	\delta \Phi_{\mu\nu}^{M\dot{N}}=0.
$$
The fact that the theory becomes non-interacting in the classical continuum limit is consistent the results of \cite{Nepomechie:1982rb} and with the no-go theorem of \cite{Henneaux:1986ht}. On the other hand, we will demonstrate in the next section that if we dimensionally reduce the theory \emph{before} taking the classical continuum limit, this gives rise to non-abelian lattice Yang-Mills theory, so the terms which are suppressed by lattice spacing in the classical theory encode Yang-Mills interactions.  Furthermore, in sections \ref{analytics} and \ref{numerics}, we will show that at nonzero lattice spacing, the quantum theory is non-trivial (for example, Wilson surfaces exhibit a volume law at strong coupling). This suggests the possibility that the continuum limit in the quantum theory may be different than the continuum limit in the classical theory due to renormalization. Since the gerbe can in principle be coupled to strings which reside on the links of the lattice, we might like to reinterpret the lattice spacing as a string scale and the action as a higher derivative effective action (or string field theory) for a non-gravitational string theory.

\section{Dimensional Reduction} \label{reduction}

In this section, we consider the dimensional reduction of the lattice gerbe theory we have introduced in preceeding sections. We will begin by dimensionally reducing a truncation of the non-abelian gerbe theory, in order to demonstrate how non-abelian lattice Yang-Mills arises in the simplest possible setting. We then describe the dimensional reduction of the full theory, which will give rise to lattice Yang-Mills theory, in addition to other fields. Formally, dimensional reduction of a continuum theory on a circle requires that a spatial direction is made compact and a harmonic (Fourier) analysis is performed on all fields along this direction. In the event that the radius of the compact direction is taken to zero, the higher modes become infnitely masive and decouple from the theory leaving an effective theory of zero modes living in the remaining dimensions. The classical action of the zero modes can be obtained directly by taking the fields to have no dependence on the compact direction, and integrating over the compact direction. For concreteness, we shall take the dimension of the lattice to be six and consider a reduction down to five dimensions.

From the discussions above regarding the classical continuum limit, one might anticipate that a dimensional reduction of the gerbe theory simply gives rise to a lower dimensional theory that includes Maxwell fields in addition to gerbes. Indeed, if we reduce along the 6'th direction, those fields that have one component along the 6'th direction are naturally interpreted as one-forms in the lower dimensional theory
$$
\left(\alpha \,b_{\mu_i\mu_6},\alpha \,C^M_{\mu_i\mu_6},\alpha\,\widetilde{C}^{\dot{M}}_{\mu_i\mu_6},\alpha\,\Phi^{M\dot{N}}_{\mu_i\mu_6}\right)\rightarrow \left(a_{\mu_i},A^M_{\mu_i},\widetilde{A}^{\dot{M}}_{\mu_i},\phi^{M\dot{N}}_{\mu_i}\right),	\qquad	i\neq 6,
$$
however, the main result of this section is that, if we dimensioanlly reduce our lattice gerbe theory and \emph{then} take the continuum limit, we find \emph{non-abelian} Yang-Mills, rather than Maxwell theory included in the lower dimensional field theory. A glimpse of this may be seen from the dimensional reduction of the gauge transformations of the fields at finite latice spacing (\ref{finitegauge}). The result is
\begin{eqnarray}
\delta a_{\mu}&=&\Delta_{\mu}\lambda,\nonumber\\
\delta A^M_{\mu}&=&\Delta_{\mu}\lambda^M+f_{NP}{}^MA_{\mu}^N\lambda^P\nonumber\\
\delta \widetilde{A}^{\dot{M}}_{\mu}&=&\alpha f_{\dot{N}\dot{P}}{}^{\dot{M}}\widetilde{A}^{\dot{N}}\lambda^{\dot{P}}_{\mu}\nonumber\\
\delta \phi^{M\dot{M}}_{\mu}&=&f_{NP}{}^M\lambda^N\phi_{\mu}^{P\dot{M}}-\alpha f_{\dot{N}\dot{P}}{}^{\dot{M}}\lambda^{\dot{N}}_{\mu}\phi_{\mu}^{M\dot{P}},\nonumber
\end{eqnarray}
where $\alpha\,\lambda^M_{\mu_6}\rightarrow \lambda^M$ under dimensional reduction. We see that $A^M_{\mu}$ transforms as a standard $SU(N)$ lattice Yang-Mills connection. Moreover, if we send $\alpha$ to zero, we have
\begin{eqnarray}\label{gauge}
\delta a_{\mu}&=&\partial_{\mu}\lambda,\nonumber\\
\delta A^M_{\mu}&=&D_{\mu}\lambda^M:=\partial_{\mu}\lambda^M+f_{NP}{}^MA_{\mu}^N\lambda^P\nonumber\\
\delta \widetilde{A}^{\dot{M}}_{\mu}&=&0\nonumber\\
\delta \phi^{M\dot{M}}_{\mu}&=&f_{NP}{}^M\lambda^N\phi_{\mu}^{P\dot{M}},
\end{eqnarray}
so that the classical continuum limit includes a conventional Yang-Mills connection $A^M_{\mu}$. We also see that $\phi^{M\dot{M}}_{\mu}$ transforms convariantly in the continuum limit and so we would expect that the continuum limit includes non-linear terms such that
\begin{equation}\label{cov}
G_{\nu\mu}^{M\dot{N}}=\partial_{\nu}A_{\mu}^{M\dot{N}}\rightarrow D_{\nu}A_{\mu}^{M\dot{N}},	\qquad	F_{\mu\nu}^M=2\partial_{[\mu}A_{\nu]}^M\rightarrow 2\partial_{[\mu}A^M_{\nu]}+f_{NP}{}^MA_{[\mu}^NA^P_{\nu]}.
\end{equation}
Such modifications would appear as cubic and quartic terms in the Lagrangian. In particular, we would expect to see cubic terms of the form $\delta_{\dot{M}\dot{N}}f_{MNP}A^{M\dot{M}}_{\mu}\partial_{\nu}A^{N\dot{N}}_{\mu}A^P_{\nu}$ and $f_{MNP}A^M_{\mu}\partial_{\nu}A^N_{\mu}A^P_{\nu}$. This is precicesly what we find in section 6.2 from the continuum limit of the reduced cubic terms in the lattice gerbe theory. Note that fields which do not have indices along the 6'th direction are not re-scaled by the lattice spacing, so their gauge transformations become linear in the classical continuum limit.

\subsection{Reduction of a Truncated Theory}

\begin{figure}  
  \centering
    \includegraphics[width=5in]{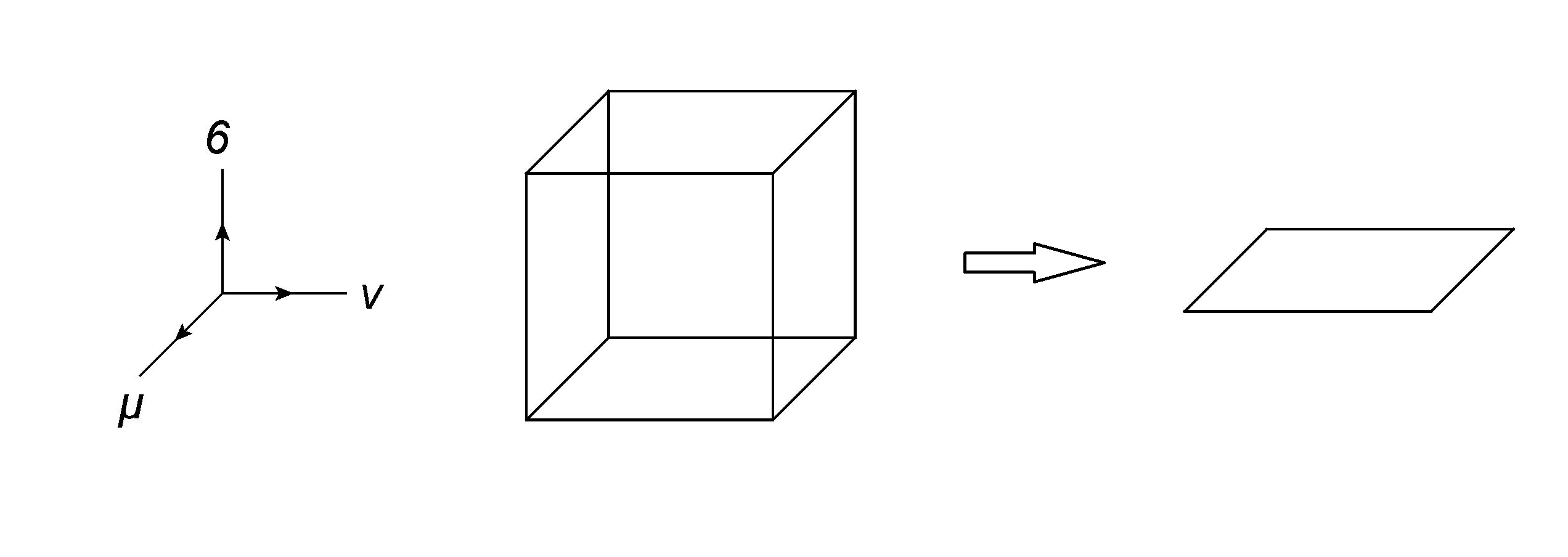}  
     \caption{In the truncated theory, dimensional reduction of a cubet along the 6th direction effectively collapses it to a plaquette.}
\label{dimred}
    \end{figure}

To illustrate the arguments more clearly it will be helpful to first consider the dimensional reduction of a truncation of the full theory. We will set out the reduction of the full theory in the next section. Let us consider a non-abelian gerbe in six dimensions, and dimensionally reduce along the 6'th direction. As we explained above, this can be accomplished by dropping dependence of the fields along the 6th direction, and summing along the 6th direction. We then obtain the following five-dimensional action
\[
S_{5d}=\beta \ell\sum_{\vec{n}}\left[\sum_{1\leq\mu<\nu<\lambda\leq5}\left(1-\frac{1}{N^{3}}\Re\Gamma_{\mu\nu\lambda}\left(\vec{n}\right)\right)+\sum_{1\leq\mu<\nu\leq5}\left(1-\frac{1}{N^{3}}\Re\Gamma_{\mu\nu}(\vec{n})\right)\right]
\]
where $\ell$ is the length of the 6th direction in lattice units, and the sum over $\vec{n}$ is over a 5d hyperplane holding the 6'th coordinate fixed. Note that there are two types of cubets: those which live in 5 dimensions, and those which have an edge in the 6th direction, as depicted in Figure \ref{dimred}. We will restrict our attention to the latter \footnote{The former simply gives a lattice gerbe theory of the kind we have been describing but in five dimensions.}. In particular, they are given by:
\begin{equation}
\Gamma_{\mu\nu}\left(\vec{n}\right)=\mathcal{W}_{\mu\nu}{}_{a_{4}\dot{a}_{2}}^{a_{3}\dot{a}_{1}}\left(\vec{n}\right)\mathcal{W}_{\nu6}{}_{\dot{a}_{7}a_{6}'}^{\dot{a}_{2}a_{5}'}\left(\vec{n}\right)\mathcal{W}_{6\mu}{}_{a_{9}'a_{3}}^{a_{6}'a_{8}}\left(\vec{n}\right)\mathcal{W}_{\mu\nu}^{\dagger}{}_{a_{8}\dot{a}_{10}}^{a_{11}\dot{a}_{7}}\left(\vec{n}\right)\mathcal{W}_{\nu6}^{\dagger}{}_{\dot{a}_{1}a_{12}'}^{\dot{a}_{10}a_{9}'}\left(\vec{n}+\vec{\mu}\right)\mathcal{W}_{6\mu}^{\dagger}{}_{a_{5}'a_{11}}^{a_{12}'a_{4}}\left(\vec{n}+\vec{\nu}\right).
\label{cubet6}
\end{equation}
We obtained equation by setting $\lambda=6$ in (\ref{cube}) and noting that $\mathcal{W}_{\mu\nu}(\vec{n})= \mathcal{W}_{\mu\nu}(\vec{n}+\vec{\lambda})$.
Recall that the face variables can be expressed in terms of gerbe fields
\[
(\mathcal{W}_{\rho\lambda})_{b\dot{b}}^{a\dot{a}}=\exp\left(i\alpha^{2}B_{\rho\lambda}{}_{b\dot{b}}^{a\dot{a}}\right).
\]
The gerbe fields can in turn be decomposed into irreducible representations of $SU(N)$ as
\[
B_{\rho\lambda}{}_{b\dot{b}}^{a\dot{a}}=b_{\rho\lambda}\delta_{b}^{a}\delta_{\dot{b}}^{\dot{a}}+C_{\rho\nu}{}_{b}^{a}\delta_{\dot{b}}^{\dot{a}}+\widetilde{C}_{\rho\nu}{}_{\dot{b}}^{\dot{a}}\delta_{b}^{a}+\Phi_{\rho\nu}{}_{b\dot{b}}^{a\dot{a}}
\]
where the components are traceless and hermitian on each pair of indices. Thus far, the analysis is completely general. In order to highlight the emergence of the Yang-Mills sector without the clutter of other fields, we now perform a truncation. This truncation is only applied here to streamline the argument, the essential result - that Yang-Mills theory arises in the reduced theory - also holds in the general case, as we shall see in the next section. The edges which lie along the 6th direction are associated with primed indices. Hence, the four face variabes which have an edge in the 6th direction, will have a pair of primed indices. To simplify our analysis further, for these face variables, we will restrict our attention to fields which are singlets with respect to unprimed or dotted indices. In other words, we will impose the truncation
\[
B_{\nu6}{}_{\dot{b}b'}^{\dot{a}a'}=\left(b_{\nu6}\delta_{b'}^{a'}+\widetilde{C}_{\nu6}{}_{b'}^{a'}\right)\delta_{\dot{b}}^{\dot{a}}, \qquad
B_{6\mu}{}_{b'b}^{a'a}=\left(b_{6\mu}\delta_{b'}^{a'}+C_{6\mu}{}_{b'}^{a'}\right)\delta_{b}^{a}.
\]
The four face variables with primed indices then reduce to
\begin{eqnarray}
\begin{array}{cc}
\mathcal{W}_{\nu6}{}_{\dot{a}_{7}a_{6}'}^{\dot{a}_{2}a_{5}'}\left({\bf n}\right)=U_{\nu}^{\dagger}{}_{a_{6}'}^{a_{5}'}\left({\bf n}\right)\delta_{\dot{a}_{7}}^{\dot{a}_{2}}, &\qquad \mathcal{W}_{6\mu}{}_{a_{9}'a_{3}}^{a_{6}'a_{8}}\left({\bf n}\right)=U_{\mu}{}_{a_{9}'}^{a_{6}'}\left({\bf n}\right)\delta_{a_{3}}^{a_{8}}\\
\mathcal{W}_{\nu6}^{\dagger}{}_{\dot{a}_{1}a_{12}'}^{\dot{a}_{10}a_{9}'}\left({\bf n}+\hat{\mu}\right)=U_{\nu}{}_{a_{12}'}^{a_{9}'}\left({\bf n}+\hat{\mu}\right)\delta{}_{\dot{a}_{1}}^{\dot{a}_{10}}, & \qquad\mathcal{W}_{6\mu}^{\dagger}{}_{a_{5}'a_{11}}^{a_{12}'a_{4}}\left({\bf n}+\hat{\nu}\right)=U_{\mu}^{\dagger}{}_{a_{5}'}^{a_{12}'}\left({\bf n}+\hat{\nu}\right)\delta{}_{a_{11}}^{a_{4}}
\end{array}\nonumber
\end{eqnarray}
where
\[
U_{\mu}{}_{b'}^{a'}=\exp\left[i\alpha^{2}\left(b_{6\mu}\delta_{b'}^{a'}+C_{6\mu}{}_{b'}^{a'}\right)\right]
\]
and we used the antisymmetry propetry in equations \ref{antcomp}. In our truncated theory, the cubet in (\ref{cubet6}) reduces to
\[
\Gamma_{\mu\nu}\left(\vec{n}\right)=N^{2}tr\left[U_{\nu}^{\dagger}\left(\vec{n}\right)U_{\mu}\left(\vec{n}\right)U_{\nu}\left(\vec{n}+\vec{\mu}\right)U_{\mu}^{\dagger}\left(\vec{n}+\vec{\nu}\right)\right]
\]
where (\ref{constraints}) has been used. Furthermore, the 5d lattice action is given by $S_{5d}=S_{5d}[{\cal W}]+S_{5d}[U]$, where $S_{5d}[{\cal W}]$ is the action for a five-dimensioanal gerbe and
\begin{equation}
S_{5d}[U]=\beta \ell\sum_{\vec{n}}\sum_{1\leq\mu<\nu\leq5}\left(1-\frac{1}{N}\Re tr\left[U_{\nu}^{\dagger}\left(\vec{n}\right)U_{\mu}\left(\vec{n}\right)U_{\nu}\left(\vec{n}+\vec{\mu}\right)U_{\mu}^{\dagger}\left(\vec{n}+\vec{\nu}\right)\right]\right),
\label{5d}
\end{equation}      
is precisely the action for lattice $U(N)$ Yang-Mills theory. Hence, dimensionally reducing the truncated theory effectively collapses those cubets with an edge along the 6th direction into plaquettes, as depicted in Figure \ref{dimred}. As we will see in the next subsection, this still holds true when we dimensionally reduce the full theory. In that case, we obtain $U(N)$ Yang-Mills theory coupled to various other fields in five dimensions. It is therefore natural to parametrize the unitary matrices in (\ref{5d}) terms of a Yang-Mills gauge field as follows:
\[
U_{\mu}{}_{b'}^{a'}=\exp\left(i\alpha A_{\mu}{}_{b'}^{a'}\right),	\qquad	A_{\mu}{}_{b'}^{a'}=\alpha\left(b_{6\mu}\delta_{b'}^{a'}+C_{6\mu}{}_{b'}^{a'}\right).
\] 
Note that we have re-scaled the fields by a power of $\alpha$, which is required by dimensional analysis. In principle, we could rescale by any multiple of $\alpha$, a point that will be elaborated upon below. Taking the classical continuum limit gives
\[
S_{5d}[U]\rightarrow\frac{1}{g^2}\int \rd^{5}x\;\text{Tr}\left(F_{\mu\nu}F^{\mu\nu}\right)
\]
where $F_{\mu\nu}{}_{b'}^{a'}=\left(\partial_{\mu}A_{\nu}-\partial_{\nu}A_{\mu}-i\left[A_{\mu},A_{\nu}\right]\right){}_{b'}^{a'}$ and $g^{-2}=R/(\beta l^2)$, giving a conventional $U(N)$ gauge theory.      

To summarize, we find the following three phenomena:
\begin{enumerate}
\item If we take the classical continuum limit of the theory prior to dimensional reduction, we obtain a noninteracting theory, as we described in section \ref{continuum}. On the other hand, if we dimensional reduce the theory and then take the classical continuum limit, we obtain an interacting theory containing $U(N)$ Yang-Mills. Hence, taking the classical continuum limit does not commute with dimensional reduction. In particular, we see that Yang-Mills interactions are encoded in the interactions of the non-abelian gerbe theory at non-zero lattice spacing.\footnote{We shall discuss a subtlety relating to this observation in section 6.3 below.}

\item In the non-abelian gerbe theory, pairs of color indices are tied to spatial indices, so tranformations of spatial indices should be accompanied by transformations of color indices. On the other hand, after dimensionally reducing along the 6th direction, the color indices of the resulting Yang-Mills fields are associated with the sixth direction, and are therefore inert under transformations of the five remaining spatial dimensions. Hence, when we take the classical continuum limit of the dimensionally reduced theory, the resulting theory is Lorentz invariant \footnote{Since we are working in Euclidean signature, this actually corresponds to rotational invariance.}. 

\item Had we dimensionally reduced the non-abelian gerbe theory along another spatial direction, the resulting Yang-Mills fields would have color indices associated with that spatial direction. We have taken the gauge group to be $U(N)$ along all spatial directions in the nonabelian lattice gerbe theory, but had we made a more general choice, dimensionally reducing along different directions would actually give Yang-Mills theories with different gauge groups, which is reminiscent of Langlands duality \cite{Kapustin:2006pk,Witten:2009at}. Since the gauge group $U(N)$ is self-dual under Langlands duality, this motivates choosing the group to be $U(N)$ along all directions in the non-abelian lattice gerbe theory.       
\end{enumerate}
Note that we have only demonstrated these properties in a truncated version of the nonabelian lattice gerbe theory. When we dimensionally reduce the full nonabelian lattice gerbe theory, we will obtain many other fields in addition to Yang-Mills. Furthermore, we have been working in the classical theory, and it is conceivable that the results may be different in the quantum theory, especially in the absence of supersymmetry. Nevertheless, it is remarkable that our simple model for a non-abelian gerbe theory naturally gives rise to non-abelian Yang-Mills theory upon dimensional reduction and hints at a potential mechanism for realizing Langlands duality from dimensional reduction. Non-abelian lattice gerbe theory may provide a promising avenue of investigation in definining the $(2,0)$ theory once self-duality and supersymmetry have been correctly incorporated.     

\subsection{Reducing the Full Theory}

To make contact with our usual picture of dimensional reduction of a classical action and also to get a feel for what happens in the most general case (i.e. without the additional truncation described above), we study the dimensional reduction of the small $\alpha$ expansion of the full theory in 6 dimensions. Consider the reduction along the direction $x^{\lambda}$. We define
$$
\alpha b_{\nu\lambda}\rightarrow a_{\nu},	\qquad	\alpha C_{\nu\lambda}^M\rightarrow A_{\nu}^M,	\qquad	\alpha C_{\lambda\nu}^M=-\alpha \widetilde{C}_{\nu\lambda}^M\rightarrow -\widetilde{A}_{\nu}^M,	\qquad	\alpha \Phi_{\mu\lambda}^{M\dot{M}}\rightarrow \phi_{\mu}^{M\dot{M}}.
$$
Dimensional reduction of the quadratic terms (\ref{cont}) is straightforward. The details are given in the Appendix B.1. Taking the continuum limit of the dimensionally-reduced quadratic terms gives the quadratic action
$$
S^{(2)}=-\frac{\widehat{\beta}_1}{12N^3}\int \rd ^5 x \text{Tr}_3{\cal H}_{\mu\nu\lambda}{\cal H}^{\mu\nu\lambda}-\frac{\widehat{\beta}_2}{12N^3}\int \rd ^5x \text{Tr}_3{\cal F}_{\mu\nu}{\cal F}^{\mu\nu}
$$
where the Lorentz indices are five dimensional, ${\cal H}_{\mu\nu\lambda}$ is defined in (\ref{H}), and ${\cal F}_{\mu\nu}$ is given by
\begin{eqnarray}\label{F}
({\cal F}_{\mu\nu})^{a\dot{a}a'}_{b\dot{b}b'}&=&F_{\mu\nu}\delta^a_b\delta^{\dot{a}}_{\dot{b}}\delta^{a'}_{b'}+F^M_{\mu\nu}(T_M)^a_b\delta^{\dot{a}}_{\dot{b}}\delta^{a'}_{b'}+\widetilde{G}^{\dot{M}}_{\mu\nu}(T_{\dot{M}})^{\dot{a}}_{\dot{b}}\delta^a_b\delta^{a'}_{b'}+\widetilde{G}_{\mu\nu}^{M'}(T_{M'})\delta^a_b\delta^{\dot{a}}_{\dot{b}}\nonumber\\
&&+G^{M\dot{M}}_{\mu\nu}(T_M)^a_b(T_{\dot{M}})^{\dot{a}}_{\dot{b}}\delta^{a'}_{b'}+G^{MM'}_{\mu\nu}(T_M)^a_b(T_{M'})^{a'}_{b'}\delta^{\dot{a}}_{\dot{b}}
\end{eqnarray}
where the $SU(N)$-irreducible components of these field are
$$
F_{\mu\nu}:=2\partial_{[\mu}a_{\nu]}, \qquad
F^M_{\mu\nu}:=2\partial_{[\mu}A^M_{\nu]},
$$
$$
\widetilde{G}^{\dot{M}}_{\mu\nu}:=\partial_{\mu}\widetilde{A}^{\dot{M}}_{\nu},\qquad
\widetilde{G}^{M'}_{\nu\mu}:=\partial_{\nu}\widetilde{A}^{M'}_{\mu},\qquad
G^{M\dot{M}}_{\mu\nu}:=\partial_{\mu}\phi^{M\dot{M}}_{\nu}
,\qquad
G^{M'M}_{\nu\mu}:=\partial_{\nu}\phi^{M'M}_{\mu}.
$$
The coupling constants in five dimensions are $\widehat{\beta}_1=\beta\alpha$ and $\widehat{\beta}_2=\beta/\alpha$. 

The cubic terms of the dimensionaly reduced lattice theory have the classical continuum limit
\begin{eqnarray}
S^{(3)}&=&-\frac{\widehat{\beta}}{12N}\int\rd^5x\; f_{MNP}\left(A^M_{\nu}\partial_{\mu}A^N_{\nu} A^P_{\mu}+A^M_{\mu}\partial_{\nu}A^N_{\nu} A^P_{\nu}\right)\nonumber\\
&&-\frac{\widehat{\beta}}{24N^2}\int\rd^5x\; f_{MNP}\left(\delta_{\dot{M}\dot{N}}\phi^{M\dot{M}}_{\nu}\partial_{\mu}\phi^{N\dot{N}}_{\nu} A^P_{\mu}+\delta_{M'N'}\phi^{MM'}_{\mu}\partial_{\nu}\phi^{NN'}_{\nu} A^P_{\nu}\right)\nonumber
\end{eqnarray}
This gives, at cubic order, the correct non-abelian completion of $S^{(2)}$ suggested by the gauge transformations (\ref{gauge}). The manifest gauge-invariance of the theory, the gauge transformations (\ref{gauge}), and the quadratic action $S^{(2)}$ are enough to infer that the full form of the continuum action is given by $S^{(2)}$ with $F^M_{\mu\nu}$, $G^{M\dot{M}}_{\mu\nu}$, and $G^{M'M}_{\nu\mu}$ as given by (\ref{cov}). Expanding the lattice gerbe action to quartic order, dimensionally reducing, and then taking the continuum limit, would provide additional confirmation of this result.

In summary, if we dimensionally reduce the $U(N) \times U(N)$ lattice gerbe theory and take the classical continuum limit, we obtain $U(N)$ Yang-Mills theory coupled to a bi-adjoint field $\phi_\nu^{M \dot{M}}$, as well as a number of other decoupled fields. It is intriguing to note that if we set the bi-adjoint field $\Phi_{\mu \nu}^{M \dot{N}}$ in the $SU(N)$ decomposition of the non-abelian gerbe to zero, dimensional reduction followed by the classical continuum limit would give rise to pure $U(N)$ Yang-Mills theory, and all the other fields would be completely decoupled.

\subsection{A subtlety regarding the continuum limit}

We have seen that the dimensional reduction of the lattice gerbe theory gives rise to a theory that includes conventional lattice Yang-Mills theory. In contrast to the continuum theory for which there is a single length scale given by the internal radius $R$ of the compact direction, the lattice theory has two natural length scales: $\alpha$ and $R=\ell\alpha$. As long as we restrict our questions to the theory at finite lattice spacing, the question of the whether we adopt the convention $\alpha\, B_{\mu\lambda}=A_{\mu}$, or $\ell\alpha\, B_{\mu\lambda}=A_{\mu}$ is largely a matter of taste, the two choices being related by the obvious field redefinition. More care is required if we wish to consider the classical continuum limit of the dimensionally reduced theory. To better understand this, consider the cubic terms for the reduced theory in which we use the alternative definition $\ell\alpha\, B_{\mu\lambda}=A_{\mu}$. These terms may then be written as
\begin{eqnarray}
S_{5d}^{(3)}=\frac{\alpha^5}{2N\ell}\sum_{\textbf{n}}\sum_{1\leq \mu<\nu\leq 5}\left(\frac{1}{g_F^2}{\cal L}_3+\frac{1}{g_H^2}\tilde{\cal L}_3\right)
\end{eqnarray}
where ${\cal L}_3$ and $\widetilde{\cal L}$ include lattice Yang-Mills cubic terms\footnote{Specicifically
$$
{\cal L}_3=f_{MNP}\left(A_{\nu}^MA_{\mu}^P\Delta_{\mu}A^N_{\nu}+\delta_{ST}\phi^{MS}_{\nu}A_{\mu}^P\Delta_{\mu}\phi^{NT}_{\mu}+(\mu\leftrightarrow\nu)\right)+...
$$
$$
\tilde{\cal L}_3=f_{MNP}\left(\widetilde{A}_{\nu}^M\widetilde{C}_{\mu\nu}^P\Delta_{\mu}\widetilde{A}^N_{\nu}+\delta_{ST}\phi^{MS}_{\nu}C_{\mu\nu}^P\Delta_{\mu}\phi^{NT}_{\mu}+(\mu\leftrightarrow\nu)\right)+...
$$
where the ellipsis denote terms of order $\alpha$.} and $g_F^{-2}=\beta/\ell\alpha$ and $g_H^{-2}=\beta\ell\alpha$. It is instructive to further reduce the theory to four dimensions. Imposing the identification $n^5\sim n^5+\alpha\tilde{\ell}$ in the 5'th direction, the reduced theory has a similar form to the five-dimensonal theory, but includes additional scalar and vector fields. The four-dimensional coupling constants for the gerbe and Yang-Mills components are
$$
\hat{g}_F=\sqrt{\frac{\ell}{\beta\tilde{\ell}}},	\qquad	\hat{g}_H=\sqrt{\frac{1}{\beta\alpha^2\ell\tilde{\ell}}}.
$$
We note that, if we set the six-dimensional coupling to $\beta=1$, then reversing the order of the dimensional reduction inverts the coupling contstant $\hat{g}_F$, which may be written as $\hat{g}_F^2=\tilde{R}/R$.

If we wish to take the continuum limit at finite $R$, it is natural to take the limit $\alpha\rightarrow 0$, $\ell\rightarrow\infty$ such that the product $\alpha\ell\rightarrow R$, is a finite constant. In this case, if we hold $\beta$ fixed in this limit, we obtain a free theory in lower dimensions. By contrast, we may consider the limit in which $\alpha\rightarrow 0$ keeping $\ell$ finite. Naively, this would give rise to a $R\rightarrow 0$ limit of a continuum theory which includes non-linear interaction terms. In summary, if we take the classical continuum limit of the theory prior to dimensional reduction, we obtain a noninteracting theory, as described in section \ref{continuum}. On the other hand, if we dimensionally reduce the theory and then take the classical continuum limit, keeping $\ell$ finite, we obtain an interacting theory containing $U(N)$ Yang-Mills. Thus, if we keep $\ell$ finite, taking the classical continuum limit does not commute with dimensional reduction. In particular, we see that Yang-Mills interactions are encoded in the interactions of the non-abelian gerbe theory at non-zero lattice spacing.

This result should not be a suprise. The cubic vertex which takes the schematic form $B\Delta B(B+\alpha\Delta B)$ is a sum of dimension seven and dimension eight operators. In a six-dimensional continuum theory such operators would be considered as a non-renormalizable interaction and discarded. What we are seeing is that, upon dimensional reduction, we have cubic interactions that include terms of the form $A^2\Delta A$ which are of dimension four and give important nonlinear contributions of the Yang-Mills theory. The importance of the lattice construction is that, by introducing a fundamental dimensionful constant (the lattice spacing $\alpha$), it is possible to sensibly incorporate such $B^2\Delta B$ terms that give rise to $A^2\Delta A$ terms in four dimensions. A similar story is expected for the quartic terms. We hope to return to the question of whether this mechanism may be realised in more general quantum theories elsewhere.

\section{Wilson Surfaces: Analytical Results} \label{analytics}

In this section, we obtain various analytical results for Wilson surfaces
in lattice gerbe theory. In section \ref{sub:Continuum-Limit}, we
compute Wilson surfaces in the classical continuum limit, and find
a volume law in three dimensions and an area law in six dimensions. In section \ref{sub:Abelian-Theory}, we consider the abelian
gerbe theory on a three-dimensional lattice, and obtain an exact expression
for Wilson surfaces for any value of the coupling. In particular,
we find a volume law for any value of the coupling in three dimensions.
In section \ref{sub:Non-abelian-theory}, we compute Wilson surfaces
in the non-abelian theory, and show that they exhibit a volume law
at strong coupling. We will verify the analytical results for the abelian theory numerically in section \ref{numerics}.

\subsection{Wilson Surfaces in the Continuum Limit\label{sub:Continuum-Limit}}

In the classical continuum limit, the lattice gerbe theory reduces
to the non-interacting theory in equation (\ref{cont}). In this section, we will compute Wilson surfaces in this non-interacting theory. We expect that this should provide a good approximation for Wilson surfaces which are much larger than the lattice spacing in the weakly coupled lattice gerbe theory, and confirm this numerically for the abelian lattice gerbe theory in section \ref{numerics}.
\begin{figure}[h] 
       \centering
       \includegraphics[width=6in]{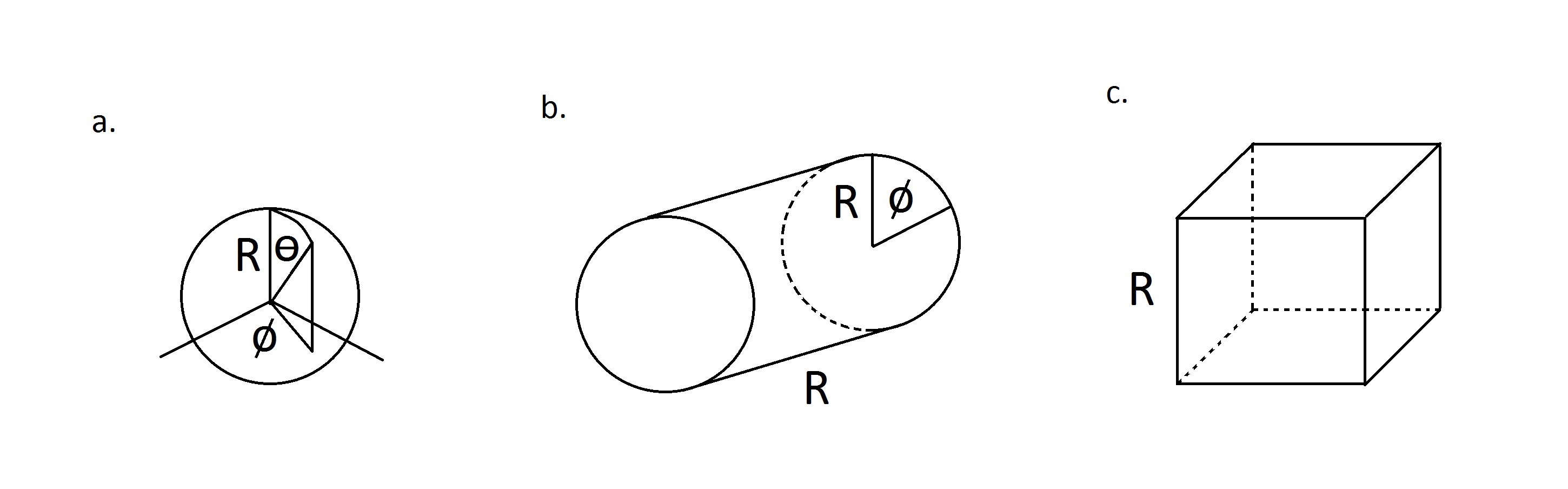}  
    \caption{Various Wilson surface geometries. It is convenient to use a spherical geometry (a) for the abelian 2-form $b_{\mu \nu}$, a cylindrical geometry (b) for adjoint field field $C^M_{\mu \nu}$ (where the length of the cylinder is along the $\nu$ direction), and a cubic geometry (c) for the bi-adjiont field $\Phi^{M \dot{N}}_{\mu \nu}$.}
    \label{surface}
    \end{figure}

Note that the action in equation  (\ref{cont}) contains three types of fields: a standard abelian 2-form $b_{\mu\nu}$, an adjoint field $C^M_{\mu\nu}$ with a Yang-Mills-like
kinetic terms, and a bi-adjiont field $\Phi^{M \dot{N}}_{\mu\nu}$ with scalar-like kinetic terms. We first compute the expectation value of a Wilson surface for the abelian 2-form \footnote{Note that a similar calculation was carried out for an abelian self-dual 2-form in \cite{Gustavsson:2004gj}.}. A natural gauge choice is 
\begin{equation}
\partial_{\mu}b^{\mu\nu}=0.\label{eq:lorentz}
\end{equation}
In this gauge, the equations of motion reduce to the Laplace equation
in Euclidean signature $\Box b_{\mu\nu}=0$. In $D\neq 2$ dimensions, the propagator for the abelian 2-form is therefore
\begin{equation}\label{prop}
\left\langle b_{\mu\nu}\left(\vec{x}\right)b_{\rho\lambda}\left(\vec{y}\right)\right\rangle =G_{\mu\nu,\rho\lambda}\left(\vec{x},\vec{y}\right)=\frac{\delta_{\mu\rho}\delta_{\nu\lambda}-\delta_{\nu\rho}\delta_{\mu\lambda}}{\left|\vec{x}-\vec{y}\right|^{D-2}},
\end{equation}
where we are neglecting numerical prefactors, which will not be important for our discussion. In the free theory, the log of the expectation value of a Wilson surface operator $\Gamma$ is simply given by computing the propagator between
two points on the surface and integrating over the location of these
points:
\begin{equation}
-\beta \ln\left\langle \Gamma\right\rangle =  \int \rd^{2}\sigma_{\mu\nu}\int \rd^{2}\sigma'_{\rho\lambda}\, G_{\mu\nu,\rho\lambda}\left(\sigma,\sigma'\right),\label{eq:abelianwilsonloop}
\end{equation}
where $\sigma,\sigma'$ are volume elements which are integrated over the surface. Since the propagator generally becomes singular when the points coincide, one
should insert a cutoff which prevents this from happening. We can
interpret this cuttoff as the lattice spacing.

For simplicity, let's take the surface to be a sphere of radius
$R$, as depicted in part $a$ Figure \ref{surface}. Let's place the sphere so that its center is located at the origin
and its north pole is located at $\theta=0$ in polar coordinates.
Since the problem is spherically symmetric, is is sufficient to set
one of the points at the north pole, integrate over the location of
the other point, and multiply the answer by the area of the sphere,
$4\pi R^{2}$. To impose a cutoff, we will take the location of the
second point to be in the range $\epsilon<\theta<\pi$. The log of
the expectation value of a spherical Wilson surface in $D\neq 2$ dimensions
is then given by
\begin{equation}
-\beta \ln\left\langle \Gamma\left[b\right]\right\rangle = 4\pi R^{2}\int_{0}^{2\pi}\rd\phi\int_{\epsilon}^{\pi}\rd\theta\frac{R^{2}\sin\theta\cos\theta}{\left(2R^{2}\left(1-\cos\theta\right)\right)^{D/2-1}}.\label{eq:w}
\end{equation}
Note that the angular cuttoff $\epsilon$ is related to the lattice
spacing $\alpha$ by $\epsilon=\alpha/R$. In three dimenions, (\ref{eq:w}) is finite as $\alpha\rightarrow0$, so that
\[
-\beta\ln\left\langle \Gamma_{3d}\left[b\right]\right\rangle  =\frac{16\pi^{2}R^{3}}{3}.
\] 
Hence, we find that the Wilson surface exhibits a volume law, in three
dimensions. In six dimensions, (\ref{eq:w}) gives 
\[
-\beta\ln\left\langle \Gamma_{6d}\left[b\right]\right\rangle =\frac{4\pi^{2}R^{2}}{\alpha^{2}}-4\pi^{2}\ln\frac{R}{\alpha}+...
\]
where the ellipsis denote constant terms, terms of order $\alpha^2/R^2$, and higher. Hence, the six-dimensional Wilson surface obeys an area law when $R\gg\alpha$.

Now let us briefly sketch how the arguments are modified if we include the adjoint and bi-adjoint sectors
in the continuum limit. The adjoint field has a gauge symmetry that we fix by an analogue for the Lorentz gauge in Yang-Mills
$$
\sum_{\mu\neq\nu}\partial^{\mu}C_{\mu\nu}^M=0.
$$
In the continuum limit the bi-adjoint fields are gauge invariant and so no gauge-fixing is required. The equations of motion for these fields reduce to
to
\[
\sum_{\rho\neq\nu}\partial_{\rho}\partial^{\rho}C_{\mu\nu}^M=0,	\qquad	\sum_{\rho\neq\mu,\nu}\partial_{\rho}\partial^{\rho}\Phi_{\mu\nu}^{M\dot{N}}=0.
\]
From these equations, we see that $\Phi_{\mu\nu}^{M\dot{N}}$
does not propagate in the $\mu\nu$ plane. Hence the bi-adjoint
fields only propagate in the $D-2$ directions, and the propagator in $D\neq 4$ dimensions is given by
\[
\left\langle \Phi_{\mu\nu}^{M\dot{M}}\left(\vec{x}\right)\Phi_{\rho\lambda}^{N\dot{N}}\left(\vec{y}\right)\right\rangle =\frac{\delta_{\mu\rho}\delta_{\nu\lambda}\delta^{MN}\delta^{\dot{M}\dot{N}}-\delta_{\mu\lambda}\delta_{\nu\rho}\delta^{M\dot{N}}\delta^{\dot{M}N}}{\left|\vec{x}-\vec{y}\right|^{D-4}},
\]
where $\vec{x}$ and $\vec{y}$ are understood to be orthogonal to the $\mu\nu$ plane. Similarly, the $C^M_{\mu\nu}$ do not propagate along the $\nu$ direction, and the propagator in $D \neq 3$ dimensions is given by
\[
\left\langle C_{\mu\nu}^{M}\left(\vec{x}\right)C_{\rho\lambda}^{N}\left(\vec{y}\right)\right\rangle =\frac{\delta_{\mu\rho}\delta_{\nu\lambda}\delta^{MN}}{\left|\vec{x}-\vec{y}\right|^{D-3}},
\]
where $\vec{x}$ and $\vec{y}$ are understood to be orthogonal to the $\nu$ direction. When $D=3$, the propagator for the adjoint field is given by
\begin{equation}
\left\langle C_{\mu\nu}^{M}\left(\vec{x}\right)C_{\rho\lambda}^{N}\left(\vec{y}\right)\right\rangle_{3d} =\delta_{\mu\rho}\delta_{\nu\lambda}\delta^{MN}\ln\left(\mu/\left|\vec{x}-\vec{y}\right|\right)
\label{cprop2}
\end{equation}
where $\mu$ is an arbitrary length scale which makes argument of the logarithm dimensionless (we could set $\mu=\alpha$ for example, but $\mu$ will not appear in our final answer, so its value does not matter). 

Since the adjoint field $C^M_{\mu\nu}$ does not propagate in the $\nu$ direction, it is convenient to consider a cylindrical Wilson surface whose axis of symmetry is along the $\nu$ direction, as depicted in part $b$ of Figure \ref{surface}. We will take the length and radius of the cylinder to be $R$. For $D \neq 3$, the Wilson surface is then given by
\[
-\beta\ln\left\langle \Gamma\left[C\right]\right\rangle =R\times2\pi R\times2\int_{\epsilon/2}^{\pi}\rd\phi\frac{R\cos\phi}{\left(2R^{2}\left(1-\cos\phi\right)\right)^{(D-3)/2}}
\]
where we imposed a cutoff by integrating along the azimuthal direction from $\epsilon/2$ to $\pi$ and doubling the final answer. For simplicity, we have ignored numerical prefactors coming from contractions of the color indices. As before, we take $\epsilon=\alpha/R$. When $D=6$, we obtain
\[
-\beta\ln\left\langle \Gamma_{6d}\left[C\right]\right\rangle  =\frac{8\pi R^{2}}{\alpha^{2}}-\frac{3\pi}{2}\ln\frac{R}{\alpha}+...
\] 
where the ellipsis denote a constant term plus terms of order $\alpha^2/R^2$ and higher. In three dimensions, we must use the propagator in equation \ref{cprop2}. We then find that the Wilson surface is not divergent and is given by
\[
-{\beta\ln\left\langle \Gamma_{3d}\left[C\right]\right\rangle }=2\pi^{2}R^{3}.
\] 

To compute  the expectation value of a Wilson surface for the bi-adjoint field, it is convenient to consider
a cubic surface of length $R$, depicted in part $c$ of Figure \ref{surface}. For a such a surface, the only non-zero contribution
to the integral in (\ref{eq:abelianwilsonloop}) will come
from propagators which connect opposite faces and are orthogonal to
each face, or propagators which begin and end on the same point. The
latter are singular, so we impose a cutoff as before. In $D$-dimensions,
the singular propagators will then have the form $\alpha^{4-D}$, where $\alpha$ is the lattice spacing. Integrating over the surface of the cube then
gives $6R^2\alpha^{4-D}$. In three dimensions, this contribution to the expectation value of the Wilson surface vanishes as $\alpha\rightarrow0$. In six dimensions, this contribution correponds
to an area law. Now let's look at the contribution which corresponds
to propagators which connect opposite faces. It is easy to see that
such a propagator has the form $R^{4-D}$. When we integrate this over each pair of faces, this gives $3 R^{6-D}$, which corresponds to a constant in six dimensions, and a volume law in three dimensions.

In summary, we find that in the classical continuum limit, the theory exhibits a volume law in three dimensions, and an area law in six dimensions (for sufficiently large Wilson surfaces).

\subsection{Abelian Wilson Surfaces on the Lattice\label{sub:Abelian-Theory}}

In this section, we will consider the abelian gerbe theory at non-zero
lattice spacing. In particular, we will show that in three dimensions,
it is possible to compute the expectation value of a Wilson surface
for any value of the coupling. From the discussion in Appendix D,
we see that it is possible to gauge-fix the face variables in two
planes to be 1. This corresponds to setting
\begin{equation}
b_{yz}=b_{zx}=0.\label{eq:3dgauge}
\end{equation}
Note that it is not possible to set the face variables in three planes
to be one, because the gauge-fixed faces would then form closed surfaces,
notably those cubets which span the three planes. With the gauge choice
in (\ref{eq:3dgauge}), the value of a cubet located at position
$\vec{n}$ reduces in the abelian theory to 
\begin{equation}
\Gamma\left(\vec{n}\right)=e^{i\alpha^3\Delta_{z}b_{xy}\left(\vec{n}\right)}\label{eq:wc}
\end{equation}
Furthermore, the action\footnote{We ignore the overall constant in the definition of the action.} is $S[{\cal W}]=\beta\sum_{\vec{n}}\Re\Gamma(\vec{n})$, where we can write $\Gamma(\vec{n})=e^{ih(\vec{n})}$, where we have introduced the dimensionless gauge-fixed field strength $h:=\alpha^{2}\Delta_{z}b_{xy}$. In 3d, it is possible to compute the expectation value of a Wilson surface $\Gamma_G$ for a general closed orientable surface $G$ in the abelian theory for any value of the coupling $\beta$. The expectation value is given by 
\[
\left\langle \Gamma_G \right\rangle=\frac{1}{\mathcal{Z}}\int {\cal D}{\cal W}\; \Gamma_G \; e^{-S[{\cal W}]},
\]
where $\Gamma_G=\prod_{k\in G}{\cal W}_k$, the product is over all faces which are located on the surface $G$. Let $V$ denote the volume enclosed by the surface, so that $\partial V=G$. In the abelian theory, the Wilson surface operator can equivalently be described by the product of all cubets inside the Wilson surface
$$
\Gamma_G=\prod_{k\in V}\Gamma_k=\prod_{k\in V}e^{ih_k}
$$
Plugging in (\ref{eq:wc}), gives
$$
{\cal Z}=\prod_{k\in\Lambda}\left(\int_{-\pi}^{\pi}\frac{\rd h_k}{2\pi}e^{\beta \cos (h_k)}\right)=\left(\int_{-\pi}^{\pi}\frac{\rd h_k}{2\pi}e^{\beta \cos (h_k)}\right)^{V_{\Lambda}/\alpha^3},
$$
where $V_{\Lambda}$ is the volume of the 3-dimensional space-time lattice, and
\begin{eqnarray}
\left\langle \Gamma_G \right\rangle &=& \frac{1}{\mathcal{Z}}\int_{-\pi}^{\pi}\prod_{k\in\Lambda}\frac{\rd h_k}{2\pi}\;\prod_{j\in G}e^{ih_j}\;\prod_{k\in\Lambda}e^{\beta \cos (h_k)}\nonumber\\
&=& \frac{1}{{\cal Z}}\prod_{k\notin G}\left(\int_{-\pi}^{\pi}\frac{\rd h_k}{2\pi}\;e^{\beta \cos (h_k)}\right)\times\prod_{k\in G}\left(\int_{-\pi}^{\pi}\frac{\rd h_k}{2\pi}\;e^{ih_k+\beta \cos (h_k)}\right)\nonumber\\
&=&\left(\frac{\int_{-\pi}^{\pi}\frac{\rd h}{2\pi}\; e^{ih+\beta\cos(h)}}{\int_{-\pi}^{\pi}\frac{\rd h}{2\pi}\; e^{\beta\cos(h)}}\right)^{V/\alpha^{3}}.\nonumber
\end{eqnarray}
The integrals are simply evaluated to be modified Bessel functions and we see that there is a volume law for all couplings
\[
-\ln\left\langle \Gamma_G \right\rangle=\frac{V}{\alpha^3}\times\ln\left(\frac{I_{0}\left(\beta\right)}{I_{1}\left(\beta\right)}\right),
\]
where $V/\alpha^3$ is the numer of cubets enclosd by the surface $G$. In section \ref{numerics}, we reproduce this result numerically using Monte Carlo techniques.

In summary, we find that the three-dimensional abelian lattice gerbe theory has a
volume law for any value of the coupling. This is analogous to lattice
electrodynamics in two dimensions, which exhibits an area law for any value of
the coupling. Furthermore, it is not difficult to see that at strong
coupling, the abelian lattice gerbe theory exhibits a volume law in
any dimension. We will prove this more generally for the non-abelian
lattice gerbe theory in the next section. In six dimensions, this
implies that the abelian theory has an area
law at weak coupling and a volume law at strong coupling. This is analogous
to lattice electrodynamics in four dimensions, which has a 
a perimeter law at weak coupling and an area law at strong coupling \cite{Creutz, Rothe}.  

\subsection{Non-abelian Wilson Surfaces at Strong Coupling\label{sub:Non-abelian-theory}}

In this section we will show that, at strong coupling, the non-abelian
lattice gerbe theory exhibits a volume law in any dimension. A special
case of this result is that the abelian lattice gerbe theory has a
volume law at strong coupling in any dimension. To prove this result,
we will have to understand precisely how to integrate over the face
variables of the lattice. In the abelian case, the relevant integral
is
\begin{equation}
\int  \rd{\cal W}\;{\cal W}{\cal W}^{\dagger}=1\label{eq:integral1}
\end{equation}
where ${\cal W}=e^{i\theta}$, $\rd {\cal W}=\rd\theta /2\pi$, and we have chosen to work with the dimensionless variables $\theta_{\mu\nu}=\alpha^2 b_{\mu\nu}$. Note that in the measure, we have divided by the volume of the gauge
group, which is $2\pi$. Now let's see how to generalize
(\ref{eq:integral1}) to the non-abelian case. To start with, we shall set the bi-adjoint components $\Phi^{M\dot{N}}_{\mu\nu}$ in (\ref{decomp}) to zero and only consider the singlet $b_{\mu\nu}$ and adjoint $(C^M_{\mu\nu},\widetilde{C}^{\dot{M}}_{\mu\nu})$ components. In this case,
\begin{equation}
{\cal W}_{a\dot{a}}^{b\dot{b}}=e^{i\theta}\;{\cal U}_{b}^{a}\;{\cal V}_{\dot{b}}^{\dot{a}}\label{eq:abelianadjoint}
\end{equation}
where ${\cal U}^a_b:=\exp(i\alpha^2C^a_b)$ and ${\cal V}^{\dot{a}}_{\dot{b}}:=\exp(i\alpha^2\widetilde{C}^{\dot{a}}_{\dot{b}})$ are $SU(N)$ matrices. It follows that, for a given face variable,
\[
\int \rd{\cal W}\;{\cal W}^{a\dot{a}}_{b\dot{b}}{\cal W}_{\;\;d\dot{d}}^{\dagger c\dot{c}}=\int \rd{\cal U}\;{\cal U}^a_b\,{\cal U}_{\;\;d}^{\dagger c}\;\int \rd{\cal V}\;{\cal V}^{\dot{a}}_{\dot{b}}\,{\cal V}_{\;\;\dot{d}}^{\dagger {\dot{c}}}
\]
where we have performed the integral over the singlet factor using
(\ref{eq:integral1}). Hence, we are left with an integral
over $SU(N)$ matrices. There are standard results for such integrals \cite{Creutz},
which appear frequently in lattice Yang-Mills
\begin{equation}\label{YM}
\int \rd U\;U_{c}^{a}\;U_{g}^{\dagger e}=\frac{1}{N}\delta_{g}^{a}\delta_{c}^{e}.
\end{equation}
Hence, we find that when the face variables have the form in (\ref{eq:abelianadjoint}),
\begin{equation}
\int \rd{\cal W}\;{\cal W}^{a\dot{a}}_{b\dot{b}}\;{\cal W}_{\;\;d\dot{d}}^{\dagger c\dot{c}}=\frac{1}{N^{2}}\delta^{a}_{d}\delta^{c}_{b}\delta^{\dot{a}}_{\dot{d}}\delta^{\dot{c}}_{\dot{b}}.\label{eq:integral2}
\end{equation}
When computing Wilson surfaces at strong coupling, the integral in
(\ref{eq:integral2}) will essentially be the only non-vanishing
integral we encounter at leading order. Using this fact, it is straightforward to prove
that Wilson surfaces exhibit a volume-law at leading order in the
strong-coupling expansion. Before doing so, note that if we include the bi-adjiont fields $\Phi^{M \dot{N}}_{\mu\nu}$ in the definition of the face-variables the integral formula in (\ref{eq:integral2}) still holds. Indeed, noting that ${\cal W}^{a\dot{a}}_{c\dot{c}}({\cal W}^{\dagger})^{c\dot{c}}_{b\dot{b}}=\delta^a_b\delta^{\dot{a}}_{\dot{b}}$, we may treat the ${\cal W}$'s as $N^2\times N^2$ matrices for the group $U(N^2)$. These results are useful in performing strong coupling calculations as we shall see below.

\begin{figure}[h] 
       \centering
       \includegraphics[width=4in]{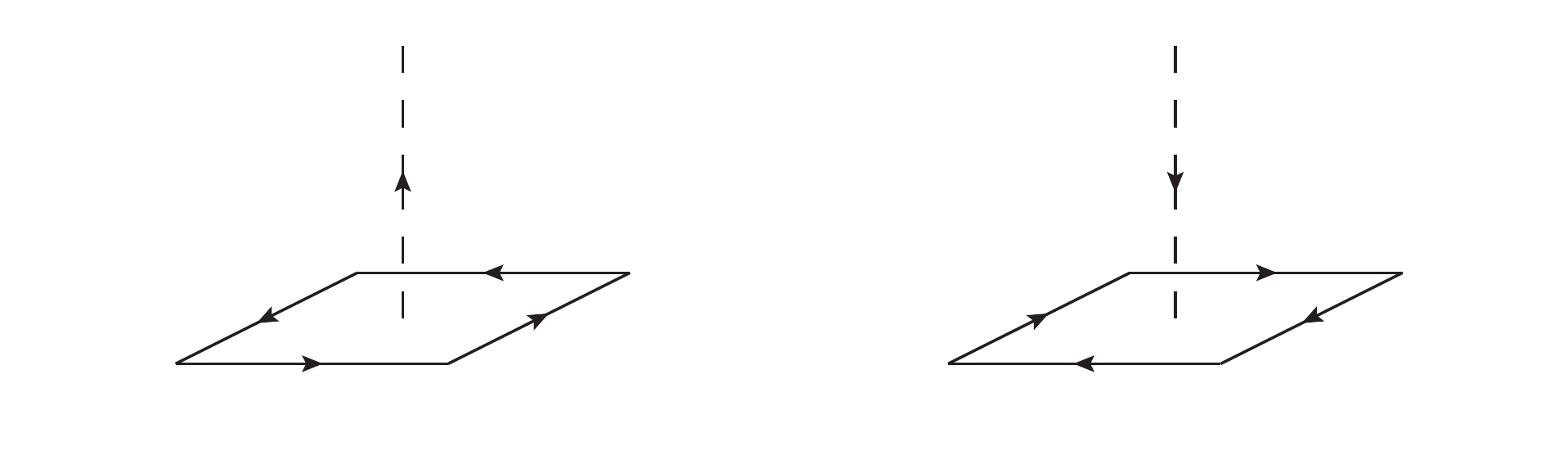}  
    \caption{Using a right-handed convention a plaquette has a natural orientation vector associated with it..}
    \label{fig:Vector}
    \end{figure}

\begin{figure}[h] 
       \centering
       \includegraphics[width=3in]{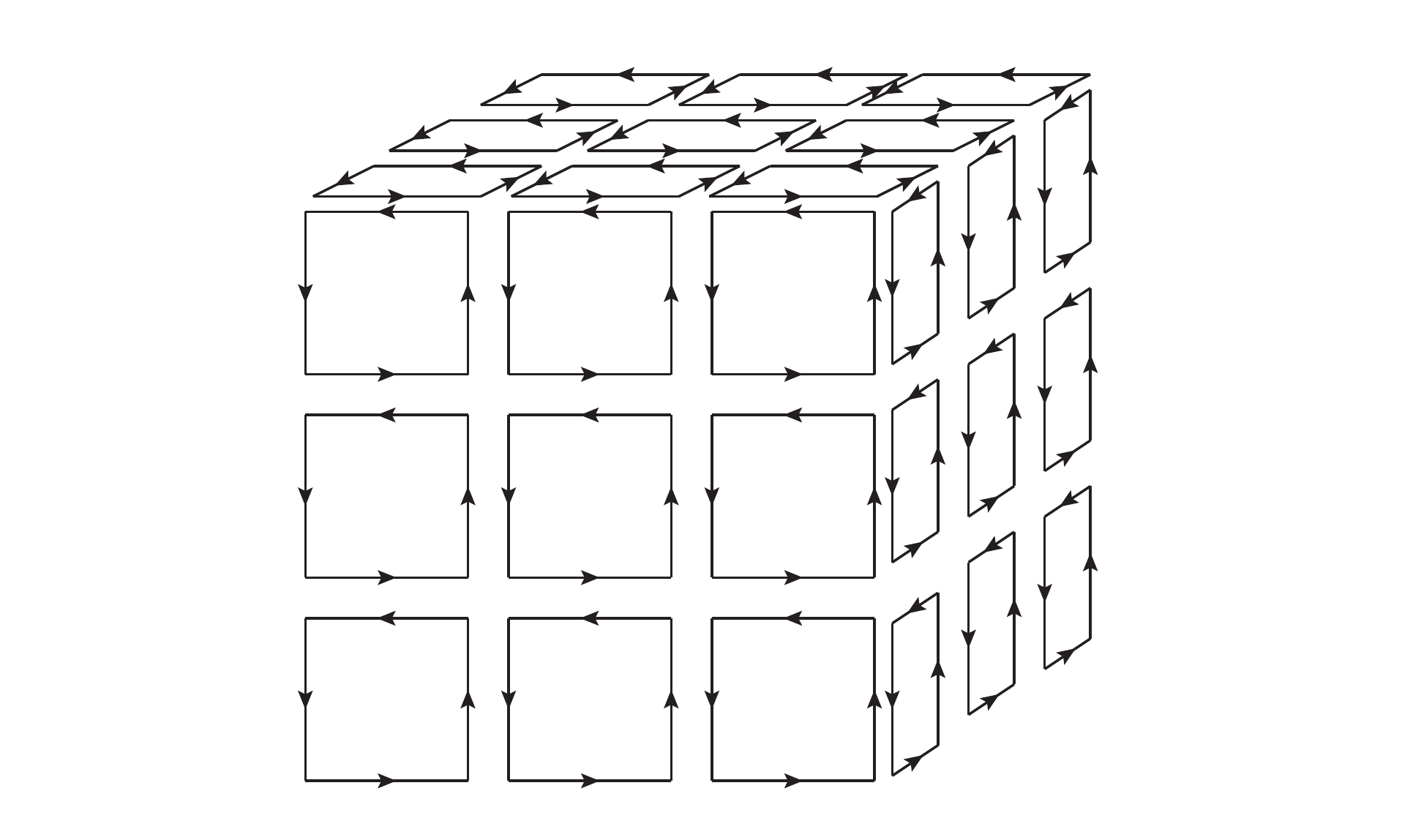}  
    \caption{A Wilson surface bounding 27 fundamental cubes. Using the right-hand convention, the vectors associated with the plaquettes of this surface are all outward-pointing.}
    \label{fig:Wilson1}
    \end{figure}

Recall that, up to an overall constant, the action is given by
$$
S[B]=\frac{\beta}{2N^3}\sum_{\mbox{\mancube}}\left(\Gamma(\mbox{\mancube})+\Gamma^{\dagger}(\mbox{\mancube})\right).
$$
Given a $D$-dimensional lattice, a given cubet lies in a three-dimensional subspace of the lattice. Within this three-dimensional subspace, a natural vector, given by a right-hand rule, may be assigned to each plaquette as shown in Figure \ref{fig:Vector}. Keeping track of these orientation vectors as we construct a fundamental cube leads us to two types of cubet (related by parity); $\Gamma(\mbox{\mancube})$ is a fundamental cube with all orientation vectors pointing outwards, whist $\Gamma^{\dagger}(\mbox{\mancube})$ is a cube with inward pointing vectors. The only gauge-invariant way to join two cubes along a common face is if the vectors on that face oppose each other (i.e. if the sum of the vectors is zero). This ensures that the orientations of the plaquettes being glued are opposite.
The expectation of an operator ${\cal O}$ is
\begin{equation}\label{O}
\langle {\cal O}\rangle=\frac{1}{{\cal Z}}\int {\cal D}{\cal W}\;{\cal O}\;e^{-S[{\cal W}]}.
\end{equation}
The operator we are interested will be the Wilson surface $\Gamma_G$ associated with the closed, oriented surface $G$ constructed from a cuboidal array of $I\times J\times K$ fundamental cubes, giving rise to a surface bounding a volume $V=IJK\alpha^3$. The generalisation to Wilson surfaces of higher genus is straightforward. The operator is given by contracting together all plaquettes ${\cal W}$ contained in the surface $G$
$$
\Gamma_G=\prod_{\Box\in G}{\cal W}(\Box)
$$
Here the product simply means the identification of edges along adjoining plaquettes as described in section 4.2. For this simple genus zero Wilson surface, there are clearly two possible orientations of the plaquettes of the surface; one in which all plaquette vectors point outwards, and the other where they all point inwards. An example of a Wilson surface $G$ bounding 27 fundamental cubes with outward pointing vectors is given in Figure \ref{fig:Wilson1}.
We then ask the question of what kind of gauge-invariant (and thus, non-zero) configurations could we have involving such a Wilson surface and fundamental cubes. Taking a $3\times 3\times 3$ cube as an example, we can place twenty-six fundamental cubes with inward pointing vectors inside the Wilson surface such that each plaquette which makes up the Wilson surface has a face of a fundamental cube attached to it. The total configuration would be gauge-invariant were it not for the six exposed faces of the cubes in the centre. These six faces have ${\cal W}$ or ${\cal W}^{\dagger}$'s appearing individually and not paired up with an appropriate ${\cal W}^{\dagger}$ or ${\cal W}$ so that the integral over ${\cal W}$ for each of the exposed faces will vanish. This can be remedied by placing a final, inward-pointing, fundamental cube in very centre of the configuration. Every plaquette in the resulting configuration is paired with another plaquette with opposite vector. Integrating over each pair of plaquettes and using (\ref{eq:integral2}) gives a non-zero result.

We have argued that most integrals over ${\cal W}$ will vanish. The exceptions are when (\ref{eq:integral2}) applies. One might fear that the expectation of the Wilson surface vanishes. We would now like to show that the configuration described above does indeed arise from our path integral expression and gives a non-trivial contribution. Expanding the exponential in powers of $\beta$ gives
$$
\langle\Gamma_G \rangle=\frac{1}{\cal Z}\int {\cal D}{\cal W}\left(\prod_{\Box_i'\in G}{\cal W}_{\Box_i'}\right)\sum_{n=0}^{\infty}c(n)\beta^n\left(\sum_{\mbox{\mancube}\in\Lambda}\Gamma+\sum_{\mbox{\mancube}\in\Lambda}\Gamma^{\dagger}\right)^n,
$$
where the numerical constants $c(n)$ are given by $c(n)=(-1/2N^3)^n/n!$. Assuming we can exchange orders of summation and integration, we use the binomial theorem to write
$$
\langle\Gamma_G\rangle=\frac{1}{\cal Z}\sum_{n=0}^{\infty}c(n)\beta^n\int {\cal D}{\cal W}\prod_{\Box_i'\in G}{\cal W}(\Box_i')\sum_{k=0}^n\left(
\begin{array}{c}
n\\
k
\end{array}
\right)\left(\sum_{\mbox{\mancube}\in\Lambda}\Gamma\right)^k\left(\sum_{\mbox{\mancube}\in\Lambda}\Gamma^{\dagger}\right)^{n-k}.
$$
As described above, we only get the above non-trivial result if we fill the interior of the Wilson  surface with fundamental cubes with inward-pointing vectors - those given by $\Gamma^{\dagger}$. Thus we are interested in the case $k=0$ above and $n=3^3=27$. More generally, for a cuboidal Wilson surface of sides $I\times J\times K$, we require $k=0$ and $n=IJK$. Denoting the volume, in lattice units, enclosed by the surface $G$ by $IJK=V/{\alpha^3}$. The contribution from this configuration is
$$
\langle \Gamma_G \rangle=\frac{1}{\cal Z}\;c(V/\alpha^3)\;\beta^{V/\alpha^3}\int {\cal D}{\cal W}\prod_{\Box_i'\in C}{\cal W}(\Box_i')\left(\sum_{\mbox{\mancube}\in C}\Gamma^{\dagger}\right)^{V/\alpha^3}+...,
$$
where the ellipsis denote higher order terms. Using the integration rule (\ref{eq:integral2}) one can evaluate the integral, giving a constant which depends on the rank of the gauge group. The configuration described above is the leading order contribution. The important point is that, to leading order, the log of expectation of the operator goes as the volume enclosed by the Wilson surface
$$
\ln\langle \Gamma_G\rangle\approx \frac{V}{\alpha^3}\times\ln\beta+...,
$$
up to overall numerical factors which we shall not be too concerned about. It is interesting to see that the leading order contribution at strong coupling follows a volume law. Drawing an analogy with the strong coupling expansion in lattice QCD, one might be tempted to introduce strings that couple to the gerbe and could form boundaries of open Wilson surfaces, with membranes playing the role of a QCD string. A volume law naively suggests confinement of such strings; however, much more work would need to be done before one could accept such an interpretation of this result. It is straightforward to adapt the conventional strong coupling expansion techniques for lattice Yang-Mills \cite{Creutz, Rothe} to calculate next to leading order corrections for lattice gerbe theory at strong coupling.

\section{Wilson Surfaces: Numerical Results} \label{numerics}

\subsection{Monte Carlo Techniques}

A major advantage of formulating gerbe theory on a lattice is
that it can be simulated on a computer. In this section, we will describe
how to numerically compute various physical observables in abelian
lattice gerbe theory, although much of this discussion can be generalized
to non-abelian gerbe theories. Many of these techniques are a strightforward
generalization of techniques used in lattice Yang-Mills theory. Let us suppose that we are working on a $D$-dimensional
hypercubic Euclidean spacetime lattice, with $L$ points in each direction.
Then there are a total of $L^{D}$ lattice points. For bosonic fields,
we can take periodic boundary conditions, so one can think of the
lattice as being $D$-dimensional torus. The expectation value of a
physical observable ${\cal O}$ is then defined as
\[
\left\langle \mathcal{O}\right\rangle ={\cal Z}^{-1}\int\mathcal{D}{\cal W}\;{\cal O}\;e^{-S[{\cal W}]}
\]
where $\mathcal{D}{\cal W}$ refers to the integral over the faces of the
lattice, $S$ is the lattice action which we defined in section \ref{naction} and ${\cal Z}$ is the partition function. Since there are
$D(D-1)/2$ faces at every point, this corresponds to a $\frac{1}{2}D(D-1) \times L^{D}$-dimensional integral. This integral corresponds to summing over all
possible gerbe configurations and is
not practical to evaluate exactly for reasonably sized lattices. On the other hand, if we can obtain
representative gerbe configurations which dominate the sum, they can be used
to obtain a good approximation to the expectation value. 

Representative configurations can be generated using Monte Carlo techinques.
Using such configurations, one can compute the expectation value of various observables of the theory. For
example, given a representative gerbe configuration, one can compute
the energy density, which is defined as the action divided by the
number of lattice points:

\[
\left\langle E\right\rangle =\frac{S}{\beta L^{d}}.
\]
From the point of view of statistical mechanics, $\beta$ can be thought of as the inverse temperature. Furthermore,
the expectation value of a Wilson surface, $W$, for a reperesentative configuration
can be obtained simply by computing the Wilson surface at various
locations in the lattice and taking the average. In section \ref{nawilson}, we describe the
general procedure for defining a closed oriented Wilson surface, and
provide explicit definitions for a $1\times1\times1$ Wilson surface
and $2\times1\times1$ Wilson surface. Once we compute the expectation
value of various different Wilson surfaces, we can then study how the logarithm of the expectation value scales as a function of
the size. We say that the theory exhibits a volume/area law if the
logarithm of the expectation value of a Wilson surface is proportional
to the volume/area. One can use similar techniques to compute
correlators of Wilson surfaces, but we will not discuss this in this
paper. 

A very efficient method for generating representative configurations
is the heat bath algorithm \cite{creutz2}. One starts with an initial configuration,
which can be stored as an array on a computer. One then replaces each
face variable ${\cal W}$ with a new value selected randomly with a weighting
given by the exponentiated action:

\begin{equation}
\rd P({\cal W})\sim \rd {\cal W}\;\exp\left(- S[{\cal W}]\right)\label{eq:probability}
\end{equation}
where$\rd {\cal W}$ refers to the variation of a single face variable (and basically corrsponds to the measure of the gauge group), while $S[{\cal W}]$ is a function
of all the face variables. If the measure of the gauge group is not known explicitly,
one can use a more general method known as the
Metropolis algorithm \cite{metropolis}. For a description of the heat bath and Metropolis algorithms in the context of lattice Yang-Mills theory, see for example \cite{Creutz}. 

\begin{figure}[htbp] 
    \centering
    \includegraphics[width=2.25in]{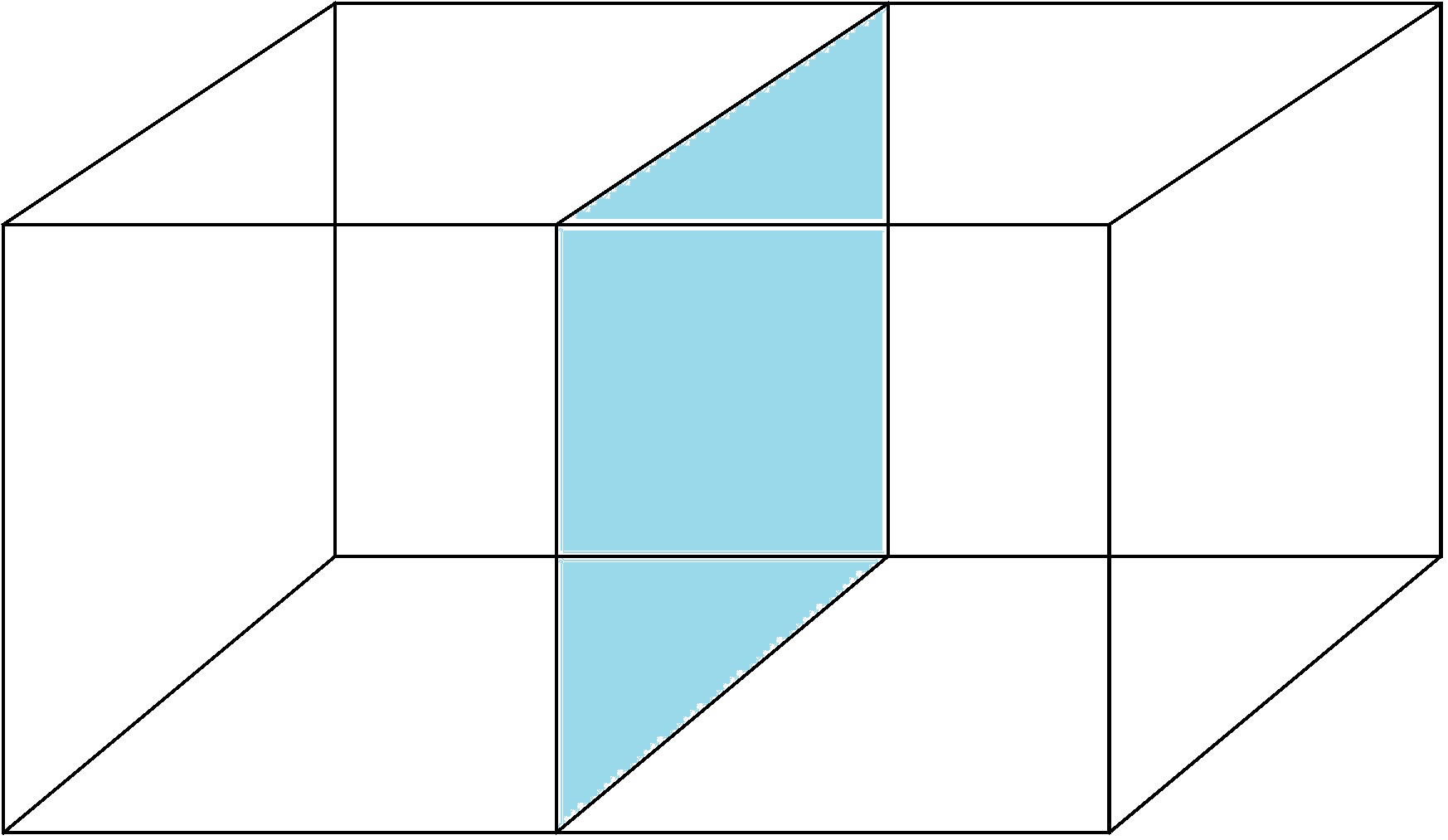}  
     \caption{A pair of cubets which share a face which is being updated using Monte Carlo. In $D$ dimensions, there are $D-2$ such pairs for each face of the lattice.}
   \label{facepairs}
    \end{figure}

We will now describe the heat bath algorithm in more detail for the case
of an abelian gerbe. For non-abelian gerbes, it is more convenient
to use the Metropolis algorithm, as we will discuss elsewhere. Let us consider varying a given face variable. Note that there
are $(D-2)$ pairs of cubets which share this face, which we label
by $\alpha=1,...,2(D-2)$. One such pair is depicted in Figure \ref{facepairs}.
If the the value of this face variable is ${\cal W}$, the terms in the action
which depend on ${\cal W}$ are given by

\[
S[{\cal W}]\sim \beta \,\Re\left({\cal W}\sum_{\alpha=1}^{2(D-2)}\widetilde{\cal W}_{\alpha}\right)
\]
where $\widetilde{\cal W}_{\alpha}$ is the product of the five other face
variables in the cubet labeled $\alpha$. Note that

\[
\sum_{\alpha=1}^{2(D-2)}\widetilde{\cal W}_{\alpha}=k\,\widetilde{\cal W}
\]
where $k=\left|\sum_{\alpha=1}^{2(D-2)}\widetilde{\cal W}_{\alpha}\right|$
and $\widetilde{\cal W}\in U(1)$. Hence, the probability density in
(\ref{eq:probability}) is 
\[
\rd P({\cal W})\sim \rd {\cal W}\exp\left(-\beta k\Re\left({\cal W}\widetilde{\cal W}\right)\right).
\]
Using the invariance of the group measure, we can absorb $\widetilde{\cal W}$
to obtain:

\[
\rd P({\cal W}\widetilde{\cal W}^{-1})\sim \rd {\cal W}\,\exp\left(-\beta \,k\,\Re({\cal W})\right)
\]
Writing ${\cal W}=e^{i\theta}$ and normalzing the probability distribution
so that it is equal to one when $\theta=0$ finally gives
\begin{equation}
\rd P({\cal W}\widetilde{\cal W}^{-1})\sim \rd\theta\,\exp\left(-\beta k\left(\cos\theta-1\right)\right).
\label{distribution}
\end{equation}
When generating a new value for the face variable ${\cal W}$, randomly generate
a $\theta\in(-\pi,\pi)$ as well as a number between 0 and 1. If the
random number is less than $\frac{\rd P}{\rd\theta}$ in (\ref{distribution}), accept this
value of theta. If not, repeat this process until a value of $\theta$
is accepted. Given a ${\cal W}$ generated in this manner, replace the face
variable with ${\cal W}\widetilde{\cal W}^{-1}$. This procedure is then repeated
for another face and so forth until all the faces of the lattice have
been updated. This corresponds to one Monte Carlo iteration.

A simple way to obtain an representative gerbe configuration is to
start with an initially random and an initially trivial configuration
and perform Monte Carlo interations on them simultaneously with the
same value of $\beta$ until their average energy densities converge
as depicted in Figure \ref{3d1}. By repeating this procedure
for various values of $\beta$, one can obtain the equilibrium energy density
as a function of temperature. 

By computing the expectation value of Wilson surfaces with different sizes, one
can determine if the theory exhibits an area law or a volume law.
If there is a phase transition from an area law to a volume law, this
will also be reflected in the convergence of the Monte Carlo algorithm.
For example, we find that the six dimensional abelian lattice gerbe theory
exhibits a first order phase transition, and that above a certain $\beta$,
the initially random and initially trivial gerbe configurations generally
converge to different values, as depicted in Figure \ref{6dthermal}. This corresponds
to one of the configurations being stuck in a metastable state. A
similar phenomenon is also encountered in abelian lattice gauge theories \cite{creutzrebbi}.
One way to overcome this difficulty is to introduce a third
gerbe configuration corresponding to an initially mixed phase such that $E_{mixed}(0)=\frac{1}{2}\left(E_{trivial}(0)+E_{random}(0)\right)=\frac{1}{2}E_{random}(0)$.
We then carry out Monte Carlo interations until the initially mixed
gerbe converges with either the initally trivial or the initially
random gerbe and use the initially mixed gerbe as our representative
configuration once convergence is achieved.

We implemented our numerical calculations for the 3d and 6d abelian gerbe theories using Mathematica (see the attached Mathematica notebooks 3dabelianheatbath.nb and 6dabelianheatbath.nb, respectively). After obtaining representative gerbe configurations for various temperatures using the heat bath algorithm, we computed the expectation value of $1\times 1 \times 1$, $2\times 1 \times 1$, and $2\times 2 \times 2$ Wilson surfaces as a function of inverse temperature. In the next two subsections, we describe our numerical results in greater detail. 

\subsection{Numerical Results in Three Dimensions}

\begin{figure}[htbp] 
    \centering
    \includegraphics[width=4in]{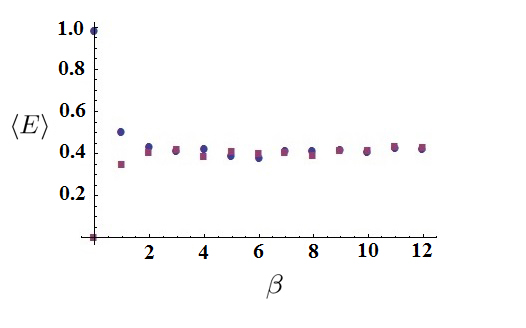}  
     \caption{Energy density as a function of Monte Carlo interations for 3d abelian gerbe theory on $10^3$ lattice. The blue points correspond to the initially random gerbe and the purple points correspond to the intially trivial gerbe.}
\label{3d1}
    \end{figure}

\begin{figure}[htbp] 
    \centering
    \includegraphics[width=4in]{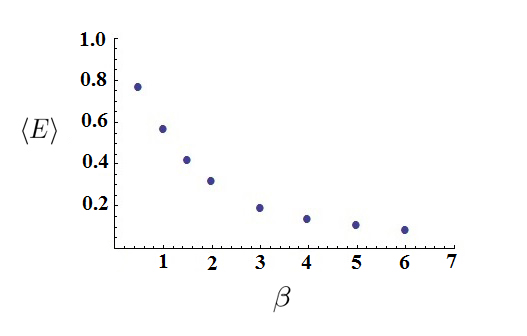}  
     \caption{Equilibrium energy density as a function of $\beta$ for 3d abelian gerbe theory.}
\label{3den}
    \end{figure}

Our results for the 3d abelian gerbe theory were obtained on a $10^{3}$
lattice. For each value of the temperature $\beta$, we started with
an initally trivial and an initially random gerbe configuration and
used the heat bath algorithm to update each gerbe configuration until
the energy densities converged, as described in the previous subsection.
We plot the energy density as a function of Monte Carlo iterations
for $\beta=1.5$ in Figure \ref{3d1}. 
Note that convergence is acheived in fewer than 20
iterations. We plot the equilbrium energy density as a function of
$\beta$ in Figure \ref{3den}. Note that the curve is smooth and does not show
any sign of a phase transition.

\begin{figure}[htbp] 
    \centering
    \includegraphics[width=6.3in]{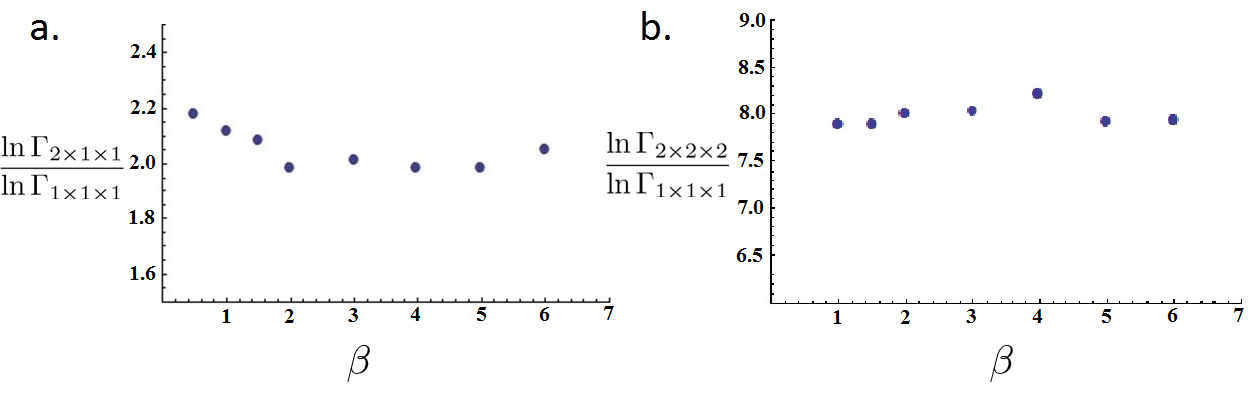}  
     \caption{In $a$, we plot $\frac{\ln\Gamma_{2\times1\times1}}{\ln\Gamma_{1\times1\times1}}$ as a function of $\beta$ in the 3d abelian gerbe theory. In $b$, we plot $\frac{\ln\Gamma_{2\times2\times2}}{\ln\Gamma_{1\times1\times1}}$ as a function of $\beta$.}
\label{3dnum}
    \end{figure}

\begin{figure}[htbp] 
    \centering
    \includegraphics[width=5in]{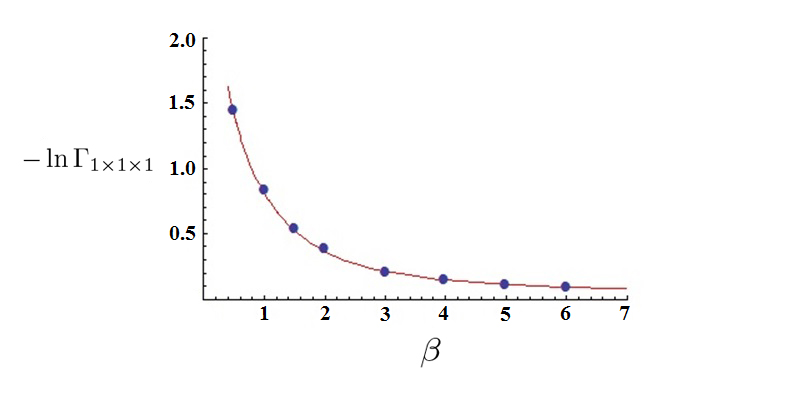}  
     \caption{$-\ln\Gamma_{1\times1\times1}$ as a function of $\beta$ for 3d abelian gerbe theory. The points were obtained using Monte Carlo techniques and the red line corresponds to the analytical result in equation \ref{eq:3d}.}
\label{3dnum2}
    \end{figure}

In section \ref{sub:Abelian-Theory}, we showed that Wilson surfaces in the 3d abelian gerbe
theory obey a volume law for any value of $\beta$:
\begin{equation}
-\ln\left\langle \Gamma_{G}\right\rangle =\frac{V}{\alpha^{3}}\times\frac{I_{0}(\beta)}{I_{1}(\beta)}\label{eq:3d}
\end{equation}
where $\Gamma_G$ is the Wilson surface operator associated with a closed oriented surface $G$, and $V$ is the volume of $G$.
We verified this result by computing $\ln\Gamma_{1\times1\times1}$, $\ln\Gamma_{2\times1\times1}$, and $\ln\Gamma_{2 \times 2\times 2}$ as a function of $\beta$. In particular, we find that $\frac{\ln\Gamma_{2\times1\times1}}{\ln\Gamma_{1\times1\times1}}\sim 2$  and $\frac{\ln\Gamma_{2\times2\times2}}{\ln\Gamma_{1\times1\times1}}\sim 8$ for any value of $\beta$, as depicted in Figure \ref{3dnum}. For an area law, we would expect these ratios to be $5/3$ and $4$, respectively. Furthermore, in Figure \ref{3dnum2}, we plot our numerical results for $-\ln\Gamma_{1\times1\times1}$ and find that the points lie precisely
on the curve predicted by equation \ref{eq:3d}. Note that our numerical
calculations generally give Wilson surfaces with a negligible imaginary
part, so we just discard the imaginary part in our results.

\subsection{Numerical Results in Six Dimensions}

We performed numerical calculations in the 6d abelian gerbe theory
using a $4^{6}$ lattice. Note that even on a lattice of this size,
it can take up to a few hours to generate an equilibrium gerbe configuration
using the heat bath alorithm on a standard desktop computer. For each
value of $\beta$, we started with three initial gerbe configurations:
an initally trivial gerbe, an initially random gerbe, and an initally
mixed gerbe whose initial energy density was roughly half of the initial
energy density of the random gerbe. 

\begin{figure}[htbp] 
    \centering
    \includegraphics[width=6.3in]{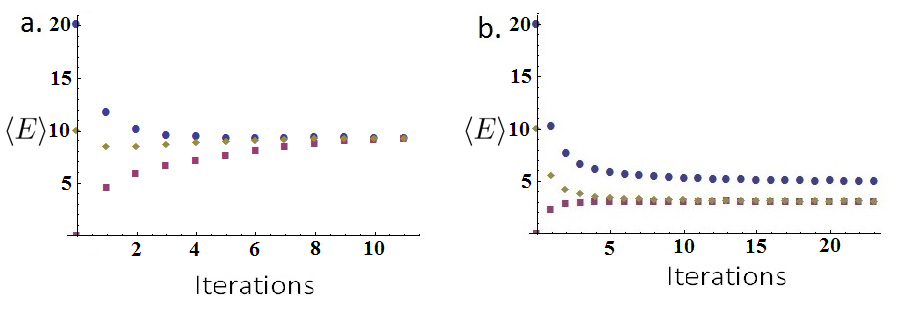}  
     \caption{Energy density as a function of iterations for 6d abelian lattice gerbe theory on a $4^6$ lattice. Blue points correspond to initally random gerbe, purple points correpond to initially trivial gerbe, and green points correspond to initially mixed gerbe. For $\beta<1.5$, all three gerbes converge, as depicted in $a$. For $\beta>1.5$, the initially trivial and initially random gerbe do not generally converge, so the phase is determined by convergence of the initially mixed gerbe. For $\beta>2$ the initially mixed gerbe always converges with the initially trivial gerbe, as depicted in $b$.}
\label{6dthermal}
    \end{figure}

Below $\beta\sim1.5$, all three gerbes converge rapidly, as depicted
in Figure \ref{6dthermal}, and the Wilson surfaces in all three gerbes exhibit a volume
law. Above $\beta\sim1.5$, however, the initially random and initally
trivial gerbe converge towards each other very slowly or not at all. Furthermore, the Wilson surfaces of the
initially random gerbe exhibits an volume law and the Wilson surfaces of the initially trivial
gerbe exhibits an area law. This behavior is suggestive of a first
order phase transition in that above a certain temperature, one of
the two gerbe configurations becomes stuck in a metastable phase.
The true phase of the theory is determined by convergence of the
initally mixed gerbe. Below $\beta\sim2$, the initially mixed gerbe
always converges with the initially random gerbe and above $\beta\sim 2$,
the initially mixed gerbe always converges with the initially trivial
gerbe, as depicted in Figure \ref{6dthermal}. 

\begin{figure}[htbp] 
    \centering
    \includegraphics[width=4in]{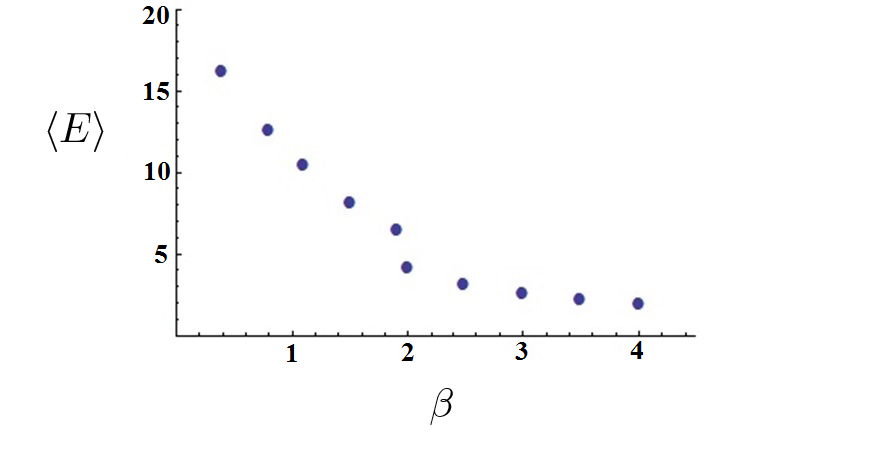}  
     \caption{Equilibrium energy density as a function of $\beta$ for 6d abelian lattice gerbe theory.}
\label{6de}
    \end{figure}

\begin{figure}[htbp] 
    \centering
    \includegraphics[width=6.3in]{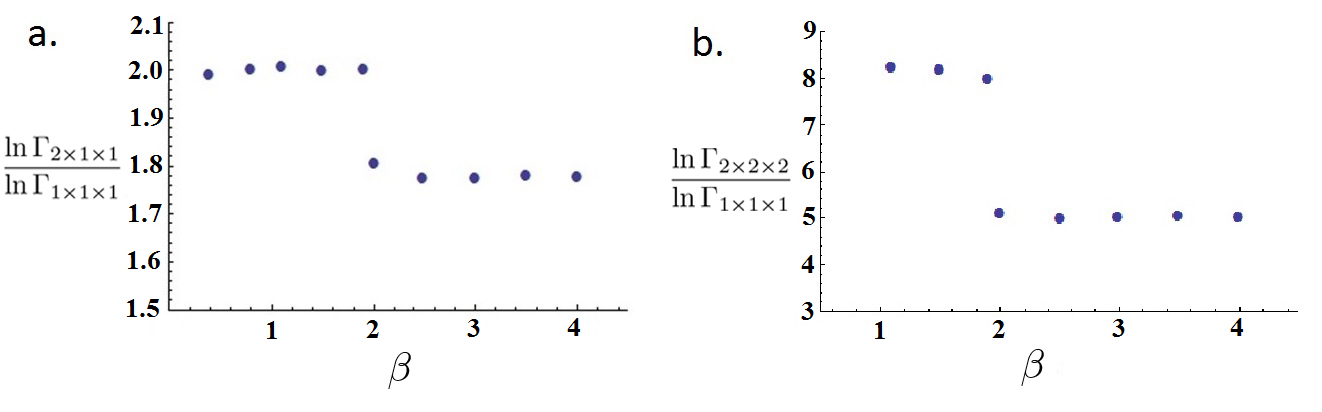}  
     \caption{In $a$, we plot $\frac{\ln\Gamma_{2\times1\times1}}{\ln\Gamma_{1\times1\times1}}$ as a function of $\beta$ in the 6d abelian gerbe theory. In $b$, we plot $\frac{\ln\Gamma_{2 \times2 \times2}}{\ln\Gamma_{1\times1\times1}}$ as a function of $\beta$}.
\label{6dratio}
    \end{figure}

After the initially mixed gerbe converges with one of the other two gerbes, we use the equilibrium configuration to compute expectation values of Wilson surfaces. We plot the equilibrium energy density as a function of $\beta$ in Figure \ref{6de}, and the ratios   $\frac{\ln\Gamma_{2\times1\times1}}{\ln\Gamma_{1\times1\times1}}$, $\frac{\ln\Gamma_{2\times2\times2}}{\ln\Gamma_{1\times1\times1}}$ as a function of $\beta$ in Figure \ref{6dratio}. From these plots, we see that the theory undegoes a first order phase transition at the ciritcal value of $\beta \sim 2$. In particular, the energy density has a small jump at $\beta \sim 2$. Furthermore, for $\beta<2$, $(\frac{\ln\Gamma_{2\times1\times1}}{\ln\Gamma_{1\times1\times1}},\frac{\ln\Gamma_{2\times 2\times 2}}{\ln\Gamma_{1\times1\times1}}) \sim (2,8)$, which is consistent with a volume law. On the other hand, for $\beta>2$, $(\frac{\ln\Gamma_{2\times1\times1}}{\ln\Gamma_{1\times1\times1}},\frac{\ln\Gamma_{2\times 2\times 2}}{\ln\Gamma_{1\times1\times1}}) \sim (1.77,5)$, which are slightly larger than what is implied by a pure area law. 

In summary, we have found that the abelian gerbe theory exhibits a volume law for all values of the coupling in three dimensions, and a phase transition from an area law at weak coupling to a volume law at strong coupling in six dimensions. These results agree with our analytical results in section \ref{analytics} as well the results of \cite{Frohlich:1982gf}.  

\section{Outlook} \label{conclusion}

In this paper, we explore a formulation of non-abelian gerbe theory, which is inspired by Wilson's formulation of lattice Yang-Mills theory. In lattice Yang-Mills theory, the basic degrees of freedom can be taken to be $U(N)$ matrices which live on the links of the lattice, and the action is obtained by adding all the Wilson loops associated with the fundamental squares of the lattice (known as plaquettes). In lattice gerbe theory, the basic degrees of freedom live on the faces of the lattice, and can be taken to transform in the bi-adjoint representation of $U(N) \times U(N)$. Furthermore, the action is obtained by adding all of the Wilson surfaces associated with the fundamental cubes of the lattice (which we refer to as cubets). In the abelian case, this construction reduces to the well-known theory of an abelian 2-form gauge field in the continuum limit. In the non-abelian case, we find that dimensional reduction does not commute with taking the classical continuum limit. Although lattice gerbe theory becomes non-interacting in the classical continuum limit, if we dimensionally reduce before taking the lattice spacing to zero, we obtain $U(N)$ Yang-Mills theory coupled to other fields. Futhermore, it is possible to compute Wilson surfaces in lattice gerbe theory using standard analytical and numerical techniques from lattice Yang-Mills. In particular, we analytically show that at strong coupling and non-zero lattice spacing, closed oriented Wilson surfaces exhibit a volume law in any spacetime dimension. Furthermore, we obtain both analytical and numerical evidence that in three dimensions, Wilson surfaces exhibit a volume law for any value of the coupling, and in six dimensions, they exhibit an area law at weak coupling and a volume law at strong coupling. 

Since the Wilson surfaces in the six-dimensional lattice gerbe theory obey an area law at weak coupling and a volume law at strong coupling, it is conceivable that in the non-abelian case, the quantum theory undegoes a second-order phase transition and has a nontrivial continuum limit associated with the conformal fixed point. In this paper, we have numerically demonstrated that the six-dimensional abelian gerbe theory undergoes a first order phase transition by computing the expectation value of Wilson surfaces and the energy density as a function of the coupling using a Monte Carlo technique known as the heatbath algorithm. It would be very interesting to extend these numerical calculations to the non-abelian theory, and to study correlators of Wilson surfaces in the non-abelian theory both analytically and numerically. If the six-dimensional non-abelian theory has a second-order phase transition, then correlation lengths should diverge as one approaches the transition point. At the transition point, correlators should have power-law behavior and one could in principle compute the critical exponent numerically.

Although the study of non-abelian gerbes is mathematically interesting in its own right, it has tremendous importance for theoretical physics. A central question in string theory is how to describe the 5-dimensional stable objects of M-theory, which are known as M5-branes. At low energies, a stack of such objects should be described by a six dimensional superconformal theory whose matter content consists of a self-dual gerbe, five scalars, and eight fermions (collectively known as a (2,0) tensor multiplet). Furthermore, when the $(2,0)$ theory is dimensionally reduced on a torus, one should  obtain maximal super-Yang-Mills theory. Given that non-abelian lattice gerbe theory naturally gives rise to $U(N)$ Yang-Mills theory (and other fields) upon dimensional reduction, this seems like a natural starting point for trying to define the $(2,0)$ theory. In particular, the next step is to incorporate self-duality and supersymmetry. Note that in Euclidean signature, if the field strength of a real abelian gerbe is self-dual, then it must vanish. On the other hand, it is possible to have a nonzero self-dual field strength if we complexify the gerbe, which is analogous to what one does to define instantons in Lorentzian signature in four dimensions \footnote{We thank Dmitri Sorokin for pointing this out to us.}. Incorporating supersymmetry is also subtle, since the lattice breaks translational symmetry. Nevertheless, it is possible to formulate supersymmetric lattice gauge theory \cite{Catterall:2005fd,Kaplan:2005ta,Catterall:2014vka}, so it would be interesting to generalize this to supersymmetric lattice gerbe theory following the approach we describe in this paper.

It has also been suggested that the M5-brane theory should correspond to a non-gravitational self-dual string theory \cite{Strominger:1995ac}. It would therefore be interesting to study if one can incorporate finite tension strings into lattice gerbe theory. Since the Wilson surfaces in our model exhibit a volume law at strong coupling, this suggests that such strings will become confined.  It would then be interesting to explore if a pair of confined strings can be described as a two dimensional membrane, generalising the QCD string. On the other hand, once self-duality and supersymmetry are incorporated into the model, Wilson surfaces are expected to obey an area law \cite{Berenstein:1998ij}. Note that the theory which describes a stack of $N$ M2-branes also has gauge group $U(N) \times U(N)$ \cite{Aharony:2008ug}. Given that the self-dual strings correspond to boundaries of M2-branes ending on M5-branes, lattice gerbe theory may provide insight into the origin the gauge group in the ABJM theory. 

In summary, we find that lattice gerbe theory provides a compelling way to define non-abelian gerbes. Since the study of lattice gerbe theory is still in its infancy, there are many exciting directions to explore.

\begin{center}
\textbf{Acknowledgements}
\end{center}

We thank Anton Kapustin, P A Marchetti, Lionel Mason, Rafael Nepomechie, Peter Orland, Soojong Rey, John Schwarz, Fumihiko Sugino, and especially Chris Hull and Dmitri Sorokin helpful discussions.  AL is supported by a Simons Postdoctoral Fellowship and would like to thank and the Aspen Center for Physics (NSF grant PHYS-1066293) and the INFN of Padova for hospitality while parts of this work were being completed. RR would like to thank the Mathematical Institute in Oxford for support and hospitality during various stages of this project.

\appendix

\section{Geometry of the Continuum Theory}

We introduce matrix-valued coordinates $(X^{\mu})^a_b$, taking values in the adjoint of $U(N_{\mu})$, where we allow for the possibility that the rank of the the group is different along different directions\footnote{Note that this is different from usual non-commutative geometry, as we have
$$
[X^{\mu},X^{\nu}]=0	\qquad	\text{if}	\qquad	\mu\neq\nu
$$
but coordinates along the same directions may not commute $[X^{\mu}_m,X^{\mu}_n]=if_{mn}{}^pX^{\mu}_p$.}. Such a construction is remeniscent of coordinates on a stack of $N$ coincident D-branes and it is suggests a possible origin in terms of a stack of five-branes\footnote{This comment really only applies to the self-dual case.}. A basis of differential forms is given by
$(\rd X^{\mu})^a_b$. It is useful to perform an $SU(N)$ decomposition
$$
(\rd X^{\mu})^a_b=\delta^a_b\rd x^{\mu}+(e^{\mu})^a_b,
$$
where $\rd x^{\mu}$ is a conventional differential form. Note that all of this takes place on the continuum, there is no lattice. Introducing a two-form
$$
B=\frac{1}{2}(B_{\mu\nu})^{a\dot{a}}_{b\dot{b}}\;(\rd X^{\mu})^a_b\wedge (\rd X^{\nu})^{\dot{a}}_{\dot{b}}
$$
where we shall take $(B_{\mu\nu})^{a\dot{a}}_{b\dot{b}}$ to be a function of $x^{\mu}$ only. We can introduce the adjoint $SU(N_{\mu})$ generators $T^M$ and $U(N)$ matrices $T^I$. We introduce the one-forms $(\rd X^{\mu})^a_b=\rd X^{\mu}_{I_{\mu}}(T^{I_{\mu}})^a_b$. We can write the two-form as follows
$$
B=\frac{1}{2}B_{\mu\nu}^{I_{\mu}J_{\nu}}\;\rd X^{\mu}_{I_{\mu}}\wedge\rd X^{\nu}_{J_{\nu}}
$$
We perform an $SU(N)$-type decomposition $\rd X^{\mu}_I=(\rd x^{\mu},e^{\mu}_{M_{\mu}})$, where $M=1,2,...N^2-1$. For example; in three dimensions, we may have $\mu=(x,y,z)$ and $M_{\mu}=(M,\dot{M},M')$, so that the basis of one-forms looks like
$$
\{\rd X^{\mu}_{I_{\mu}}\}=\{\rd x,\rd y,\rd z, e^x_{M},e^y_{\dot{M}},e^z_{M'}\},
$$
where $M=1,...,N^2-1$, $\dot{M}=1,...,\dot{N}^2-1$ and $M'=1,...,N'^2-1$, so that the $\rd X^{\mu}_{I_{\mu}}$ span a $D=N^2+\dot{N}^2+N'^2$ dimensional vector space. Note that, if $N_{\mu}=N$ is the same for all $\mu$, then $\rd X^{\mu}_{I_{\mu}}$ may be thought of as a $d\times N^2$ matrix and so may be written as
$$
\rd X^{\mu}_{I_{\mu}}=\left(
\begin{array}{cccc}
\rd x^1 & \rd x^2 & ... & \rd x^d \\
e^1_1 & e^2_1 & ... & e^d_1 \\
\vdots & \vdots & & \\
e^1_{N^2-1} & e^2_{N^2-1} & ... & e^d_{N^2-1}
\end{array}
\right)
$$
For clarity of notation, we shall write $\rd X^{\mu}_{I_{\mu}}$ as $\rd X^{\mu}_I$ from now on, where it is to be understood that the $I$ index is tied to the $\mu$ index. The two-form may be written as
\begin{equation}
B=\frac{1}{2}b_{\mu\nu}\;\rd x^{\mu}\wedge\rd x^{\nu}+\frac{1}{2}C^M_{\mu\nu}\;e^{\mu}_M\wedge\rd x^{\nu}+\frac{1}{2}\widetilde{C}^M_{\nu\mu}\;\rd x^{\nu}\wedge e^{\mu}_M+\frac{1}{2}\Phi_{\mu\nu}^{MN}\;e_M^{\mu}\wedge e_N^{\nu}.
\end{equation}
Antisymmetry of the wedge product means that
$$
b_{\mu\nu}=-b_{\nu\mu},	\qquad	C^M_{\mu\nu}=-\widetilde{C}^M_{\nu\mu},	\qquad		\Phi^{MN}_{\mu\nu}=-\Phi^{NM}_{\nu\mu}.
$$
We shall assume that the components are independent of $e^{\mu}_M$ and only depend on the $\rd x^{\mu}$, and so the field strength is given by
\begin{equation}
\rd B=\frac{1}{2}\partial_{[\lambda}b_{\mu\nu]}\rd x^{\lambda}\wedge\rd x^{\mu}\wedge \rd x^{\nu}+\partial_{[\lambda|}C^M_{\mu|\nu]}\rd x^{\lambda}\wedge e^{\mu}_M\wedge\rd x^{\nu}+\frac{1}{2}\partial_{\lambda}\Phi^{MN}_{\mu\nu}\rd x^{\lambda}\wedge e^{\mu}_M\wedge e^{\nu}_N.
\end{equation}
We can write this as
$$
{\cal H}=H+F_{\mu}^M\wedge e^{\mu}_M+\frac{1}{2}G_{\mu\nu}^{MN}\wedge e^{\mu}_M\wedge e^{\nu}_N.
$$
The above construction may be related to the explicit field strength we had for the continuum limit of the lattice theory. As an example, let us take $D=3$ and the one-form basis is $\rd x,\rd y,\rd z, e^x_M,e^y_{\dot{M}},e^z_{M'}$, so that
\begin{eqnarray}
{\cal H}&=&H+F_x^M\wedge e^x_M+F_y^{\dot{M}}\wedge e^y_{\dot{M}}+F_z^{M'}\wedge e^z_{M'}\nonumber\\
&&+\frac{1}{2}G_{xy}^{M\dot{M}}\wedge e^x_M\wedge e^y_{\dot{M}}+\frac{1}{2}G_{yz}^{\dot{M}M'}\wedge e^y_{\dot{M}}\wedge e^z_{M'}+\frac{1}{2}G_{zx}^{M'M}\wedge e^z_{M'}\wedge e^x_M,
\end{eqnarray}
where no summation in $\{x,y,z\}$ is implied.

In general, we can write this field strength in terms of the matrix-valued coordinates
$$
{\cal H}=\frac{1}{6}({\cal H}_{\mu\nu\lambda})^{a\dot{a}a'}_{b\dot{b}b'}(\rd X^{\mu})^b_a\wedge (\rd X^{\nu})^{\dot{b}}_{\dot{a}}\wedge (\rd X^{\lambda})^{b'}_{a'}.
$$
Using an appropriate volume form, the action (\ref{cont}) can be constructed from this generalised form\footnote{To construct an action we need a metric so that we can raise/lower indices. We introduce the generalised metric $g:\rd X\times\rd X\rightarrow \R$ as
\begin{eqnarray}
\rd s^2&=&g_{\mu\nu}^{IJ}\rd X^{\mu}_I\otimes \rd X^{\nu}_J\nonumber\\
&=&g_{\mu\nu}\rd x^{\mu}\otimes\rd x^{\nu}+g^M_{\mu\nu}e_M^{\mu}\otimes \rd x^{\nu} + \tilde{g}_{\nu\mu}^M\rd x^{\mu}\otimes e^{\nu}_M+g^{MN}_{\mu\nu}e^{\mu}_M\otimes e^{\nu}_N,
\end{eqnarray}
where the components have the symmetries
$$
g_{\mu\nu}=g_{\nu\mu},	\qquad	g^M_{\mu\nu}=\tilde{g}^M_{\nu\mu},	\qquad		g^{MN}_{\mu\nu}=g^{NM}_{\nu\mu}.
$$
$g_{\mu\nu}$ is a conventional Reimannian metric and $g_{\mu\nu}^{MN}$ is something a little more akin to the non-geometric four-component tensor that appears in the (4,0) multiplet.}. Similar considerations may be given to one-forms $A=a_{\mu}\rd x^{\mu}+A_{\mu}^Me^{\mu}_M$, the result of which is a generalised two-form. Such two-forms play a role in the dimensional reduction of the theory.  It is interesting to see that something akin to non-commutative geometry plays a role in this construction and the resulting theory is not Lorentz-invariant, just like more conventional non-commutative space-times. Despite this lack of Lorentz invariance in the general theory, it should be stressed that it is still possible to find Lorentz-invariant truncations and solutions to the theory. We can efficiently truncate the theory to a Lorentz-invariant sub-sector by setting to zero all of the $e_M^{\mu}$ terms and keeping only those components that couple to the (singlet) $\rd x^{\mu}$'s.

\section{Expansion of the Classical Action}

In this appendix we provide details of the expansion of the classical action in powers of the lattice spacing.

\subsection{Quadratic Terms}

In this appendix, we use the variables $\theta_i$ defined in section 4.2. Pugging (\ref{W}) into (\ref{G}) and keeping terms only of second in $\theta$ gives (after a couple of pages of algebra)
$$
\Gamma_2=-\frac{N}{2}\sum_{i=1}^6(\theta_i,\theta_i)-N\sum_{i=1}^3(\theta_i,\theta_{i+3})-\sum_{i<j\neq i+3}\langle\theta_i,\theta_j\rangle
$$
Where we have introduced the inner products
$$
A^{a\dot{a}}_{b\dot{b}}B^{b\dot{b}}_{a\dot{a}}:=(A, B),	\qquad	A^{a\dot{c}}_{b\dot{c}}B^{b\dot{d}}_{a\dot{d}}:=\langle A, B\rangle.
$$
We start with the first two terms which may be written as
$$
-\frac{N}{2}\sum_{i=1}^6(\theta_i,\theta_i)-N\sum_{i=1}^3(\theta_i,\theta_{i+3})=-\frac{N}{2}\sum_{i=1}^3(\theta_i+\theta_{i+3},\theta_i+\theta_{i+3})
$$
The remaining terms are
$$
-\sum_{i<j\neq i+3}\langle\theta_i,\theta_j\rangle=-\sum_{i<j}\langle\theta_i+\theta_{i+3},\theta_j+\theta_{j+3}\rangle
$$
where $i,j=1,2,3$ in the sum on the right hand side.

Note that
$$
(\theta_1)^{a\dot{a}}_{b\dot{b}}+(\theta_4)^{a\dot{a}}_{b\dot{b}}=\alpha^3\Delta_{\mu}(B_{\nu\lambda})^{a\dot{a}}_{b\dot{b}}
$$
and similarly for other contributions, so $\Gamma_2$ depends only on derivatives of $B$. It is helpful to define
$$
h_i:=\theta_i+\theta_{i+3}, 	\qquad	i=1,2,3.
$$
We then see that
\begin{equation}\label{G2}
\Gamma_2=-\frac{N}{2}\sum_i(h_i,h_i)-\sum_{i<j}\langle h_i,h_j\rangle.
\end{equation}
Using the standard normalisation
$$
\text{Tr}(T_MT_N)=\frac{1}{2}\delta_{MN}
$$
it can be shown that, under the $SU(N)$ decomposition (\ref{decomp}), the quadratic part of the action (\ref{G2}) becomes
$$
\frac{\beta}{N^3}\sum_{\mbox{\mancube}}{\cal V}_2=-\frac{\beta}{6}\sum_{\textbf{n}}\alpha^6\sum_{1\leq \mu<\nu<\lambda\leq 6}\left(\frac{1}{2}(H_{\mu\nu\lambda})^2+\frac{1}{4N}\sum_{i=1}^3(F^i_{\mu\nu\lambda})^2+\frac{1}{8N^2}\sum_{i=1}^3(G^i_{\mu\nu\lambda})^2\right)
$$
where the components are given is section 5.1.

\subsection{Cubic Terms}

Given two face variables, we can construct two products that define a map from two face variables to a new face variable, as
$$
(\theta_i\circ\theta_j)^{a\dot{a}}_{b\dot{b}}:=(\theta_i)^{a\dot{a}}_{c\dot{c}}(\theta_j)^{c\dot{c}}_{b\dot{b}}	\qquad	(\theta_i*\theta_j)^{\dot{a}a'}_{\dot{b}b'}:=(\theta_i)^{a\dot{a}}_{b\dot{b}}(\theta_j)^{a'b}_{b'a}
$$
The $\circ$ product can be used to combine two face variables that live in the same plane (up to lattice translation), whilst $*$ can be used to combine two faces that lie in intersecting planes. Note that
$$
\text{Tr}_2\;\theta_i\circ\theta_j=(\theta_i,\theta_j),	\qquad	\text{Tr}_2\;\theta_i*\theta_j=\langle\theta_i,\theta_j\rangle\;.
$$
The first of the novel non-abelian interaction terms appears at order $\theta^3$. $\Gamma_3$ includes terms of the form
$$
(\theta_i\circ\theta_j,\theta_k)=(\theta_i,\theta_j\circ\theta_k)=(\theta_i)^{a\dot{a}}_{b\dot{b}}(\theta_j)^{b\dot{b}}_{c\dot{c}}(\theta_k)^{c\dot{c}}_{a\dot{a}},	\qquad	\langle\theta_i\circ\theta_j,\theta_k\rangle=(\theta_i)^{a\dot{a}}_{e\dot{e}}(\theta_j)^{e\dot{e}}_{b\dot{a}}(\theta_k)^{e'b}_{e'a},
$$
$$
(\theta_i,\theta_j*\theta_k)=(\theta_i*\theta_j,\theta_k)=(\theta_i)^{a\dot{a}}_{b\dot{b}}(\theta_j)^{a'b}_{b'a}(\theta_i)^{\dot{b}b'}_{\dot{a}a'},
$$
Performing the $SU(N)$ decomposition of the face variables (\ref{decomp}) and using the results
$$
\text{Tr}(T_MT_NT_P)=\frac{1}{4}(if_{MNP}+d_{MNP}),	\qquad	\text{Tr}(T_MT_N)=\frac{1}{2}\delta_{MN},
$$
where $d_{MNP}$ s totally symmetric and $f_{MNP}$ is totally antisymmetric. The imaginary parts of these terms may be written as
\begin{eqnarray}
\Im(\theta_i\circ\theta_j,\theta_k)&=&\frac{N}{4}f_{MNP}\theta_i^M\theta_j^N\theta^P_k+\frac{N}{4}f_{\dot{M}\dot{N}\dot{P}}\theta^{\dot{M}}_i\theta^{\dot{N}}_j\theta^{\dot{P}}_k
\nonumber\\
&&-\frac{1}{16}(f_{MNP}\,d_{\dot{M}\dot{N}\dot{P}}+d_{MNP}\,f_{\dot{M}\dot{N}\dot{P}})\theta^{M\dot{M}}_i\theta^{N\dot{N}}_j\theta^{P\dot{P}}_k\;,\nonumber
\end{eqnarray}
$$
\Im\langle\theta_i\circ\theta_j,\theta_k\rangle=\frac{N^2}{4}f_{MNP}\theta_i^M\theta_j^N\tilde{\theta}^P_k+\frac{N}{8}\delta_{\dot{M}\dot{N}}f_{MNP}\theta^{M\dot{M}}_i\theta^{N\dot{N}}_j\tilde{\theta}^P_k\;,
$$
$$
\Im(\theta_i*\theta_j,\theta_k)=0
$$
With a little work, the real part of the cubic term may be written as ${\cal V}_3=V_3+\widetilde{V}_3$ where
$$
V_3=\frac{N^2}{2}\alpha^7\left(f_{MNP}K^{MNP}+f_{\dot{M}\dot{N}\dot{P}}K^{\dot{M}\dot{N}\dot{P}}+f_{M'N'P'}K^{M'N'P'}\right)
$$
where
\begin{eqnarray}\label{K1}
K^{MNP}&=&C_{\nu\lambda}^M\Delta_{\mu}C^N_{\nu\lambda}\left(C^P_{\mu\lambda}+\frac{\alpha}{2}\Delta_{\nu}C^P_{\mu\lambda}\right)+C_{\mu\lambda}^M\Delta_{\nu}C^N_{\mu\lambda}\left(C^P_{\nu\lambda}+\frac{\alpha}{2}\Delta_{\mu}C^P_{\nu\lambda}\right)\nonumber\\
K^{\dot{M}\dot{N}\dot{P}}&=&-\widetilde{C}_{\nu\lambda}^{\dot{M}}\Delta_{\mu}\widetilde{C}^{\dot{N}}_{\nu\lambda}\left(C^{\dot{P}}_{\mu\nu}+\frac{\alpha}{2}\Delta_{\lambda}C^{\dot{P}}_{\mu\nu}\right)+C_{\mu\nu}^{\dot{M}}\Delta_{\lambda}C^{\dot{N}}_{\mu\nu}\left(\widetilde{C}^{\dot{P}}_{\nu\lambda}+\frac{\alpha}{2}\Delta_{\mu}\widetilde{C}^{\dot{P}}_{\nu\lambda}\right)\nonumber\\
K^{M'N'P'}&=&-C_{\lambda\mu}^{M'}\Delta_{\nu}C^{N'}_{\lambda\mu}\left(\widetilde{C}^{P'}_{\mu\nu}+\frac{\alpha}{2}\Delta_{\lambda}\widetilde{C}^{P'}_{\mu\nu}\right)-C_{\nu\mu}^{M'}\Delta_{\lambda}C^{N'}_{\nu\lambda}\left(\widetilde{C}^{P'}_{\mu\lambda}+\frac{\alpha}{2}\Delta_{\nu}\widetilde{C}^{P'}_{\mu\lambda}\right)\nonumber\\
\end{eqnarray}
Similarly
$$
\widetilde{V}_3=\frac{N}{4}\alpha^7\left(f_{MNP}\widetilde{K}^{MNP}+f_{\dot{M}\dot{N}\dot{P}}\widetilde{K}^{\dot{M}\dot{N}\dot{P}}+f_{M'N'P'}\widetilde{K}^{M'N'P'}\right)
$$
where
\begin{eqnarray}\label{K2}
\widetilde{K}^{MNP}&=&\delta_{\dot{M}\dot{N}}\Phi_{\nu\lambda}^{M\dot{M}}\Delta_{\mu}\Phi^{N\dot{N}}_{\nu\lambda}\left(C^P_{\mu\lambda}+\frac{\alpha}{2}\Delta_{\nu}C^P_{\mu\lambda}\right)+\delta_{M'N'}\Phi_{\mu\lambda}^{MM'}\Delta_{\nu}\Phi^{NN'}_{\mu\lambda}\left(C^P_{\nu\lambda}+\frac{\alpha}{2}\Delta_{\mu}C^P_{\nu\lambda}\right)\nonumber\\
K^{\dot{M}\dot{N}\dot{P}}&=&-\delta_{MN}\Phi_{\nu\lambda}^{M\dot{M}}\Delta_{\mu}\Phi^{N\dot{N}}_{\nu\lambda}\left(C^{\dot{P}}_{\mu\nu}+\frac{\alpha}{2}\Delta_{\lambda}C^{\dot{P}}_{\mu\nu}\right)+\delta_{M'N'}\Phi_{\mu\nu}^{\dot{M}M'}\Delta_{\lambda}\Phi^{\dot{N}N'}_{\mu\nu}\left(\widetilde{C}^{\dot{P}}_{\nu\lambda}+\frac{\alpha}{2}\Delta_{\mu}\widetilde{C}^{\dot{P}}_{\nu\lambda}\right)\nonumber\\
K^{M'N'P'}&=&-\delta_{MN}\Phi_{\lambda\mu}^{M'M}\Delta_{\nu}\Phi^{N'N}_{\lambda\mu}\left(\widetilde{C}^{P'}_{\mu\nu}+\frac{\alpha}{2}\Delta_{\lambda}\widetilde{C}^{P'}_{\mu\nu}\right)-\delta_{\dot{M}\dot{N}}\Phi_{\nu\mu}^{\dot{M}M'}\Delta_{\lambda}\Phi^{\dot{N}N'}_{\nu\lambda}\left(\widetilde{C}^{P'}_{\mu\lambda}+\frac{\alpha}{2}\Delta_{\nu}\widetilde{C}^{P'}_{\mu\lambda}\right)\nonumber\\
\end{eqnarray}
The cubic part of the action may then be written as
$$
\frac{\beta}{N^3}\sum_{\mbox{\mancube}}{\cal V}_3=\frac{\beta}{6N^3}\sum_{\textbf{n}}\sum_{1\leq \mu<\nu<\lambda\leq 6}\left(V_3+\widetilde{V}_3\right).
$$

\section{Dimensional Reduction of Cubic Terms}

We introduce the redefinition $\alpha^2B_{\mu\lambda}=\alpha A_{\mu}$. The cubic terms (\ref{K1}) and (\ref{K2}) become

\begin{eqnarray}
\alpha^3 K^{MNP}&=&A_{\nu}^M\Delta_{\mu}A^N_{\nu}\left(A^P_{\mu}+\frac{\alpha}{2}\Delta_{\nu}A^P_{\mu}\right)+A_{\mu}^M\Delta_{\nu}A^N_{\mu}\left(A^P_{\nu}+\frac{\alpha}{2}\Delta_{\mu}A^P_{\nu}\right)\nonumber\\
\alpha^2 K^{\dot{M}\dot{N}\dot{P}}&=&-\widetilde{A}_{\nu}^{\dot{M}}\Delta_{\mu}\widetilde{A}^{\dot{N}}_{\nu}C^{\dot{P}}_{\mu\nu}\nonumber\\
\alpha^2 K^{M'N'P'}&=&-\widetilde{A}_{\mu}^{M'}\Delta_{\nu}\widetilde{A}^{N'}_{\mu}\widetilde{C}^{P'}_{\mu\nu}\nonumber
\end{eqnarray}
and
\begin{eqnarray}
\alpha^3 \widetilde{K}^{MNP}&=&\delta_{\dot{M}\dot{N}}\phi_{\nu}^{M\dot{M}}\Delta_{\mu}\phi^{N\dot{N}}_{\nu}\left(A^P_{\mu}+\frac{\alpha}{2}\Delta_{\nu}A^P_{\mu}\right)+\delta_{M'N'}\phi_{\mu}^{MM'}\Delta_{\nu}\phi^{NN'}_{\mu}\left(A^P_{\nu}+\frac{\alpha}{2}\Delta_{\mu}A^P_{\nu}\right)\nonumber\\
\alpha^2 K^{\dot{M}\dot{N}\dot{P}}&=&-\delta_{MN}\phi_{\nu}^{M\dot{M}}\Delta_{\mu}\phi^{N\dot{N}}_{\nu}C^{\dot{P}}_{\mu\nu}\nonumber\\
\alpha^2 K^{M'N'P'}&=&-\delta_{MN}\phi_{\mu}^{MM'}\Delta_{\nu}\phi^{NN'}_{\mu}\widetilde{C}^{P'}_{\mu\nu}\nonumber
\end{eqnarray}
where we have assumed that the fields do not depend on the $\lambda$-direction. We note that $K_{MNP}$ gives \emph{precisely} the cubic term in lattice Yang-Mills theory. We rescale the coupling and take the limit
$$
{\cal V}^c_3:=\lim_{\alpha\rightarrow 0}:\frac{1}{\alpha^4}{\cal V}_3
$$
whilst taking the number of lattice points to infinity (the classical continuum limit). In this limit the only terms that survive are
\begin{eqnarray}
{\cal V}^c_3&=&\frac{N^2}{2}f_{MNP}\left(A_{\nu}^M\partial_{\mu}A^N_{\nu}A^P_{\mu}+A_{\mu}^M\partial_{\nu}A^N_{\mu}A^P_{\nu}\right)\nonumber\\
&&+\frac{N}{4}f_{MNP}\left(\delta_{\dot{M}\dot{N}}\phi_{\nu}^{M\dot{M}}\partial_{\mu}\phi^{N\dot{N}}_{\nu}A^P_{\mu}+\delta_{M'N'}\phi_{\mu}^{MM'}\partial_{\nu}\phi^{NN'}_{\mu}A^P_{\nu}\right)\nonumber
\end{eqnarray}
The terms in the first line give the cubic contribution to Yang-Mills, whilst the second line gives the intercation term arising from the covariant coupling of $\phi$ to the Yang-Mills connection $A$.

\section{Gauge Fixing}\label{gaugefix}

Gauge fixing is not needed for lattice theories as the integral over all gauges does not contain any divergences so, in contrast to the continuum limit, the quantum theory exists even without gauge fixing. However, a judicous gauge choice can simplify calculations. To this end, we introduce a gauge fixing techique for lattice gerbe theory.

Let $P(\cal W)$ be some polynomial in the face variables. The expectation
value is
\[
\left\langle P({\cal W})\right\rangle=\frac{1}{{\cal Z}}\int {\cal D}{\cal W}\;P({\cal W})\;e^{-S[{\cal W}]}
\]
where the integral is over all face variables in the lattice. The measure is defined as
$$
{\cal D}{\cal W}=\prod_{\Box}\rd {\cal W}_{\Box}
$$
denoting a product over all plaquettes in the latice and ${\cal Z}$ is the partition function such that $\left\langle 1 \right\rangle=1$. We will demonstrate that the expectation value is unchanged if we do not
integrate over one of the face variables. To do so, define a delta
functional on the face variable
\[
\int \rd{\cal W}\;\delta ({\cal W},\omega)\;f({\cal W})=f(\omega),
\]
where the integral is over the value of a single face variable. Suppose
that we would like to compute the expectation value of $P$ while
holding the face variable ${\cal W}_{\mu\nu}\left(\vec{n}\right)$ fixed
to the value $\omega$. This is given by 
\[
I(P,\omega)={\cal Z}^{-1}\int {\cal D}{\cal W}\;\delta\left({\cal W},\omega\right)\;e^{-S[\cal W]}\;P({\cal W}).
\]
If we integrate $I(P,\omega)$ over $\omega$, this gives the original expectation
value 
\begin{equation}
\left\langle P\right\rangle=\int \rd \omega\;I(\omega,P).\label{eq:pi}
\end{equation}
Since $I(P,\omega)$ is invariant under gauge transformations. It follows that $I(P,\omega)$ is actually independent of $\omega$, so
(\ref{eq:pi}) implies that 
\[
I(P,\omega)=\left\langle P\right\rangle.
\]
This process can be repeated to fix more face variables. In particular,
we can fix any set of face variables as long as this set does not
contain a closed surface, since a closed surface is a gauge invariant
object, so it cannot be set to an arbitrary value using gauge transformations.


\begin{thebibliography}{99}

\bibitem{Perry:1996mk} 
  M.~Perry and J.~H.~Schwarz,
  ``Interacting chiral gauge fields in six-dimensions and Born-Infeld theory,''
  Nucl.\ Phys.\ B {\bf 489}, 47 (1997)
  [hep-th/9611065].

\bibitem{Pasti:1997gx} 
  P.~Pasti, D.~P.~Sorokin and M.~Tonin,
  ``Covariant action for a D = 11 five-brane with the chiral field,''
  Phys.\ Lett.\ B {\bf 398}, 41 (1997)
  [hep-th/9701037].

\bibitem{Aharony:2008ug} 
  O.~Aharony, O.~Bergman, D.~L.~Jafferis and J.~Maldacena,
  ``N=6 superconformal Chern-Simons-matter theories, M2-branes and their gravity duals,''
  JHEP {\bf 0810}, 091 (2008)
  [arXiv:0806.1218 [hep-th]].

\bibitem{ArkaniHamed:2001ie} 
  N.~Arkani-Hamed, A.~G.~Cohen, D.~B.~Kaplan, A.~Karch and L.~Motl,
  ``Deconstructing (2,0) and little string theories,''  JHEP {\bf 0301}, 083 (2003)  [hep-th/0110146].  

\bibitem{Lambert:2010wm} 
  N.~Lambert and C.~Papageorgakis,
  ``Nonabelian (2,0) Tensor Multiplets and 3-algebras,''
  JHEP {\bf 1008}, 083 (2010)
  [arXiv:1007.2982 [hep-th]].

\bibitem{Douglas:2010iu} 
  M.~R.~Douglas,
  ``On D=5 super Yang-Mills theory and (2,0) theory,''
  JHEP {\bf 1102}, 011 (2011)
  [arXiv:1012.2880 [hep-th]].

\bibitem{Lambert:2010iw} 
  N.~Lambert, C.~Papageorgakis and M.~Schmidt-Sommerfeld,
  ``M5-Branes, D4-Branes and Quantum 5D super-Yang-Mills,''
  JHEP {\bf 1101}, 083 (2011)
  [arXiv:1012.2882 [hep-th]].

\bibitem{Ho:2011ni} 
  P.~-M.~Ho, K.~-W.~Huang and Y.~Matsuo,
  ``A Non-Abelian Self-Dual Gauge Theory in 5+1 Dimensions,''
  JHEP {\bf 1107}, 021 (2011)
  [arXiv:1104.4040 [hep-th]].

\bibitem{Chu:2012um} 
  C.~-S.~Chu and S.~-L.~Ko,
  ``Non-abelian Action for Multiple Five-Branes with Self-Dual Tensors,''
  JHEP {\bf 1205}, 028 (2012)
  [arXiv:1203.4224 [hep-th]].

\bibitem{Samtleben:2012mi} 
  H.~Samtleben, E.~Sezgin, R.~Wimmer and L.~Wulff,
  ``New superconformal models in six dimensions: Gauge group and representation structure,''
  PoS CORFU {\bf 2011}, 071 (2011)
  [arXiv:1204.0542 [hep-th]].

\bibitem{Saemann:2012uq} 
  C.~Saemann and M.~Wolf,
  ``Non-Abelian Tensor Multiplet Equations from Twistor Space,''
  arXiv:1205.3108 [hep-th].

\bibitem{Bonetti:2012st} 
  F.~Bonetti, T.~W.~Grimm and S.~Hohenegger,
  ``Non-Abelian Tensor Towers and (2,0) Superconformal Theories,''
  JHEP {\bf 1305}, 129 (2013)
  [arXiv:1209.3017 [hep-th]].

\bibitem{Kim:2012tr} 
  H.~-C.~Kim and K.~Lee,
  ``Supersymmetric M5 Brane Theories on R x CP2,''
  JHEP {\bf 1307}, 072 (2013)
  [arXiv:1210.0853 [hep-th]].

\bibitem{Samtleben:2012fb} 
  H.~Samtleben, E.~Sezgin and R.~Wimmer,
  ``Six-dimensional superconformal couplings of non-abelian tensor and hypermultiplets,''
  JHEP {\bf 1303}, 068 (2013)
  [arXiv:1212.5199 [hep-th]].

\bibitem{Saemann:2013pca} 
  C.~Saemann and M.~Wolf,
  ``Six-Dimensional Superconformal Field Theories from Principal 3-Bundles over Twistor Space,''
  arXiv:1305.4870 [hep-th].

\bibitem{Tachikawa:2011ch} 
  Y.~Tachikawa,
  ``On S-duality of 5d super Yang-Mills on $S^1$,''
  JHEP {\bf 1111}, 123 (2011)
  [arXiv:1110.0531 [hep-th]].

\bibitem{Kim:2011mv} 
  H.~-C.~Kim, S.~Kim, E.~Koh, K.~Lee and S.~Lee,
  ``On instantons as Kaluza-Klein modes of M5-branes,''
  JHEP {\bf 1112}, 031 (2011)
  [arXiv:1110.2175 [hep-th]].

\bibitem{Young:2011aa} 
  D.~Young,
  ``Wilson Loops in Five-Dimensional Super-Yang-Mills,''
  JHEP {\bf 1202}, 052 (2012)
  [arXiv:1112.3309 [hep-th]].

\bibitem{Kim:2012ava} 
  H.~-C.~Kim and S.~Kim,
  ``M5-branes from gauge theories on the 5-sphere,''
  JHEP {\bf 1305}, 144 (2013)
  [arXiv:1206.6339 [hep-th]].

\bibitem{Kallen:2012zn} 
  J.~Kallen, J.~A.~Minahan, A.~Nedelin and M.~Zabzine,
  ``$N^3$-behavior from 5D Yang-Mills theory,''
  JHEP {\bf 1210}, 184 (2012)
  [arXiv:1207.3763 [hep-th]].

\bibitem{Kim:2012qf} 
  H.~-C.~Kim, J.~Kim and S.~Kim,
  ``Instantons on the 5-sphere and M5-branes,''
  arXiv:1211.0144 [hep-th].

\bibitem{Bern:2012di} 
  Z.~Bern, J.~J.~Carrasco, L.~J.~Dixon, M.~R.~Douglas, M.~von Hippel and H.~Johansson,
  ``D = 5 maximally supersymmetric Yang-Mills theory diverges at six loops,''
  Phys.\ Rev.\ D {\bf 87}, 025018 (2013)
  [arXiv:1210.7709 [hep-th]].

\bibitem{Papageorgakis:2014dma} 
  C.~Papageorgakis and A.~B.~Royston,
  ``Revisiting Soliton Contributions to Perturbative Amplitudes,''
  arXiv:1404.0016 [hep-th].

\bibitem{Gaiotto:2009we} 
  D.~Gaiotto,
  ``N=2 dualities,''
  JHEP {\bf 1208}, 034 (2012)
  [arXiv:0904.2715 [hep-th]].

\bibitem{Mandelstam:1974pi} 
  S.~Mandelstam,
  ``Vortices and Quark Confinement in Nonabelian Gauge Theories,''
  Phys.\ Rept.\  {\bf 23}, 245 (1976).

\bibitem{Polyakov:1975rs} 
  A.~M.~Polyakov,
  ``Compact Gauge Fields and the Infrared Catastrophe,''
  Phys.\ Lett.\ B {\bf 59}, 82 (1975).

\bibitem{Polyakov:1976fu} 
  A.~M.~Polyakov,
  ``Quark Confinement and Topology of Gauge Groups,''
  Nucl.\ Phys.\ B {\bf 120}, 429 (1977).

\bibitem{'tHooft:1977hy} 
  G.~'t Hooft,
  ``On the Phase Transition Towards Permanent Quark Confinement,''
  Nucl.\ Phys.\ B {\bf 138}, 1 (1978).

\bibitem{'tHooft:1981ht} 
  G.~'t Hooft,
  ``Topology of the Gauge Condition and New Confinement Phases in Nonabelian Gauge Theories,''
  Nucl.\ Phys.\ B {\bf 190}, 455 (1981).

\bibitem{Seiberg:1994rs} 
  N.~Seiberg and E.~Witten,
  ``Electric - magnetic duality, monopole condensation, and confinement in N=2 supersymmetric Yang-Mills theory,''
  Nucl.\ Phys.\ B {\bf 426}, 19 (1994)
  [Erratum-ibid.\ B {\bf 430}, 485 (1994)]
  [hep-th/9407087].

\bibitem{Wilson:1974sk} 
  K.~G.~Wilson,
  ``Confinement of Quarks,''
  Phys.\ Rev.\ D {\bf 10}, 2445 (1974).

\bibitem{Frohlich:1982gf} 
  J.~Frohlich and T.~Spencer,
  ``Massless Phases And Symmetry Restoration In Abelian Gauge Theories And Spin Systems,''
  Commun.\ Math.\ Phys.\  {\bf 83}, 411 (1982).

\bibitem{Omero:1982hp} 
  C.~Omero, P.~A.~Marchetti and A.~Maritan,
  ``Gauge Differential Form Theories On The Lattice,''
  J.\ Phys.\ A {\bf 16}, 1465 (1983).

\bibitem{Creutz} 
M. Creutz, "Quarks, gluons, and lattices. Cambridge," Uk: Univ. Pr. ( 1983) 169 P. ( Cambridge Monographs On Mathematical Physics).

\bibitem{Rothe}
H.J. Rothe, "Lattice gauge theories: an introduction," World Sci.Lect.Notes Phys., 74:1–605, 2005.

\bibitem{Nepomechie:1982rb} 
  R.~I.~Nepomechie,
  ``Approaches To A Nonabelian Antisymmetric Tensor Gauge Field Theory,''
  Nucl.\ Phys.\ B {\bf 212}, 301 (1983).

\bibitem{Orland:1982fv} 
  P.~Orland,
  ``Frustrating Lattice Qcd,''
  Phys.\ Lett.\ B {\bf 122}, 78 (1983).

\bibitem{Orland:1984pt} 
  P.~Orland,
  ``Disorder, Frustration And Semiclassical Calculations In Lattice Gauge Theories,''
  Imperial/TP/83-84/49.

\bibitem{Orland:1984bi} 
  P.~Orland,
  ``Frustration And Dual Superconductivity In Lattice Gauge Theories,''
  In *Argonne 1984, Proceedings, Gauge Theory On A Lattice: 1984*, 305-310

\bibitem{Rey:2010uz}
  S.~-J.~Rey and F.~Sugino,
  ``A Nonperturbative Proposal for Nonabelian Tensor Gauge Theory and Dynamical Quantum Yang-Baxter Maps,''
  arXiv:1002.4636 [hep-th].

\bibitem{Orland:1981ku} 
  P.~Orland,
  ``Instantons and Disorder in Antisymmetric Tensor Gauge Fields,''
  Nucl.\ Phys.\ B {\bf 205}, 107 (1982).

\bibitem{Orland:1994qt} 
  P.~Orland,
  ``Extrinsic curvature dependence of Nielsen-Olesen strings,''
  Nucl.\ Phys.\ B {\bf 428}, 221 (1994)
  [hep-th/9404140].

\bibitem{Quevedo:1996uu} 
  F.~Quevedo and C.~A.~Trugenberger,
  ``Phases of antisymmetric tensor field theories,''
  Nucl.\ Phys.\ B {\bf 501}, 143 (1997)
  [hep-th/9604196].

\bibitem{Polyakov:1996nc} 
  A.~M.~Polyakov,
  ``Confining strings,''
  Nucl.\ Phys.\ B {\bf 486}, 23 (1997)
  [hep-th/9607049].

\bibitem{Diamantini:1996vf} 
  M.~C.~Diamantini, F.~Quevedo and C.~A.~Trugenberger,
  ``Confining string with topological term,''
  Phys.\ Lett.\ B {\bf 396}, 115 (1997)
  [hep-th/9612103].

\bibitem{Hitchin:1999fh}
  N.~J.~Hitchin,
  ``Lectures on special Lagrangian submanifolds,''
  math/9907034.

\bibitem{Henneaux:1986ht}
  M.~Henneaux and C.~Teitelboim,
  ``P Form Electrodynamics,''
  Found.\ Phys.\  {\bf 16} (1986) 593.

\bibitem{Kapustin:2006pk} 
  A.~Kapustin and E.~Witten,
  ``Electric-Magnetic Duality And The Geometric Langlands Program,''
  Commun.\ Num.\ Theor.\ Phys.\  {\bf 1}, 1 (2007)
  [hep-th/0604151].

\bibitem{Witten:2009at} 
  E.~Witten,
  ``Geometric Langlands From Six Dimensions,''
  arXiv:0905.2720 [hep-th].

\bibitem{Gustavsson:2004gj} 
  A.~Gustavsson,
  ``Conformal anomaly of Wilson surface observables: A Field theoretical computation,''
  JHEP {\bf 0407}, 074 (2004)
  [hep-th/0404150].

\bibitem{creutz2} 
M. Creutz. "Monte Carlo Study of Quantized SU(2) Gauge Theory." Phys. Rev., D21:2308–2315, 1980.

\bibitem{metropolis}
N. Metropolis, A.W. Rosenbluth, M.N. Rosenbluth, A.H. Teller, and E. Teller. "Equation of state calculations by fast computing machines." J.Chem.Phys., 21:1087–1092, 1953. 

\bibitem{creutzrebbi}
Michael Creutz, Laurence Jacobs, and Claudio Rebbi. "Monte Carlo Study of Abelian Lattice Gauge Theories." Phys. Rev., D20:1915, 1979

\bibitem{Catterall:2005fd} 
  S.~Catterall,
  ``Lattice formulation of N=4 super Yang-Mills theory,''  JHEP {\bf 0506}, 027 (2005)  [hep-lat/0503036].  

\bibitem{Kaplan:2005ta} 
  D.~B.~Kaplan and M.~Unsal,
  ``A Euclidean lattice construction of supersymmetric Yang-Mills theories with sixteen supercharges,''
  JHEP {\bf 0509}, 042 (2005)
  [hep-lat/0503039].

\bibitem{Catterall:2014vka} 
  S.~Catterall, D.~Schaich, P.~H.~Damgaard, T.~DeGrand and J.~Giedt,
  ``N=4 Supersymmetry on a Space-Time Lattice,''
  arXiv:1405.0644 [hep-lat].

\bibitem{Strominger:1995ac} 
  A.~Strominger,
  ``Open p-branes,''
  Phys.\ Lett.\ B {\bf 383}, 44 (1996)
  [hep-th/9512059].

\bibitem{Berenstein:1998ij} 
  D.~E.~Berenstein, R.~Corrado, W.~Fischler and J.~M.~Maldacena,
  ``The Operator product expansion for Wilson loops and surfaces in the large N limit,''
  Phys.\ Rev.\ D {\bf 59}, 105023 (1999)
  [hep-th/9809188].

\end{thebibliography}
\end{document}